\documentclass[prb,twocolumn,notitlepage,superscriptaddress,longbibliography]{revtex4-2}
\usepackage{amsmath,amsfonts,amssymb,amsthm,epsfig,array}
\usepackage{dsfont}
\usepackage{slashed}
\usepackage{graphics}
\usepackage{float}
\usepackage{verbatim}
\usepackage{color}
\usepackage[dvipsnames]{xcolor}
\usepackage[hidelinks,colorlinks,linkcolor=blue,
citecolor=blue,urlcolor=blue]{hyperref}
\usepackage[titletoc,title]{appendix}
\usepackage{multirow}
\usepackage{bibentry}

\newcommand{\bsl}[1]{\boldsymbol{#1}}

\newcommand{\ii}{\mathrm{i}}

\newcommand{\bra}[1]{\langle #1|}
\newcommand{\ket}[1]{|#1 \rangle}

\newcommand{\Tr}{\mathop{\mathrm{Tr}}}
\renewcommand{\Re}{\mathop{\mathrm{Re}}}
\renewcommand{\Im}{\mathop{\mathrm{Im}}}

\newcommand{\eqnref}[1]{Eq.\,\eqref{#1}}
\newcommand{\figref}[1]{Fig.\,\ref{#1}}

\newcommand{\secref}[1]{Sec.\,\ref{#1}}
\newcommand{\appref}[1]{Appendix.\,\ref{#1}}
\newcommand{\refcite}[1]{Ref.\,[\onlinecite{#1}]}

\newcommand{\mat}[1]{\left(\begin{matrix}#1\end{matrix}\right)}

\newcommand{\eq}[1]{\begin{equation} #1 \end{equation}}

\newcommand{\eqa}[1]{\begin{align}\begin{split} #1 \end{split}\end{align}}
\let\oldAA\AA
\renewcommand{\AA}{\text{\normalfont\oldAA}}
\newcommand{\sgn}[1]{\text{sgn}(#1)}
\newcommand{\ie}{{\emph{i.e.}}}
\newcommand{\eg}{{\emph{e.g.}}}
\newcommand{\TR}{\mathcal{T}}
\newcommand{\cc}{\mathcal{K}}
\usepackage[mathscr]{euscript}

\begin{document}
\title{Dynamical Piezomagnetic Effect in Time-Reversal-Invariant Weyl Semimetals with Axionic Charge-Density Waves}
\author{Jiabin Yu}
\affiliation{Department of Physics, the Pennsylvania State University, University Park, PA 16802}
\affiliation{Condensed Matter Theory Center, Department of Physics, University of Maryland, College Park, Maryland 20742, USA}
\author{Benjamin J. Wieder}
\affiliation{Department of Physics, Massachusetts Institute of Technology, Cambridge, MA 02139, USA}
\affiliation{Department of Physics, Northeastern University, Boston, MA 02115, USA}
\affiliation{Department of Physics, Princeton University, Princeton, New Jersey 08544, USA}
\author{Chao-Xing Liu}
\email{cxl56@psu.edu}
\affiliation{Department of Physics, the Pennsylvania State University, University Park, PA 16802}
\begin{abstract}
Charge-density waves (CDWs) in Weyl semimetals (WSMs) have been shown to induce an exotic axionic insulating phase in which the sliding mode (phason) of the CDW acts as a dynamical axion field, giving rise to a large positive magneto-conductance [Wang et al., Phys. Rev. B 87, 161107(R) (2013); Roy et al., Phys. Rev. B 92, 125141 (2015); J. Gooth et al., Nature 575, 315 (2019)].
In this work, we predict that dynamical strain can induce a bulk orbital magnetization in time-reversal- (TR-) invariant WSMs that are gapped by a CDW.
We term this effect the ``dynamical piezomagnetic effect" (DPME).
Unlike in [J. Gooth et al., Nature 575, 315 (2019)], the DPME introduced in this work occurs in a bulk-constant (\ie, static and spatially homogeneous in the bulk) CDW, and does not rely on fluctuations, such as a phason.
By studying the low-energy effective theory and a minimal tight-binding (TB) model, we find that the DPME originates from an effective valley axion field that couples the electromagnetic gauge field with a strain-induced pseudo-gauge field.
In particular, whereas the piezoelectric effects studied in previous works are characterized by 2D Berry curvature, the DPME represents the first example of a \emph{fundamentally 3D} strain effect originating from the Chern-Simons 3-form.
We further find that the DPME has a discontinuous change at a critical value of the phase of the CDW order parameter.
We demonstrate that, when there is a jump in the DPME, the surface of the system undergoes a topological quantum phase transition (TQPT), while the bulk remains gapped.
Hence, the DPME provides a bulk signature of the boundary TQPT in a TR-invariant Weyl-CDW.
\end{abstract}
\maketitle

\tableofcontents

\section{Introduction}
\label{sec:intro}

The last four decades have witnessed a paradigm shift in condensed matter physics driven by the discovery of the geometric phase and topology of electronic wave functions~\cite{Hasan2010TI,Qi2010TITSC}.
The search for experimentally observable response signatures of bulk nontrivial topology has emerged as central to the advancement of solid-state physics and material science.
In two-dimensional (2D) systems, the Berry curvature in the momentum space not only provides an essential contribution towards quantized topological response effects but also provides non-negligible contributions to various non-quantized physical phenomena.
As an example of a quantized topological response, integrating the Berry curvature over the 2D (magnetic) first Brillouin zone (1BZ) has been shown to yield a topological invariant called the first Chern number, which determines the quantized Hall conductance in the quantum Hall effect or the quantum anomalous Hall effect (QAHE)~\cite{Klitzing1980IQHE,TKNN, Haldane1988QAH, Chang2013QAH}.
A non-quantized example is the intrinsic anomalous Hall effect in a variety of ferromagnetic metals, which is given by the integration of the Berry curvature over the occupied states~\cite{Nagaosa2010AHE}.
Other non-quantized response effects include the expression for orbital magnetic moments in terms of the Berry curvature~\cite{Xiao2005BerryDOS,Xiao2006BerryATET,Xiao2010BerryRMP} and the nonlinear Hall effect induced from a Berry curvature dipole~\cite{Sodemann2015BCDipole,Zhang2018BCDipoleWSM,Xu2018BCDipoleWTe2,Ma2019NonlinearHall,
Kang2019NonlinearAHEWTe2}.
More recently, it has been demonstrated that the piezoelectric response can also be related to the Berry curvature~\cite{Martin1972Piezo,Vanderbilt2000BPPZ,Vaezi2013StrainGraphene,Droth2016PETBN,Rostami2018PiE}, and can have a discontinuous change across a TQPT in 2D TR-invariant systems~\cite{Yu2019PETTQPT}.
A natural question to ask is whether a strain-induced response can be related to the topological property of three-dimensional (3D) systems.

In 3D systems, a celebrated topological response originates from effective axion electrodynamics through the action~\cite{Qi2008TFT,Essin2009AI}
\eq{
\label{eq:S_theta}
S_{eff,\theta} = \frac{e^2}{32\pi^2} \int dt d^3 r  \theta \varepsilon^{\mu\nu\rho\delta}  F_{\mu\nu}F_{\rho\delta}\ ,
}
where (and for the remainder of this work unless specified otherwise) we choose units in which $\hbar=c=1$, and notation in which duplicated indices are summed over for notational simplicity.
In \eqnref{eq:S_theta}, $F_{\mu\nu}=\partial_\mu A_\nu-\partial_\nu A_\mu$ is the field strength of the electromagnetic $U(1)$ gauge field $A_\mu$, and
$\theta$ is the effective axion field.
The bulk average value of $\theta$ in a 3D gapped crystal with vanishing Hall conductivity, labeled as $\theta^{bulk}$, is determined by the Chern-Simons 3-form~\cite{Qi2008TFT} instead of the 2D Berry curvature, and it is only well defined modulo $2\pi$.
The winding of $\theta^{bulk}$ in a fourth dimension gives the second Chern number~\cite{Qi2008TFT}.
As $\theta^{bulk}$ has the same transformation properties as $\bsl{E}\cdot\bsl{B}$, any symmetry that flips the sign of $\bsl{E}\cdot\bsl{B}$ (\ie, ``axion-odd" symmetries) can quantize $\theta^{bulk}$ to $0$ or $\pi$ modulo $2\pi$, providing a symmetry-protected $Z_2$ indicator of axionic bulk topology in 3D insulators~\cite{Qi2008TFT,Essin2009AI,Wang2010TRIAxion,Hughes2011NTRInver,
Turner2012NTRInv,Fang2012TIPG,Varjas2015AI,Varnava2018AI,Schindler2018HOTI,
Wieder2018AXIFragile,Xu2019AxionEuInAs,Ahn2019C2T,Varnava2019AIWannier,
Sun2020C4TAI}.
Various axion-odd symmetries can also provide simplified expressions for evaluating $\theta^{bulk}$, which are consistent in a 3D insulator with vanishing Hall conductivity; thereby the mismatch among the expressions indicates the existence of the gapless points or a nonzero Hall conductivity~\cite{Yu2020AxionGapless}.
A direct physical consequence of nonzero $\theta^{bulk}$ is the magnetoelectric effect~\cite{Qi2008TFT}, where an external electric (magnetic) field induces a magnetization (polarization) in the parallel direction; if quantized by an axion-odd symmetry, the effect is called the topological magnetoelectric effect~\cite{Qi2008TFT,Zhang2018AIMnBi2Te4,Vishwanath2019MnBiTeAI}.
Other physical consequences of a non-zero $\theta^{bulk}$ include the surface half QAHE~\cite{Qi2008TFT}, a giant magnetic-resonance-induced current~\cite{Yu2019MRAI,Wang2020AnisotropicTME,Liu2020DynamicalAxionMR}, the topological magneto-optical effect~\cite{Wu2016THzTI,Okada2016THzQAH,Dziom2017THzTI} (especially its exact quantization~\cite{Wu2016THzTI}), the zero-Hall plateau state~\cite{Wang2015AI,Mogi2017AI,Xiao2018AI}, and the image magnetic monopole~\cite{Qi2009AIMono}.

In this work, we explore the role of a ``valley-separated" variant of an effective axion field in the physical response induced by dynamical strain.
In particular, we propose that dynamical homogeneous strain can induce a nonzero bulk-uniform magnetization in a class of 3D TR-invariant insulators that have vanishing total $\theta^{bulk}$ (thus vanishing magnetoelectric response) but have a non-quantized and tunable $\theta$ per valley.
We term the strain-induced magnetoelectric response the DPME.
Specifically, for the DPME to be relevant to experimental probes, we require a 3D TR-invariant insulator to have (i) low-energy physics that are well captured by a pair of TR-related valleys in the 1BZ, and (ii) non-vanishing valley axion fields, despite exhibiting an overall trivial $\theta^{bulk}$.
As we will demonstrate below, because the DPME originates from a valley axion field, the DPME is related to the bulk 3D topological properties, and in particular, a discontinuous change in the DPME serves as a direct signature of boundary TQPT.

Following \refcite{Zhang2013WSMCDWAxion,Roy2015WSMCDWAxion,Gooth2019WSMCDWAxion}, we recognize that the requirements for observing the DPME are satisfied by a TR-invariant WSM with a bulk-constant CDW.
We emphasize that in this work, the CDW order parameter is taken to be constant (\ie\ static and homogeneous) in the bulk of the system, in contrast to \refcite{Zhang2013WSMCDWAxion,Roy2015WSMCDWAxion,Gooth2019WSMCDWAxion}, which focused on bulk fluctuations of the CDW order parameter.
By studying the low-energy effective theory of a TR-invariant Weyl-CDW, we find that the phase of the CDW order parameter determines the valley axion field, leading to effective \emph{valley} axion electrodynamics, whereas the CDW wavevector induces a valley layered QAHE.
Importantly, TR symmetry restricts that the valley axion field and valley layered QAHE sum to zero when taken over two TR-related valleys, as TR symmetry reverses the signs of $\theta$ and the Hall conductivity.
On the other hand, dynamical strain can act as a pseudo-gauge field in WSMs~\cite{Cortijo2015StrainWSM,Cortijo2016WSMStrain,Franz2016ChiAnoStrainDSMWSM,
You2016WSMAxoinString,
Roy2018WSMStrain,Soto2018WSMStrain,Peri2019StrainWSM,
Behrends2019WSMStrain,
Munoz2019WSMStrain,Heidari2020WSMStrain,Ilan2020NatRevPseudoGauge,
Sukhachov2020PhononPseudogauge},
which was previously proposed as a means of studying the chiral anomaly of WSMs.
In this work, we find that pseudo-gauge fields can further couple to the electromagnetic field and the (valley) axion field in the effective action in the presence of a CDW.
According to the effective action, the analogues of the valley axion electrodynamics and the valley layered QAHE for the pseudo-gauge field are the DPME and the piezoelectric response, respectively, which are non-vanishing after summing over valleys.

Crucially, the bulk response coefficient of the DPME is the average value of valley axion field that is determined by the Chern-Simons 3-form.
As the Chern-Simons 3-form can only exist in three or higher dimensions, the DPME has a completely different origin compared to the previously studied piezoelectric effects, which instead are due to the 2D Berry curvature~\cite{Martin1972Piezo,Vanderbilt2000BPPZ,Vaezi2013StrainGraphene,
Droth2016PETBN,Rostami2018PiE,Yu2019PETTQPT}.
As a result, the DPME cannot be realized by trivially
stacking of 2D systems with nontrivial piezoelectric effects.
In this sense, the generalization from 2D piezoelectricity to the 3D DPME is analogous to that from the 2D quantum spin-Hall insulator to the 3D TR-invariant strong topological insulator.
Specifically, the 3D TR-invariant strong topological insulator cannot be constructed from a simple 3D stacking of the 2D quantum spin-Hall insulators, and was instead discovered through more involved theoretical efforts~\cite{Hasan2010TI,Qi2010TITSC}.
Hence, the DPME \emph{cannot} be viewed as a simple 3D generalization of 2D piezoelectricity --- the DPME is instead a intrinsically 3D effect.

To test the low-energy results against the UV completion, we construct a minimal TB model for a TR-invariant WSM with varying values of the CDW order parameters.
We specifically demonstrate that a CDW can drive a minimal TR-invariant WSM into a weak topological insulator (WTI) phase with a nontrivial weak $Z_2$ index, originating from the odd valley Chern number induced by the CDW wavevector.
By evaluating the bulk average value of the valley axion field at various phases of the CDW order parameter, we numerically verify the relation between the phase of the CDW $\phi$ and the valley axion field $\theta$ predicted by the low-energy theory.
We further show that dynamical strain generically induces a nonuniform current distribution, which can be split into a uniform background current and a nonuniform part.
The uniform background current gives rise to the piezoelectric effect, which is a trivial stacking of the 2D piezoelectric effect discussed in \refcite{Yu2019PETTQPT}.
Conversely, we find that the nonuniform part of the current is mainly localized at the surfaces, and possesses opposite signs on two surfaces perpendicular to the CDW wavevector (provided a large enough sample), which induces the bulk orbital magnetization that corresponds to the intrinsically 3D DPME.

In addition, we demonstrate the existence of discontinuous changes in the DPME by continuously varying the bulk CDW phase in the TB model.
We show that the discontinuous change originates from gap closings on certain system boundaries (appearing as two TR-related 2D gapless Dirac cones on one surface), which reverses the direction of the corresponding surface currents and thus dramatically changes the bulk orbital magnetization.
We note that, in this work, we employ the high-energy convention in which 2D and 3D Dirac cones have two and four components, respectively.
This convention is notably distinct from the recent condensed-matter works~\cite{Kane20152DDSM,Wieder20172DDSM,Wieder2018WallPaperFermion}, in which both 2D and 3D fourfold degeneracies are termed ``Dirac cones."
We observe that the gap closing on each Weyl-CDW surface takes the same form as the 2D $Z_2$ TQPT (\ie, the TQPT between a 2D TR-protected $Z_2$ trivial insulator phase and a 2D TR-protected $Z_2$ nontrivial topological insulator (TI) phase) at generic momenta, and is unaccompanied by a bulk gap closing; hence the surface gap closing, in the presence of two valleys, is a boundary TQPT that changes the relative surface $Z_2$ index.
When the gap closings on different surfaces happen simultaneously, we can attribute the DPME jump to a $2\pi$ change in the valley axion field induced by the surface $Z_2$ TQPT, through the bulk-boundary correspondence of the low-energy effective theory.
Our findings suggest that jumps in the DPME can serve as bulk signatures of boundary TQPTs.

This work is organized as follows.
First, we will present an intuitive picture for the DPME in \secref{sec:intuitive_picture}.
Then, we will describe a low-energy theory for the DPME in \secref{sec:LE}, which we will then verify with a TB model in \secref{sec:TB}.
In \secref{sec:Boundary_TQPT_DPME_Jump}, we will elucidate the relationship between the jump of the DPME and the boundary TQPT.
Finally, we will conclude in \secref{sec:conclusion} by introducing a proposal for measuring the DPME in experiment.

\section{Intuitive Picture}
\label{sec:intuitive_picture}

\begin{figure}[t]
    \centering
    \includegraphics[width=\columnwidth]{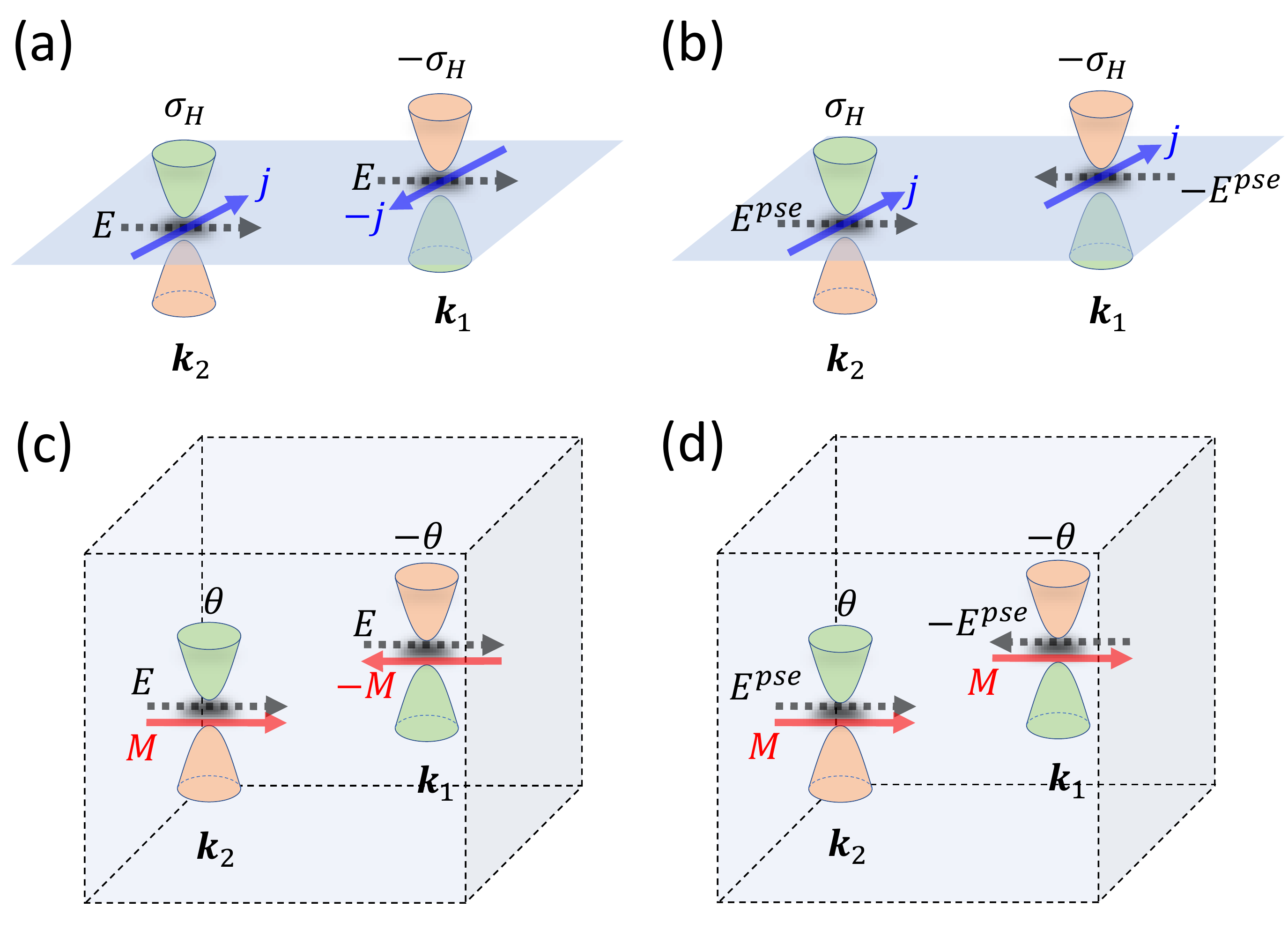}
    \caption{
    2D and 3D response effects in insulators with two valleys.
    In (a) and (b), two 2D gapped Dirac cones at $\bsl{k}_{1,2}$ are related by TR symmetry.
    The two massive 2D Dirac cones provide opposite and canceling contributions to the Hall conductance $\sigma_H$.
    In (c) and (d), two 3D gapped Dirac cones at $\bsl{k}_{1,2}$ are related by TR symmetry.
    The two 3D Dirac cones have complex masses and have opposite axion fields $\theta$, due to TR symmetry.
    }
    \label{fig:intuitive}
\end{figure}

Before presenting supporting analytic and numerical calculations, we will first provide an intuitive picture of the DPME, in comparison with the Berry-curvature contribution to the piezoelectric response~\cite{Martin1972Piezo,Vanderbilt2000BPPZ,Vaezi2013StrainGraphene,
Droth2016PETBN,Rostami2018PiE,Yu2019PETTQPT}.
We start with a 2D TR-invariant insulator whose low-energy physics is described by two 2D gapped Dirac cones (regarded as two valleys below), as shown in \figref{fig:intuitive}(a).
Because the two valleys are related by TR symmetry, there must be an oppositely-signed valley Hall conductance within each valley (determined by the integral of the Berry curvature over each valley), and the Hall currents induced by a uniform electric field must point in opposite directions and exactly cancel.
However, a non-vanishing charge current can still be generated by applying a \emph{pseudo}-electric field that points in opposite direction in each valley, which causes the induced Hall currents to point in the same directions and add up to a nonzero total current (\figref{fig:intuitive}(b)).
The pseudo-electric field can be generated by a dynamical homogeneous strain tensor $u$ as $E^{pse}\sim \dot{u}$~\cite{Ilan2020NatRevPseudoGauge,Franz2016ChiAnoStrainDSMWSM}, and the resultant total Hall current $j\sim \dot{u}$ characterizes the piezoelectric effect~\cite{Martin1972Piezo,Vanderbilt2000BPPZ,Vaezi2013StrainGraphene,Droth2016PETBN,Rostami2018PiE}.
With regards to TR symmetry, a static external electric field preserves TR symmetry whereas a static external pseudo-electric field breaks TR symmetry, explaining the dramatic differences in the current response.
Although the piezoelectric effect is not quantized in one insulating phase, its nonzero discontinuous \emph{change} across a 2D TR-invariant TQPT (\eg\ a 2D $Z_2$ TQPT~\cite{Kane2005Z2}) is proportional to the change of a topological invariant~\cite{Yu2019PETTQPT} and thus serves as a new experimental probe of the topology (or more precisely a change in topology).

Although the strain also acts as a pseudo-electric field in the 3D DPME, the 3D DPME is fundamentally distinct from a trivial stacking of the 2D piezoelectric effect, as the 3D DPME originates from 3D axion field instead of the Hall conductance.
To see this, consider a TR-invariant 3D gapped system whose low-energy physics is captured by two 3D gapped Dirac cones with complex masses (also regarded as two valleys below), as shown in \figref{fig:intuitive}(c) and (d).
Because the TR symmetry maps one valley to the other, a single Dirac cone is not necessarily TR-invariant, and, in the absence of additional crystal symmetries, generically has an unpinned effective axion field $\theta$.
For a finite-sized configuration of the 3D system with a fully gapped TR-invariant 2D boundary, TR symmetry requires that the two Dirac cones have opposite $\theta$ angles, implying that an external electric field would induce opposite magnetizations in the two valleys (\figref{fig:intuitive}(c)), which sum to zero.
In contrast, a pseudo-electric field, which can be induced by dynamical homogeneous strain and points in opposite directions at each valley, would induce the same magnetizations at the two valleys (\figref{fig:intuitive}(d)), resulting in a nonzero total magnetization $M\propto E^{pse}\sim \dot{u}$.
We term this effect the DPME.
The DPME is different from the conventional piezomagnetic effect~\cite{Dzialoshinskii1958PM} because the magnetization of the former is proportional to the time derivative of the strain tensor $\dot{u}$, whereas the magnetization of the latter is directly proportional to the strain tensor $u$.
Crucially, the bulk average value of the effective axion field $\theta$ is determined by the integrals of the Chern-Simons 3-form~\cite{Qi2008TFT,Essin2009AI,Wang2010TRIAxion} $\epsilon^{ijl}\Tr[\mathcal{A}_i\partial_{k_j}\mathcal{A}_l+\ii \frac{2}{3}\mathcal{A}_i \mathcal{A}_j \mathcal{A}_l]d^3k$, where $\mathcal{A}$ is the non-Abelian Berry connection, and the Chern-Simons 3-form can only exist in three or higher dimensions.
Therefore, unlike the 2D piezoelectric effects, the $\theta$-induced 3D DPME originates from the Chern-Simons 3-form instead of the Berry curvature, meaning that the DPME is intrinsically 3D and cannot be given by trivial stacking systems with a 2D piezoelectric effect.

\section{Low-Energy Effective Theory of TR-Invariant WSMs with Axionic CDWs}
\label{sec:LE}

In this section, we will provide a low-energy theory for the DPME in TR-invariant WSMs with axionic CDWs.
We would like to emphasize that the derivation below is not confined to Weyl-CDWs, and can be generalized to any TR-invariant system with valley axion fields.

\subsection{Minimal Model}

\begin{figure}[t]
    \centering
    \includegraphics[width=\columnwidth]{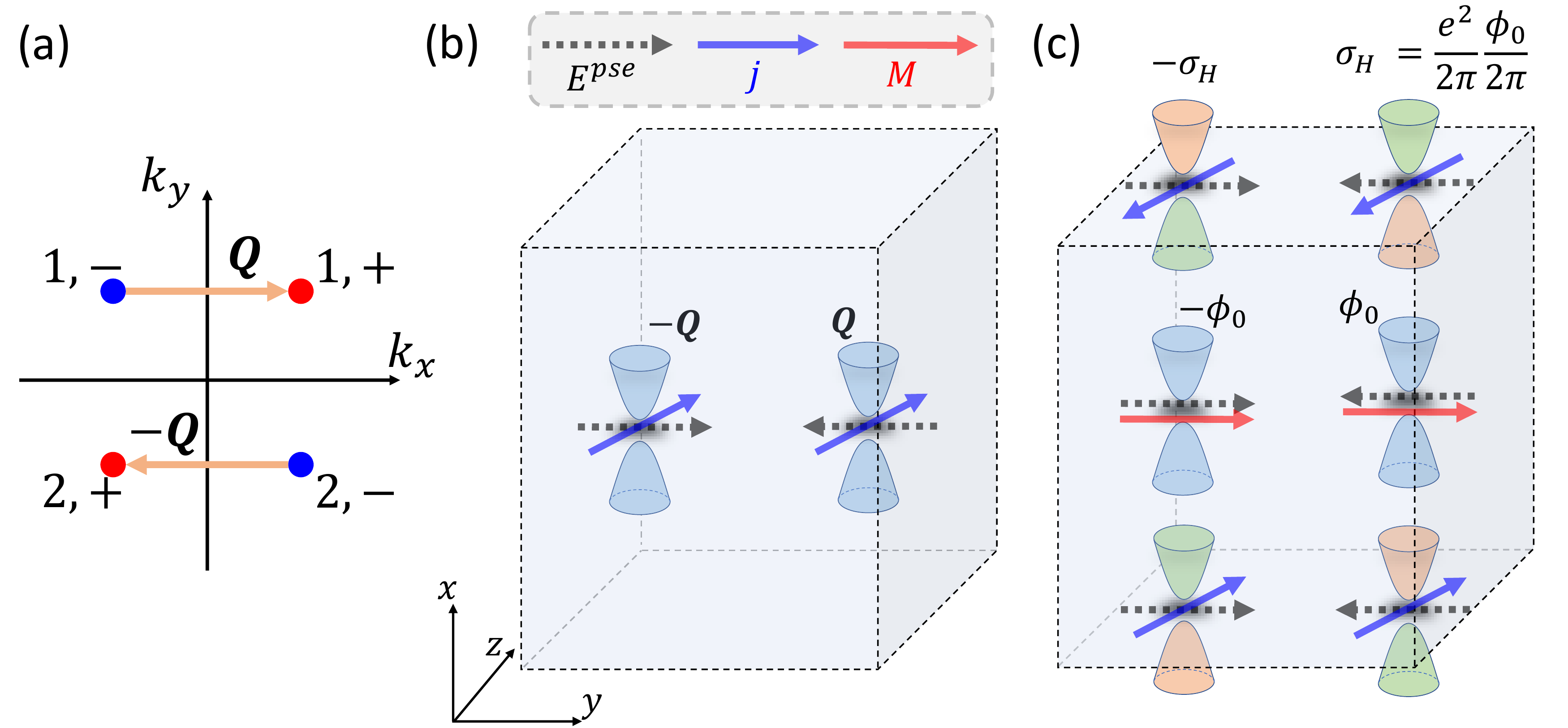}
    \caption{
    (a) The projection of four Weyl points on the $k_x-k_y$ plane in a TR-invariant minimal WSM.
    The arrows in (a) indicate the projection of the CDW wavevectors $\bsl{Q}$.
    (b) The low-energy piezoelectric current induced by the CDW wavevector.
    The dashed box above (b) indicates that the black dashed arrows, the blue solid arrows, and the red solid arrows in (b) and (c) respectively represent the pseudo-electric field, current, and magnetization.
    (c) The DPME induced by the phase of the CDW order parameter.
    In (c), we have chosen open boundary conditions for only the surfaces perpendicular to $x$; the two $x$-normal surfaces are gapped and symmetry-preserving.
    The two bulk Dirac cones have opposite bulk phases of the CDW order parameter.
    Through the bulk-boundary correspondence, this implies that the two valleys have opposite surface Hall conductances.
    }
    \label{fig:LE_Action}
\end{figure}

A WSM phase can only emerge in systems that either break TR symmetry (magnetic materials) or break inversion symmetry (non-centrosymmetric crystals).
CDWs in magnetic WSMs have been previously studied in numerous works, including \refcite{Zhang2013WSMCDWAxion,Roy2015WSMCDWAxion}.
In this work, we focus on CDWs in TR-invariant WSMs, which can be realized in non-centrosymmetric crystals.
Since two Weyl points related by TR symmetry share the same chirality, and because the total chirality of the whole system must vanish~\cite{Nielsen1981ChialFermionDoubling}, then there must be four Weyl points in a minimal model of a TR-invariant WSM.
\figref{fig:LE_Action}(a) schematically shows a distribution of four Weyl points in a minimal TR-invariant WSM, where for simplicity, we have enforced an additional mirror symmetry that flips $y$, labeled as $m_y$, such that all four Weyl points are symmetry-related.
The momenta of the four Weyl points take the form
\eq{
\label{eq:WP_pos}
\bsl{k}_{a,\alpha}=(-1)^{a-1}(\alpha k_{0,x}, k_{0,y}, \alpha k_{0,z})\ ,
}
where $\alpha=\pm$ indicates the relative chirality of the Weyl points, and $a=1,2$ is termed the ``valley index."
TR symmetry, labeled as $\TR$, relates $\bsl{k}_{1,\alpha}$ to $\bsl{k}_{2,\alpha}$ with the same chirality index $\alpha$, while the mirror $m_y$ changes the chirality of Weyl points and relates $\bsl{k}_{1,\alpha}$ to $\bsl{k}_{2,-\alpha}$.
We would like to emphasize that, although we keep $m_y$ in the following derivation for simplicity, mirror symmetry is not essential for the physics discussed below.

Through a unitary transformation of the bases and by rotating and rescaling axes, we can always transform the low-energy Lagrangian of the four Weyl points into the following form~\cite{Zhang2013WSMCDWAxion}
\eq{
\label{eq:L_WP}
\mathcal{L}_{a,\alpha}=\psi^\dagger_{t,\bsl{r},a,\alpha}\left[ \ii \partial_t-\alpha\sum_{i}v_i (-\ii\partial_{i}-k_{a,\alpha,i}) \sigma_i\right]\psi_{t,\bsl{r},a,\alpha}
\ ,
}
where $\psi_{t,\bsl{r},a,\alpha}$ is a two-component field for the two bands that form the Weyl point at $\bsl{k}_{a,\alpha}$, $\sigma_{0,x,y,z}$ are the Pauli matrices, $v_i$ indicates the Fermi velocity along the $i$ direction, and $t$ and $\bsl{r}$ are time and position, respectively.
In this work, we will for simplicity focus on the case in which $i=x,y,z$ are the three laboratory directions.
Following the derivation in \refcite{Zhang2013WSMCDWAxion,Burkov2012ChiralAnomaly}, we keep $\bsl{k}=0$ as the momentum-space origin of all fermion fields, as this choice naturally includes the Weyl-point-induced valley Hall effect in the effective action, as discussed below.
Throughout this section on the low-energy theory, we adopt a proper rescaling of the space and fields to cancel the Fermi velocities as elaborated in \appref{app:S_eff}; the Fermi velocities will later be restored for comparison to the TB model in \secref{sec:TB}.

As mentioned above, both $\TR$ and $m_y$ change the valley index $a$ of the fields in \eqnref{eq:L_WP}, while $m_y$ ($\TR$) changes (preserves) the chirality index $\alpha$.
Thus, we can always choose the bases to represent $\TR$ and $m_y$ as
$\ii\sigma_y \cc$ and $-\ii\sigma_y$ for the band index, respectively, where $\cc$ is complex conjugation.
According to the above symmetry representation for $\TR$ and $m_{y}$, a symmetry-preserving mean-field CDW term that couples two Weyl points of the same valley index $a$ can be written as~\cite{Zhang2013WSMCDWAxion}
\eq{
\mathcal{L}_{CDW}=-\sum_{a} m_a(\bsl{r}) e^{\ii (-1)^{a-1} \bsl{Q}\cdot \bsl{r} } \psi^\dagger_{a,+}\psi_{a,-}+h.c.\ ,
}
where the $t,\bsl{r}$ dependence of $\psi$ is implied, and where $\bsl{Q}=\bsl{k}_{1,+}-\bsl{k}_{1,-}=-(\bsl{k}_{2,+}-\bsl{k}_{2,-})$ is the CDW wavevector, as shown in \figref{fig:LE_Action}(a).
Throughout this work, we will include the spatial dependence of the CDW order parameter $m_a(\bsl{r})$, while keeping the order parameter time-independent (\ie\ static).
TR symmetry requires that $m_1(\bsl{r})=m_2(\bsl{r})^*\equiv m(\bsl{r})$, and $m_y$ symmetry requires that $m_1^*(m_y\bsl{r})=m_2(\bsl{r})$.
In general, $m(\bsl{r})=|m(\bsl{r})|e^{\ii \phi(\bsl{r})}$ is complex, and $|m(\bsl{r})|$ and $\phi(\bsl{r})$ are the magnitude and phase of the CDW order parameter, respectively.
As discussed in \secref{sec:intro}, we consider the case where $m(\bsl{r})$ is equal to a complex constant mass $m_0=|m_0|e^{\ii \phi_0}$ in the bulk throughout the work, \ie, $|m(\bsl{r})|=|m_0|$ and $\phi(\bsl{r})=\phi_0$ for $\bsl{r}$ in the bulk.
The underlying interaction that gives rise to the bulk CDW is discussed in \appref{app:CDW_MF} at the mean-field level.
Nevertheless, $m(\bsl{r})$, as well as $|m(\bsl{r})|$ and $\phi(\bsl{r})$, can still have spatial dependence if the sample size is finite.
Inspired by \refcite{Qi2008TFT}, we set $|m(\bsl{r})|\rightarrow \infty$ and $\phi(\bsl{r})= 0$ for $\bsl{r}$ deep in the vacuum.
Different gapped and symmetry-preserving boundaries can then be represented by different ways of smoothly connecting the bulk and vacuum limits of $|m(\bsl{r})|$ and $\phi(\bsl{r})$.
We next introduce the $\gamma$ matrices
\eq{
\gamma^\mu=(\tau_x\sigma_0,-\ii \tau_y \bsl{\sigma})_\mu\ ,\ \gamma^5=\ii \gamma^0 \gamma^1 \gamma^2 \gamma^3\ ,
}
where $\mu=0,1,2,3$ and $\tau_{0,x,y,z}$ are Pauli matrices for the chirality index $\alpha$.
Using the above definitions of the $\gamma$ matrices, we can rewrite the CDW term as
\eq{
\label{eq:L_CDW_MF}
\mathcal{L}_{CDW}=-\sum_{a} |m(\bsl{r})| \overline{\psi}_{a} e^{-\ii \Phi_a(\bsl{r}) \gamma^5}\psi_{a}\ ,
}
where $\overline{\psi}_a=\psi^\dagger_a\gamma^0$, and $\Phi_a(\bsl{r})=(-1)^{a-1} (\phi(\bsl{r}) +\bsl{Q}\cdot \bsl{r})$.
For the remainder of this work, the spatial dependence of $|m|, \phi, \Phi_a$ will be implicit, and we will suppress the explicit dependencies on $\bsl{r}$ for notational simplicity.

In order to elucidate the strain-induced linear response, we next introduce an electron-strain coupling for normal strain (\ie\ stretch or compression along a specified axis) along the $z$ direction, labeled as $u_{zz}(t)$.
We require that the strain be adiabatic, homogeneous, and infinitesimal.
Enforcing TR and mirror symmetries, the most general form of the leading-order electron-strain coupling reads
\eqa{
\label{eq:L_str}
&\mathcal{L}_{str}=\sum_{a}\overline{\psi}_{a} [-\xi_0\gamma^0+(-1)^{a}(\gamma^1\gamma^5 \xi_x+\gamma^2 \xi_y\\
& +\gamma^3 \gamma^5 \xi_z)]\psi_{a}u_{zz} \ ,
}
where the time dependence of $u_{zz}$ is implied, and where the parameters $\xi_{0,x,y,z}$ are material-dependent.
In \eqnref{eq:L_str}, we do not include the effects of strain that couple different Weyl points, as Weyl-point coupling strain is necessarily proportional to $|m_0| u_{zz}$, and because $|m_0|$ is typically small in real materials.
The detailed procedure of adding the electron-strain coupling is shown in \appref{app:el-str}.
We set the strain to be uniform throughout all of space, such that the gapped and symmetry-preserving boundary is implemented by the spatial dependence of the CDW order parameter $m$, as opposed to an inhomogeneous strain field.

Summing up \eqnref{eq:L_WP}, \eqnref{eq:L_CDW_MF}, and \eqnref{eq:L_str} and including the $U(1)$ gauge field coupling for the electromagnetic field, we arrive at the total low-energy Lagrangian $\mathcal{L}=\sum_{a} \mathcal{L}_a$ with
\eqa{
\label{eq:L_tot}
&\mathcal{L}_a= \overline{\psi}_a \left[\ii (\slashed{\partial}+\ii e \slashed{\widetilde{A}_a}-\ii \slashed{\partial} \varphi_{a}- \ii \slashed{A}_{a,5}\gamma^5) - |m| e^{-\ii \Phi_a \gamma^5}\right]\psi_{a}\ ,
}
where $\slashed{\partial}=\gamma^\mu\partial_\mu$, $\slashed{\widetilde{A}_a}=\gamma^\mu \widetilde{A}_{\mu,a}$, $\slashed{A}_{a,5}=\gamma^\mu A_{a,5,\mu}$, $\varphi_a=(\bsl{k}_{a,+}+\bsl{k}_{a,-})\cdot \bsl{r}/2$, and the metric is chosen as  $(-,+,+,+)$.
$\mathcal{L}_a$ describes a massive 3D Dirac fermion that couples to a valley-dependent $U(1)$ gauge field $\widetilde{A}_{a}$ and a valley-dependent chiral gauge field $A_{a,5}$, and $\Phi_a$ is the mass phase of the Dirac fermion.
In terms of $u_{zz}$ and the CDW wavevector $\bsl{Q}$, the valley-dependent chiral gauge field is given by
\eq{
A_{a,5,\mu}=(-1)^{a-1}\partial_{\mu} (\bsl{Q}\cdot\bsl{r}/2)+(-1)^a u_{zz}(0, \xi_x, 0, \xi_z)_\mu \ .
}
The valley-dependent $U(1)$ gauge field takes the form
\eq{
\label{eq:A_a_expression}
\widetilde{A}_{a,\mu}=A_{\mu}+\frac{u_{zz}}{e}(\xi_0 , 0, (-1)^{a-1}\xi_y ,0)_\mu\ ,
}
which contains the physical gauge field $A_\mu$ and the pseudo-gauge field induced by the strain $u_{zz}$.
In particular, the $y$ component of the pseudo-gauge field can provide a pseudo-electric field that points in opposite directions in each of the two valleys
\eq{
\label{eq:E_pse}
\bsl{E}^{pse}_a=(-1)^a \frac{\xi_y}{e} \dot{u}_{zz} \bsl{e}_y\ .
}
As we will show below, all the nontrivial leading-order linear response comes from the pseudo-electric field in \eqnref{eq:E_pse}.

\subsection{Effective Action}

The low-energy response to $A$ and $u_{zz}$ can be derived from the total effective action $S_{eff}=\sum_a S_{eff,a}$, which takes the form
\eq{
\label{eq:Z_a}
e^{\ii S_{eff,a}}=\int D\overline{\psi}_a D\psi_a \exp\left[\ii \int dt d^3 r \mathcal{L}_a\right] \ ,
}
where the measure of the functional integral is in real space
\eq{
\label{eq:Dpsi_a}
\int D\overline{\psi}_a D\psi_a \propto \prod_{t,\bsl{r}}\int d\overline{\psi}_{t,\bsl{r},a}d \psi_{t,\bsl{r},a}\ .
}
We note that \eqnref{eq:L_WP}, as well as \eqnref{eq:Z_a}, are only exact when the momenta of the fermion fields in \eqnref{eq:Dpsi_a} are restricted near the Weyl points (or equivalently the momentum deviation from the corresponding Weyl point is below a finite momentum cutoff $\Lambda$).
In \appref{app:S_eff}, we demonstrate, however, that the correction to the response of interest brought by a finite $\Lambda$ is of the order $O(|m_0|^2/\Lambda^2)$, which is negligible owing to $|m_0|\ll \Lambda$ in realistic materials. 
Since the focus of this work is on the low-energy response of the system, we can limit $\Lambda\rightarrow \infty$ and take the functional integration (\eqnref{eq:Dpsi_a}) over the entire $\bsl{k}\in \mathds{R}^3$, following \refcite{Burkov2012ChiralAnomaly,Zhang2013WSMCDWAxion,Roy2015WSMCDWAxion,
Gooth2019WSMCDWAxion}.

The physical $U(1)$ gauge field $A$ and the strain tensor $u_{zz}$ in \eqnref{eq:L_tot} are treated as fixed backgrounds, meaning that we neglect their dynamics.
Under this assumption, $\mathcal{L}_a$ has local valley $U(1)$ gauge invariance, \ie, invariance under
$
\psi_{a}\rightarrow \psi_{a} e^{\ii e \Gamma_a(t,\bsl{r})}
$ and
$
\widetilde{A}_{a,\mu}\rightarrow \widetilde{A}_{a,\mu}- \partial_\mu \Gamma_a(t,\bsl{r}),
$
where $\Gamma_{a}$ is a valley-dependent scalar function and the corresponding transformation on $\overline{\psi}_a$ is implicit here (and will remain implicit and for the reminder of this work).
The valley $U(1)$ gauge invariance corresponds to a separate vector current conservation for each valley; as different valleys are decoupled in \eqnref{eq:L_tot}, we preserve the current conservation in each valley against all orders of quantum correction, which is reasonable as long as valleys are well defined.
As a result, the measure of the functional integral (\eqnref{eq:Dpsi_a}) is invariant under $\psi_a\rightarrow e^{\ii e \Gamma_a}\psi_a$, and $S_{eff,a}$ is gauge invariant  $S_{eff,a}[\widetilde{A}_{a,\mu}- \partial_\mu \Gamma_a]=S_{eff,a}[\widetilde{A}_{a,\mu}]$.
The valley $U(1)$ gauge invariance of $S_{eff,a}[\widetilde{A}_{a,\mu}]$ allows us to perform a gauge transformation $\widetilde{A}_{a,\mu}\rightarrow \widetilde{A}_{a,\mu}+ \partial_\mu \varphi_a/e$ to cancel the $\varphi_a$ term in \eqnref{eq:L_tot} without changing the form of $S_{eff,a}$, resulting in
\eq{
\label{eq:L_tot_sim}
\mathcal{L}_a= \overline{\psi}_a \left[\ii (\slashed{\partial}+\ii e \slashed{\widetilde{A}_a}- \ii \slashed{A}_{a,5}\gamma^5) -|m| e^{-\ii \Phi_a \gamma^5}\right]\psi_{a}\ .
}
$\mathcal{L}_a$ also has an effective Lorentz invariance, which we also preserve against quantum corrections and take to be a symmetry of the effective action $S_{eff,a}$.

We can further perform a chiral gauge transformation $\psi_a\rightarrow e^{\ii \Phi_a \gamma^5/2}\psi_a$  to cancel the phase of the Dirac mass in \eqnref{eq:L_tot_sim}.
Owing to the valley $U(1)$ gauge invariance and the effective Lorentz invariance, Fujikawa's method suggests that the Jacobian of the measure (\eqnref{eq:Dpsi_a}) would contain a topologically nontrivial factor~\cite{Srednicki2007QFT,Bertlmann2000Anomalies}, which enters into the effective action as
\eq{
\label{eq:S_eff_a_gen}
S_{eff,a}= \int dt d^3 r  \frac{e^2}{16\pi^2} \frac{\Phi_a}{2} \varepsilon^{\mu\nu\rho\delta}  \widetilde{F}_{a,\mu\nu}\widetilde{F}_{a,\rho\delta}+ ...\ ,
}
where $\widetilde{F}_{a,\mu\nu}=\partial_\mu \widetilde{A}_{a,\nu}-\partial_\nu \widetilde{A}_{a,\mu}$, and ``..." includes all other terms.
In this work, we only consider the leading-order linear response to $A$ and $u_{zz}$.
Through an explicit evaluation of Feynman diagrams in \appref{app:S_eff}, we find that the only leading-order linear response contained in ``..." is the trivial correction to the permittivity and permeability in the material, which can be absorbed into the Maxwell term of $A$.
Hence, all nontrivial leading-order linear responses come from the first term of \eqnref{eq:S_eff_a_gen}.
After omitting all of the higher-order and trivial terms, we can split \eqnref{eq:S_eff_a_gen} into three parts by using \eqnref{eq:A_a_expression}:
\eqa{
\label{eq:S_eff_a_leading}
S_{eff,a}= S_{eff,a,\theta}+S_{eff,a,\Sigma}+S_{eff,a,u_{zz}}\ .
}
$S_{eff,a,\theta}$ is the action for valley-separated axion electrodynamics
\eq{
S_{eff,a,\theta} = \frac{e^2}{32\pi^2} \int dt d^3 r  \theta_a \varepsilon^{\mu\nu\rho\delta}  F_{\mu\nu}F_{\rho\delta}\ ,
}
where the valley axion field is given by the phase of the CDW order parameter as
\eq{
\label{eq:theta_a}
\theta_a=(-1)^{a-1} \phi\ .
}
$S_{eff,a,\Sigma}$ takes the form of a Chern-Simons theory
\eq{
S_{eff,a,\Sigma}=\frac{1}{2} \int dt d^3 r\ \Sigma_{H,a,i} \varepsilon^{i\mu\nu\rho} A_\mu \partial_{\nu} A_\rho\ ,
}
and describes a valley layered QAHE with a valley Hall conductivity given by the CDW wavevector $\bsl{Q}$
\eq{
\label{eq:sigma_H_a}
\bsl{\Sigma}_{H,a}=(-1)^a   \frac{\bsl{Q}}{2\pi} \frac{e^2}{2\pi}\ .
}
Because the two valleys are related to each other by TR symmetry, then $\bsl{\Sigma}_{H,a}$ ($a=1,2$) take opposite values in each of the two valleys and add to a net-zero total Hall conductivity, coinciding with the vanishing Chern number required by the TR symmetry.
In general, an odd-integer valley layered QAHE indicates that a gapped Weyl-CDW with well-defined valleys is in a WTI phase if all higher energy bands are topologically trivial, as discussed and numerically confirmed in \secref{sec:TB}.
The last term $S_{eff,a,u_{zz}}$ describes the strain-induced effect and reads
\eq{
\label{eq:S_eff_a_uzz}
S_{eff,a,u_{zz}}= \frac{e \xi_y}{4\pi^2}\int dt d^3 r   (\phi+\bsl{Q}\cdot\bsl{r})\dot{u}_{zz} B_y\ ,
}
where $B_y=F_{31}$.
The effective action (\eqnref{eq:S_eff_a_leading}) is one central result of this work.
We emphasize that the validity of \eqnref{eq:S_eff_a_leading} replies on the fact that the system must be gapped everywhere -- including the boundary -- if the system is finite-sized~\cite{Qi2008TFT}.

\subsection{Piezoelectric Effect and DPME}

Because $S_{eff,a,\theta}$ and $S_{eff,a,\Sigma}$ in \eqnref{eq:S_eff_a_leading} have opposite signs in each of the two valleys, then the contributions of $S_{eff,a,\theta}$ and $S_{eff,a,\Sigma}$ sum to zero in the total effective action.
Hence, the total effective action only includes $S_{eff,a,u_{zz}}$, which can be rewritten as
\eq{
\label{eq:S_eff_tot_leading}
S_{eff}=\int dt d^3 r\sum_a \bsl{A}\cdot\left[(-\frac{e^2}{4\pi^2}\nabla \theta_a +\bsl{\Sigma}_{H,a})\times \bsl{E}^{pse}_a\right]\ .
}
\eqnref{eq:S_eff_tot_leading} indicates that the total action relies on a nonzero electron-strain coupling, implying that the response of the action characterizes the deviation of the electron from the homogeneous deformation of the sample~\cite{Vanderbilt2000BPPZ}.

The total current derived from $S_{eff}$ can be decomposed into two parts
\eq{
\label{eq:j_eff}
 \bsl{j}=\frac{\delta S_{eff}}{\delta \bsl{A}}=\bsl{j}_{PE}+\bsl{j}_{M}\ .
}
$\bsl{j}_{PE}$ is the total low-energy valley Hall current induced by the pseudo-electric field
\eqa{
\label{eq:j_PE_eff}
\bsl{j}_{PE}= \sum_a \bsl{\Sigma}_{H,a}\times \bsl{E}^{pse}_{a},
}
which, as required by $m_y$ symmetry, lies in the $xz$ plane.
\eqnref{eq:E_pse} and \eqnref{eq:sigma_H_a} together imply that the pseudo-electric field and the low-energy Hall conductivity both have opposite signs in the two valleys, such that the induced Hall currents add constructively to give a total nonzero value. (See \figref{fig:LE_Action}(b).)
In \eqnref{eq:j_PE_eff}, $\bsl{j}_{PE}$ is thus the bulk-uniform piezoelectric current induced by the CDW wavevector $\bsl{Q}$, where the piezoelectric coefficient is given by
\eq{
\label{eq:LE_chi}
\chi_{izz}=\frac{\partial j_{PE,i}}{\partial \dot{u}_{zz}}= \frac{e}{2\pi^2} \xi_y \left(\bsl{Q}\times \bsl{e}_y\right)_i\ .
}
Hence, $\bsl{j}_{PE}$ can be understood as a 3D stack of 2D valley Hall systems in which each layer exhibits the 2D piezoelectric effect discussed in previous literature~\cite{Vaezi2013StrainGraphene,Droth2016PETBN,Rostami2018PiE,Yu2019PETTQPT}.
We would like to emphasize that \eqnref{eq:LE_chi} only includes the low-energy contribution to the piezoelectric current, while the high-energy contribution to the piezoelectric current is typically also present in realistic materials.
Nevertheless, the high-energy contributions to the piezoelectric current should also be uniform in the bulk of the system.

In \eqnref{eq:j_eff}, $\bsl{j}_{M}$ takes the form of a magnetization current
\eq{
\label{eq:j_M_eff}
\bsl{j}_{M}= \nabla\times \bsl{M}
}
in which the total orbital magnetization $\bsl{M}$ is induced by a pseudo-electric field through the valley axion field
\eq{
\bsl{M}=-\sum_a  \frac{e^2}{4\pi^2} \theta_a   \bsl{E}^{pse}_a\ .
}
Physically, \eqnref{eq:j_M_eff} can be understood from the bulk-boundary correspondence as follows.
First, given a gapped and symmetry-preserving boundary, the CDW phase $\phi$ smoothly changes from a constant value $\phi_0$ in the bulk to zero in the vacuum, implying that the magnetization current $\bsl{j}_{M}$ is localized on the boundary.
According to the bulk-boundary correspondence of the axion field, the surface valley Hall conductance (along the normal direction of the surface) should take the form $\sigma_{H,a}=\frac{e^2}{2\pi} \frac{\theta_a^{bulk}}{2\pi}$ on any surface, where $\theta_a^{bulk}=(-1)^{a-1}  \phi_0$.
Hence, the surface-localized magnetization current $\bsl{j}_{M}$ is simply the surface Hall current induced by the pseudo-electric field, as shown in \figref{fig:LE_Action}(c).
The surface current generates a uniform bulk magnetization of the form
\eq{
\label{eq:M_bulk}
\bsl{M}^{bulk}=-\sum_a \frac{e^2}{2 \pi}\frac{\theta_a^{bulk}}{2\pi} \bsl{E}^{pse}_{a}=\frac{e \xi_y }{2\pi^2 }\phi_0 \dot{u}_{zz} \bsl{e}_y   \ ,
}
which is the DPME proposed in this work, as illustrated in \figref{fig:intuitive}(d).
Unlike the piezoelectric current in \eqnref{eq:j_PE_eff} originating from the 2D valley Hall conductance, $j_m$ in \eqnref{eq:M_bulk} originates from the \emph{fundamentally 3D} bulk valley axion field, and is fundamentally different from \eqnref{eq:j_PE_eff}.

\section{Minimal TB Model for TR-invariant WSM}
\label{sec:TB}

\begin{figure}[t]
    \centering
    \includegraphics[width=\columnwidth]{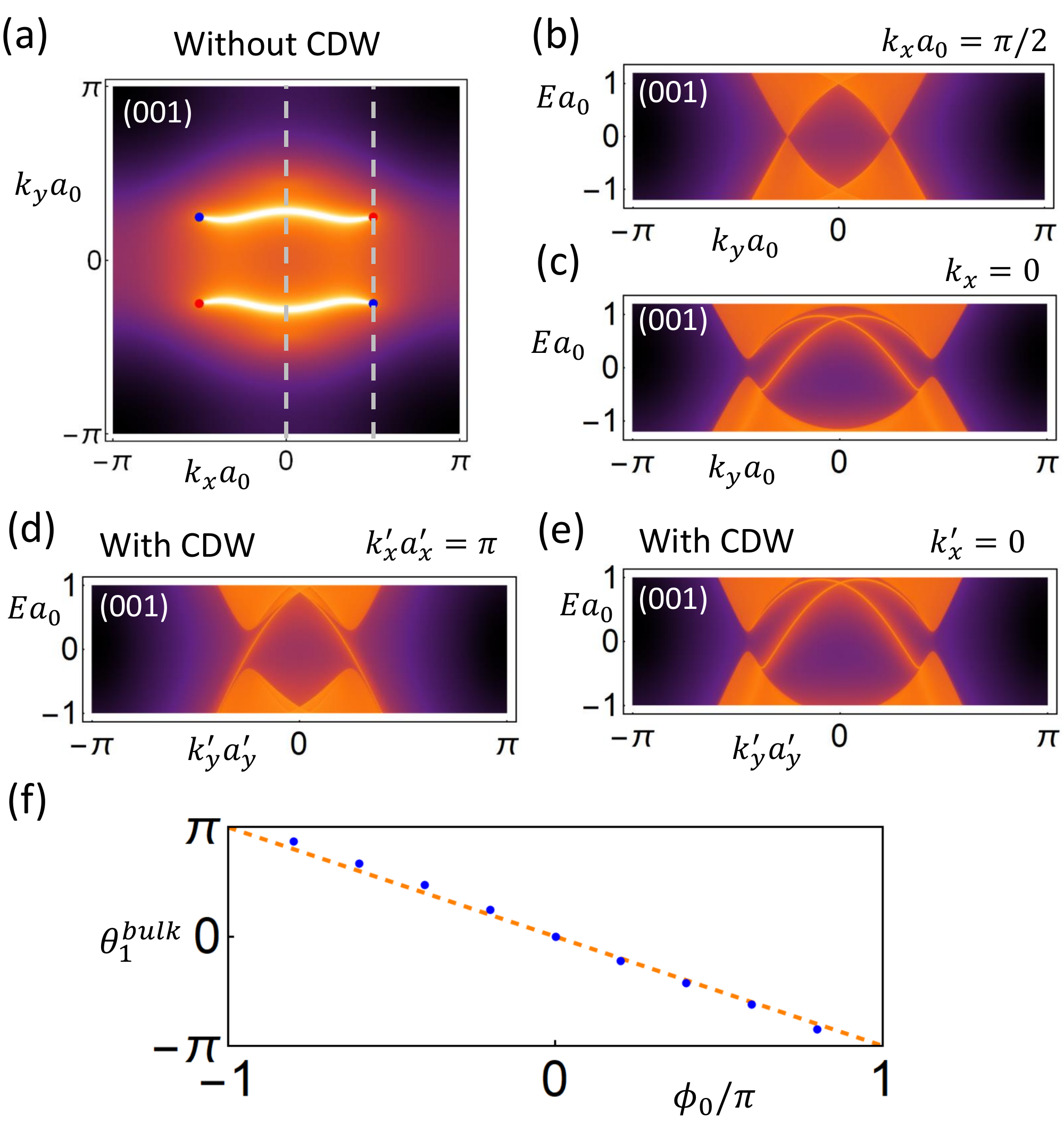}
    \caption{
    (a) Density of states on the (001) surface of the TB model $H_{TB,0}$ in the absence of a CDW.
    The red and blue dots in (a) indicate the projections of four Weyl points, and the bright lines are the topological surface Fermi arcs.
    The (red) blue dots in (a) indicate Weyl points with chiral charge ($-1$) 1.
    (b,c) Surface spectral function along the gray dashed lines in (a) at $k_x a_0=\pi/2$ and $k_x=0$, respectively.
    (d,e) The surface spectral function along $k_x' a_x'=\pi$ and $k_x'=0$ on the (001) surface of the TB model in the presence of a CDW ($H_{TB,0}+H_{TB,CDW}$) for $\arg(\mu_1 +\ii \mu_2)=\pi/4$, respectively.
    (f) The bulk value of the axion field as a function of $\phi_0$ in the $a=1$ valley.
    The blue dots and orange dashed line in (f) have respectively been obtained from the TB model ($H_{TB,0}+H_{TB,CDW}$) and the effective action \eqnref{eq:S_eff_a_leading}, respectively.
    }
    \label{fig:TB_Dispersion}
\end{figure}

The analysis in \secref{sec:LE} is based on low-energy effective field theory.
It is natural to ask whether our low-energy prediction of a DPME in TR-invariant Weyl-CDWs remains valid in the presence of high-energy bands (or equivalently in a UV completion).
Furthermore, the low-energy analysis in \secref{sec:LE} relies on ``valley'' quantum numbers; however, valley index is neither a generic symmetry of TB models, nor a symmetry of real solid-state materials.
To address these questions, we will construct a minimal TB model of a TR-invariant Weyl-CDW and compute the bulk-average value of the valley axion field and the DPME, which we will compare to those predicted by the effective action.

Prior to the onset of a CDW, we begin with an orthorhombic lattice, in which we choose for simplicity the lattice constants to be $a_x=a_y=a_z=a_0$.
We then consider there to be two sublattices in each unit cell.
We next place a Kramers pair of spinful $s$ orbitals on one sublattice and a Kramers pair of spinful $p_y$ orbitals on the other sublattice, resulting in a four-component basis $c^\dagger_{\bsl{k},i,s}$, where $i=1,2$ is the sublattice index and $s=\pm$ is the spin index.
We then construct a four-band TB model $H_{TB,0}$ that preserves TR and $m_y$ symmetries, which has the following form
\eq{
\label{eq:H_TB_0}
H_{TB,0}=\sum_{\bsl{k}}c^\dagger_{\bsl{k}} h_{TB,0}(\bsl{k}) c_{\bsl{k}}\ ,
}
where
\eqa{
& h_{TB,0}(\bsl{k})\\
&=\frac{1}{a_0}\left[ d_1 \tau_z\sigma_0+d_2\tau_y\sigma_0+d_3\tau_x\sigma_x+d_4\tau_x\sigma_z+d_5\tau_y\sigma_x\right]\ ,
}
in which $\tau_{0,x,y,z}$ and $\sigma_{0,x,y,z}$ have been redefined to respectively act on the sublattice and spin indices, and where explicit expressions for $d_{1,2,3,4,5}$ are provided in \appref{app:TB}.
Throughout this work, we will take the inverse of the lattice constant without strain $1/a_0$ as the unit of energy, which occurs because we have employed a convention in which $\hbar=c=1$.
Using parameters specified in \appref{app:TB}, $H_{TB,0}$ hosts four Weyl points at
\eq{
\bsl{k}_{a,\alpha} = (-1)^{a-1}(\alpha\frac{\pi}{2 a_0}, \frac{\pi}{4 a_0}, 0 )\ ,
}
that are related by TR and $m_{y}$ symmetries.
The projection of the bulk Weyl points onto the (001) surface is shown with blue and red dots in \figref{fig:TB_Dispersion}(a), where a (red) blue dot  indicates a Weyl point with a chiral charge ($-1$) 1.
The topological surface Fermi arcs that connect Weyl points with opposite chiralities appear as bright curves in \figref{fig:TB_Dispersion}(a), and the linear dispersion of the (001)-projecting bulk bands from the Weyl points is shown in \figref{fig:TB_Dispersion}(b).
On the $k_x=0$ plane, $H_{TB,0}$ is gapped in the bulk and exhibits a nontrivial TR-protected 2D $Z_2$ index.
The nontrivial $Z_2$ topology is indicated by the appearance of gapless helical modes along $k_x=0$ on the (001) surface, as shown in \figref{fig:TB_Dispersion}(c).

We next add a CDW term that preserves TR and $m_y$ symmetries into the TB model, where the CDW coupling takes the form
\eqa{
 \label{eq:TB_CDW}
& H_{TB,CDW}=\sum_{\bsl{k}} c^\dagger_{\bsl{k}+(\frac{\pi}{a_0},0,0)} \left[-\ii \mu_1 \sin(k_x a_0) M_1(k_y,k_z)\right. \\
&\left. +\mu_2 M_2(k_y,k_z)\right] c_{\bsl{k}}\ ,
}
in which $M_1$ and $M_2$ are Hermitian matrices whose explicit forms are provided in \appref{app:TB}, and where $\mu_1$ and $\mu_2$ are real scalar parameters.
\eqnref{eq:TB_CDW} suggests that the CDW term contains two channels that are characterized by two real coupling constants, $\mu_1$ and $\mu_2$.
Throughout this work, we will set $|\mu_1 +\ii \mu_2|=0.3/a_0$ for all of the numerical calculations for the TB model in the presence of the CDW.
Unlike \refcite{Sehayek2020CDWWeyl}, we do not study the microscopic origin of the CDW order parameter in this work, as the main goal of introducing the TB model is simply to provide a UV completion of the low-energy theory on which our analysis is rigorously based.
Owing to the lattice-commensurate nature of the CDW in $H_{TB,CDW}$, $H_{TB,0}+H_{TB,CDW}$ has reduced (but not fully relaxed) lattice translation symmetry, where the new lattice constants of the modulated cell are given by $a_x'=2 a_0$, $a_y'=a_0$, and $a_z'=a_0$.
The CDW backfolds two Weyl points of the same valley index onto the same momentum in the reduced 1BZ
\eq{
\bsl{k}_a'=(-1)^{a-1}(\frac{\pi}{a_x'},\frac{\pi}{4 a_y'},0)
}
to form an unstable 3D Dirac fermion, which then becomes gapped.
The gap induced by the CDW is reflected by the appearance of a bulk gap at $k_x' a_x'=\pi$ in \figref{fig:TB_Dispersion}(d), which stands in contrast to the gapless (WSM) bulk in \figref{fig:TB_Dispersion}(b).

\subsection{Weak $Z_2$ Topological Insulator Phase}

While the bulk of the Weyl-CDW phase is gapped, \figref{fig:TB_Dispersion}(d) demonstrates the existence of the gapless helical edge modes along $k_x' a_x'=\pi$ on the (001) surface.
As shown in \figref{fig:TB_Dispersion}(e), there are also gapless helical edge modes on the (001) surface at $k_x'=0$.
Hence, the $k_x'=0$ and $k_x' a_x'=\pi$ planes both exhibit nontrivial TR-protected $Z_2$ topology.
This indicates that the model $H_{TB,0}+H_{TB,CDW}$ is in a WTI phase characterized by a nontrivial weak $Z_2$ index vector $(\nu_x,\nu_y,\nu_z)=(1,0,0)$, where $\nu_x,\nu_y,\nu_z$ are the weak $Z_2$ indices in the $k_{x}'=0$, $k_{y}'=0$, and $k_{z}'=0$ planes, respectively.

The nontrivial WTI index vector can be understood from the odd-integer valley layered QAHE $S_{eff,a,\Sigma}$ in the effective action \eqnref{eq:S_eff_a_leading}, provided that the high-energy bands are trivial.
To see this, we first project the TB model into each of the two valleys.
From this, we see that $H_{TB,0}+H_{TB,CDW}$ reproduces \eqnref{eq:L_WP} and \eqnref{eq:L_CDW_MF} with
\eq{
\label{eq:TB_v}
v_x=\sqrt{2}\ ,\ v_y=-2\ ,\ v_z=\frac{1}{2}\ .
}
and
\eq{
\label{eq:TB_phi_0}
m_0=(\mu_1+\ii \mu_2)e^{-\ii 2\varphi} \Rightarrow \phi_0=\arg(\mu_1+\ii \mu_2)-2\varphi\ ,
}
where $\varphi$ is the $U(1)$ gauge degree of freedom of the eigenvectors (further discussed in \appref{app:TB}).
We emphasize that, at this stage, we have not yet incorporated the effects of dynamical strain, which will be added in the next section.
After restoring the Fermi velocities for \eqnref{eq:sigma_H_a} (\appref{app:S_eff}), we can use \eqnref{eq:TB_v} and $\bsl{Q}=(\pi/a_0,0,0)=(2\pi/a_x',0,0)$ to derive the valley Hall conductivity induced by the CDW wavevector, which we find to be given by
\eq{
\bsl{\Sigma}_{H,a}=(-1)^{a-1}  \frac{e^2}{2\pi}\left(\frac{1}{a_x'},0,0\right)\ .
\label{eq:valleyHallInducedBy}
}
\eqnref{eq:valleyHallInducedBy} implies that the two valleys as a set contribute two counterpropagating chiral edge modes for each $k_{x}'$-indexed plane in the 1BZ.
The two couterpropagating chiral modes exhibit a TR-protected crossing at $k_x'=0$ and $k_x'=\pi/a_x'$, indicating the presence of a nontrivial weak $Z_2$ index $\nu_{x}=1$.

We pause to compare the WTI phase of $H_{TB,0}+H_{TB,CDW}$ (\eqnref{eq:H_TB_0} and \eqnref{eq:TB_CDW}) to the WTI Dirac-CDW phase in \refcite{Wieder2020AxionCDWWeyl}.
In \refcite{Wieder2020AxionCDWWeyl}, which was revised to include TR-symmetric semimetal-CDWs during the final stages of preparing this work, the authors study a TR-invariant Dirac semimetal that is gappd by a CDW.
Specifically, in \refcite{Wieder2020AxionCDWWeyl}, two 3D Dirac points become coupled by the CDW order parameter, and the gapped Dirac-CDW additionally respects spatial inversion symmetry when the phase of the CDW order parameter $\phi_0=0,\pi$.
The authors of \refcite{Wieder2020AxionCDWWeyl} find that $\phi_0=0,\pi$ correspond to two distinct WTI phases (with nontrivial weak $Z_2$ indices) that differ by a fractional lattice translation in the modulated cell.
Although the model in \refcite{Wieder2020AxionCDWWeyl} appears to be similar to $H_{TB,0}+H_{TB,CDW}$ with the two valleys moved to the same momentum, the $m_y$ symmetry enforced in this work is essentially different from the inversion symmetry at $\phi_0=0,\pi$ in \refcite{Wieder2020AxionCDWWeyl}.
This can be seen by recognizing that $m_y$ in this work is a symmetry of the Weyl-CDW for \emph{all} values of $\phi_0$, including $\phi_0\neq 0,\pi$.
We do note that $H_{TB,0}+H_{TB,CDW}$ has an effective spinless-$m_x$ symmetry at $\arg(\mu_1 +\ii \mu_2)=0, \pi$ (see \appref{app:TB}).
However, the spinless $m_x$ symmetry is an artifact of our simple TB model, which only occurs because all of the Weyl points are located at the $k_z=0$ plane.
Because the effective action \eqnref{eq:L_tot} allows the four Weyl points to move away from the $k_z=0$ plane -- thus breaking any form of $m_x$ symmetry (effective or physical), the artificial spinless $m_x$ symmetry of our TB model does not affect any of the conclusions of and analysis performed in this work.

\subsection{Valley Axion Field}

In addition to the valley QAH term $S_{eff,a,\Sigma}$, the term $S_{eff,a,\theta}$ in \eqnref{eq:S_eff_a_leading} indicates that the bulk average value of the valley axion field is
\eq{
\label{eq:theta_bulk_RFV_LE}
\theta_a^{bulk}= (-1)^{a-1} \sgn{v_x v_y v_z} \phi_0 =(-1)^{a} \phi_0
}
after restoring the Fermi velocities (\appref{app:S_eff}).
To compare \eqnref{eq:theta_bulk_RFV_LE} with the TB model, we first set $\theta_a^{bulk}=0$ at $\arg(\mu_1+\ii \mu_2)=0$ as the reference for the TB model.
We are required to set a reference for the evaluation of $\theta_a^{bulk}$, because the valley axion field is defined over an open manifold, such that its bulk average value is reference-dependent.
Hence, we may only evaluate the change of $\theta_a^{bulk}$ relative to a reference value.
Because the low-energy result \eqnref{eq:theta_bulk_RFV_LE} suggests that $\theta_a^{bulk}=0$ for $\phi_0=0$, we further set the U(1) degree of freedom $\varphi$ in \eqnref{eq:TB_phi_0} to $\varphi=0$ in order to match the zero points of $\theta_a^{bulk}$ in the low-energy result and in the TB model.
We may then use $\phi_0=\arg(\mu_1+\ii \mu_2)$ for the TB model and evaluate $\theta_a^{bulk}$ with~\cite{Qi2008TFT}
\eqa{
&\theta_a^{bulk}(\phi_0)=-\int_0^{\phi_0} d\phi_0\int_a \frac{d^3 k'}{16\pi} \varepsilon^{i_1 i_2 i_3 i_4} \Tr[\mathcal{F}_{i_1 i_2} \mathcal{F}_{i_3 i_4}]\ ,
\label{eq:thetaValleyforTBProj}
}
where $\bsl{k}'$ is only integrated over the half of the reduced 1BZ that contains the $a$th valley.
In \eqnref{eq:thetaValleyforTBProj},
\eq{
\mathcal{F}_{i_1 i_2}=\partial_{k_{i_1}'}\mathcal{A}_{i_2}-\partial_{k_{i_2}'}\mathcal{A}_{i_1}-\ii [\mathcal{A}_{i_1},\mathcal{A}_{i_2}]
}
is the non-Abelian Berry curvature of the non-Abelian Berry connection $[\mathcal{A}_i]_{n m}=-\ii \bra{u_{n,\bsl{k}'}}\partial_{k_i'} \ket{u_{m,\bsl{k}'}}$, and $i_{1,2,3,4}$ takes values from $1$ to $4$ where $k'_4=\phi_0$.
Although $\theta_a^{bulk}$ can be equivalently evaluated based on the Chern-Simons 3-form~\cite{Qi2008TFT}, this method would require to choose the gauge and $\bsl{k}$-space boundary condition very carefully, and thereby we use gauge-invariant \eqnref{eq:thetaValleyforTBProj} in this work.

The resulting numerical computation of $\theta_a^{bulk}$ for the $a=1$ valley is shown in \figref{fig:TB_Dispersion}(f), and extremely closely matches the value $\theta_1^{bulk}=-\phi_0$ expected from the low-energy action.
However, there is still a quantitative deviation between the low-energy and TB results, which occurs because the bulk valley axion field is not defined over a closed manifold, and is thus not quantized, implying that high-energy degrees of freedom (which are necessarily present in a solid-state material) can drive the value away from the low-energy result.
Nevertheless, the relatively small deviation between the TB and low-energy results in \figref{fig:TB_Dispersion}(f) suggests that the effect of high-energy modes on the valley axion field is small in the TB model employed in this work.

\begin{figure*}[t]
    \centering
    \includegraphics[width=1.9\columnwidth]{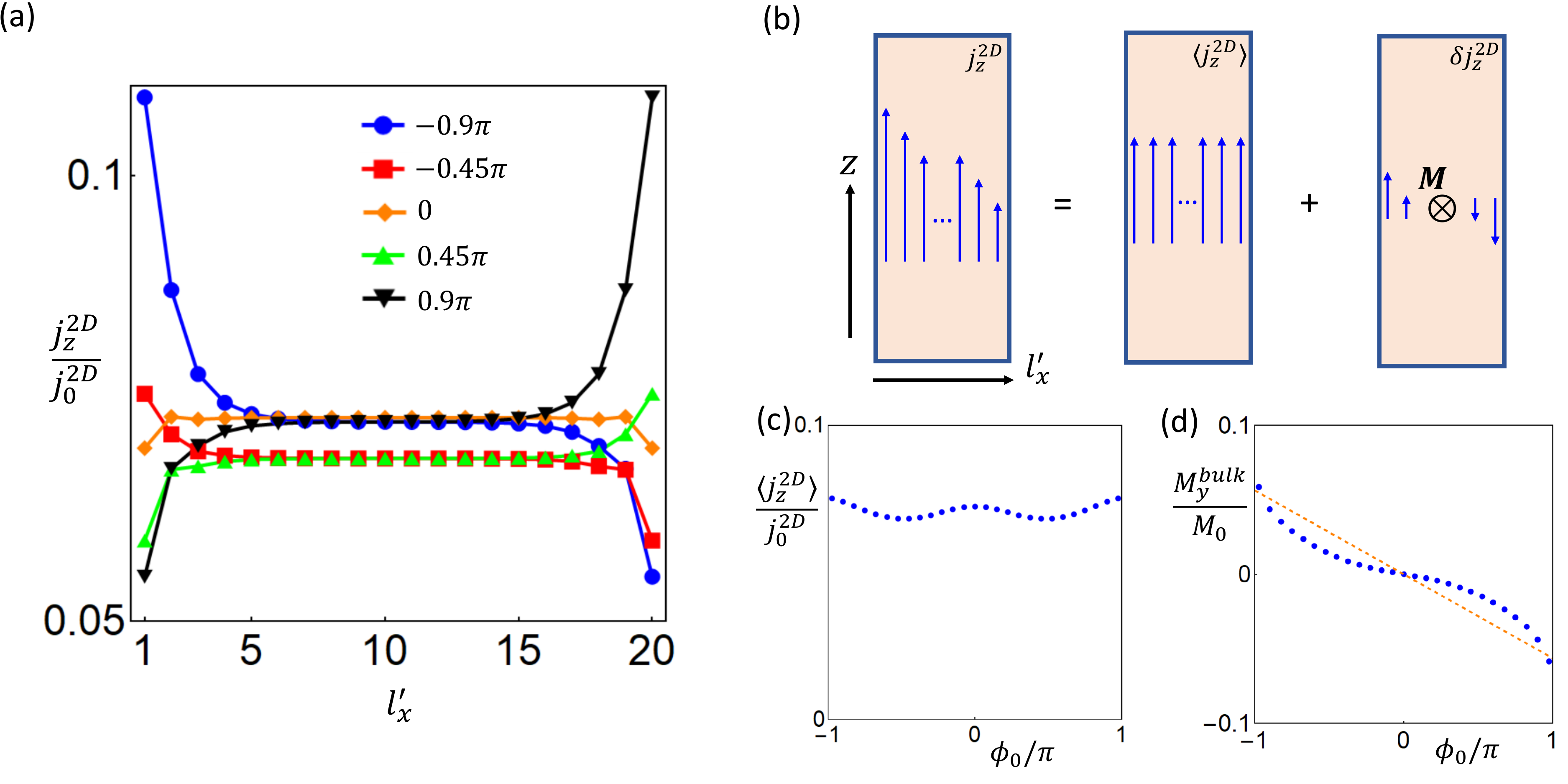}
    \caption{
    (a) The layer distribution of the 2D strain-induced $z$-directed current density in the slab configuration of $H_{TB,0}+H_{TB,CDW}$ for several typical values of $\phi_0$.
    (b) The current distribution can be split into a uniform background current and the nonuniform magnetization current.
    (c) The spatial average of the 2D current density as a function of $\phi_0$, where $j^{2D}_0= e \dot{u}_{zz} /a_0$.
    (d) The $\phi_0$ dependence of the bulk magnetization along $y$, where the blue dots and orange dashed line indicate data obtained from the TB model $H_{TB,0}+H_{TB,CDW}$ and the effective action (\eqnref{eq:S_eff_a_uzz}), respectively.
    }
    \label{fig:TB_response}
\end{figure*}

\subsection{Piezoelectric Effect and the DPME}

Finally, we use the TB model to verify the strain-induced piezoelectric effect and the DPME described by the term $S_{eff,a,u_{zz}}$ in \eqnref{eq:S_eff_a_leading}.
We incorporate the effects of strain into the TB model by adding a prefactor of $(1- \frac{(\Delta r_z)^2}{(\Delta \bsl{r})^2} u_{zz})$ for each hopping term in $H_{TB,0}$, where $\Delta\bsl{r}$ is the displacement of the hopping~\cite{Li2016StrainTMD} (see \appref{app:el-str} and \appref{app:TB} for further details).
As a result, the low-energy projection of the extra strain term implies that the added strain reproduces \eqnref{eq:L_str} with
\eq{
\label{eq:TB_xi}
\xi_0=\xi_z=0\ ,\ \xi_x=\frac{1}{\sqrt{2} a_0}\ ,\ \xi_{y}=-\frac{1}{\sqrt{2} a_0}\ .
}
Because the TB model is a WTI with the weak $Z_2$ index vector $(1,0,0)$, we consider a slab configuration with $N$ layers perpendicular to $x$ with periodic boundary conditions along $y$ and $z$.
The slab can be viewed as a quasi-2D system, and we may therefore calculate the 2D piezoelectric tensor of the slab using~\cite{Vanderbilt2000BPPZ}
\eqa{
\label{eq:PET_A}
\chi_{izz}^{2D}=e\int \frac{d^2k'}{(2\pi)^2} \sum_{n\in\ occupied}\partial_{u_{zz}}\left. \mathcal{A}_{n,i}^{slab}(k_y',k_z')\right|_{u_{zz}\rightarrow 0}\ ,
}
where $\mathcal{A}_{n,i}^{slab}=-\ii \bra{\varphi_{k_y',k_z',n}} \partial_{k_i'}\ket{\varphi_{k_y',k_z',n}} $ and $\ket{\varphi_{\bsl{k}',n}}$ is the periodic part of the Bloch states of the slab Hamiltonian in the presence of strain.
By inserting a projection operator onto each layer of the slab, we can then derive the 2D piezoelectric tensor for each layer, which we label $\chi_{izz}^{2D}(l_x')$, where $l_x'=1,2,...,N$ is the layer index (see \appref{app:TB} for details).
The 2D current density induced by the infinitesimal dynamical strain for each layer is then given by $j_{i}^{2D}(l_x')=\chi_{izz}^{2D}(l_x') \dot{u}_{zz}$.
Using this method, we next calculate the 2D $z$-directional current density of each layer for varying values of $\phi_0$ and $N=20$ by setting $\varphi=0$ in \eqnref{eq:TB_phi_0} and using $\phi_0=\arg(\mu_1+\ii \mu_2)$.
We note that the current along the y direction conversely vanishes in our numerics at each value of $\phi_{0}$, due to the bulk $m_y$ symmetry at each value of $\phi_0$.

In \figref{fig:TB_response}(a), we plot the current density distribution $j_{z}^{2D}(l_x')$ for $\phi_0=-0.9 \pi, -0.45 \pi, 0, 0.45 \pi,-0.9 \pi$.
As schematically shown in \figref{fig:TB_response}(b), we can decompose the current density distribution into a uniform background current $\langle j_{z}^{2D} \rangle$ (averaged over the layer index) and a layer-dependent part $\delta j_z^{2D}(l_x')=j_{z}^{2D}(l_x')-\langle j_{z}^{2D} \rangle$.
The uniform background current $\langle j_{z}^{2D} \rangle$ characterizes the uniform piezoelectric response, and, as shown in \figref{fig:TB_response}(c), the piezoelectric current is nearly independent of $\phi_0$, as expected from the low-energy expression \eqnref{eq:j_M_eff}.

On the other hand, the layer-dependent contribution to the layer current density  $\delta j_z^{2D}(l_x')$ is asymmetrically distributed.
Specifically, $\delta j_z^{2D}(l_x')$ exhibits opposite signs near the two surfaces, resulting in a bulk magnetization $M^{bulk}_y$.
To calculate the bulk magnetization, we treat the $l_x'$th layer as a uniform 2D system that is infinite in the $y,z$ directions but finite in the $x$ direction as $x\in [(l_x'-1) a_x',l_x' a_x']$.
From this, we then express the 3D current density as
\eq{
\label{eq:j_3D_z}
j^{3D}_z(x)= \sum_{l_x'} \Theta_{l_x'}(x) j_z^{2D}(l_x')/a_x'\ ,
}
where $\Theta_{l_x'}(x)=1$ for $x\in [(l_x'-1) a_x',l_x' a_x']$ and $\Theta_{l_x'}(x)=0$ otherwise.
We next take the magnetization at the center of the sample $x_0= a_x'N/2$ derived from the Biot-Savart law to be the bulk magnetization, which yields
\eqa{
\label{eq:M_bulk_j_gen}
M^{bulk}_y=-\frac{1}{2}\sum_{l_x'} \delta  j_z^{2D}(l_x') \sgn{l_x'-\frac{N+1}{2}}\ ,
}
where the uniform background contribution naturally vanishes.
In \eqnref{eq:M_bulk_j_gen}, we have chosen $\sgn{0}=0$, because, when $N$ is odd, there exists an $l_x'=(N+1)/2$ layer with a vanishing current contribution.

Using \eqnref{eq:j_3D_z}, we plot $M^{bulk}_y$ as a function of $\phi_0$ in \figref{fig:TB_response}(d).
To compare with the TB result, we restore the Fermi velocity for the low-energy expression \eqnref{eq:M_bulk} and substitute \eqnref{eq:TB_v} and \eqnref{eq:TB_xi} into \eqnref{eq:M_bulk}, from which we obtain
\eq{
\label{eq:M_bulk_TB_LE}
\frac{\bsl{M}_{bulk}}{M_0}=\frac{a_0 \xi_y \sgn{v_x v_y v_z}}{2\pi^2 v_y}\phi_0  \bsl{e}_y  =-\frac{1}{4 \pi^2 \sqrt{2} }\phi_{0}  \bsl{e}_y\ ,
}
where $M_0=e \dot{u}_{zz} /a_0$.
As shown in \figref{fig:TB_response}(d), the TB and low-energy results are of the same order of magnitude (the deviation is smaller than 70\% of the low-energy result).
This agrees with our earlier determination that the high-energy modes have relatively small effects on the valley axion field (\figref{fig:TB_Dispersion}(f)).
In particular, the TB and low-energy results in \figref{fig:TB_response}(c) match extremely well as $\phi_0$ approaches $\pm \pi$ (the deviation is smaller than 7\% of the low-energy result).
As discussed below, the agreement between the TB and low-energy results can be attributed to the TB model exhibiting boundary gap closings at exactly $\phi_0=\pm \pi$.

\section{Boundary TQPT and DPME Jump}
\label{sec:Boundary_TQPT_DPME_Jump}

\begin{figure}[t]
    \centering
    \includegraphics[width=\columnwidth]{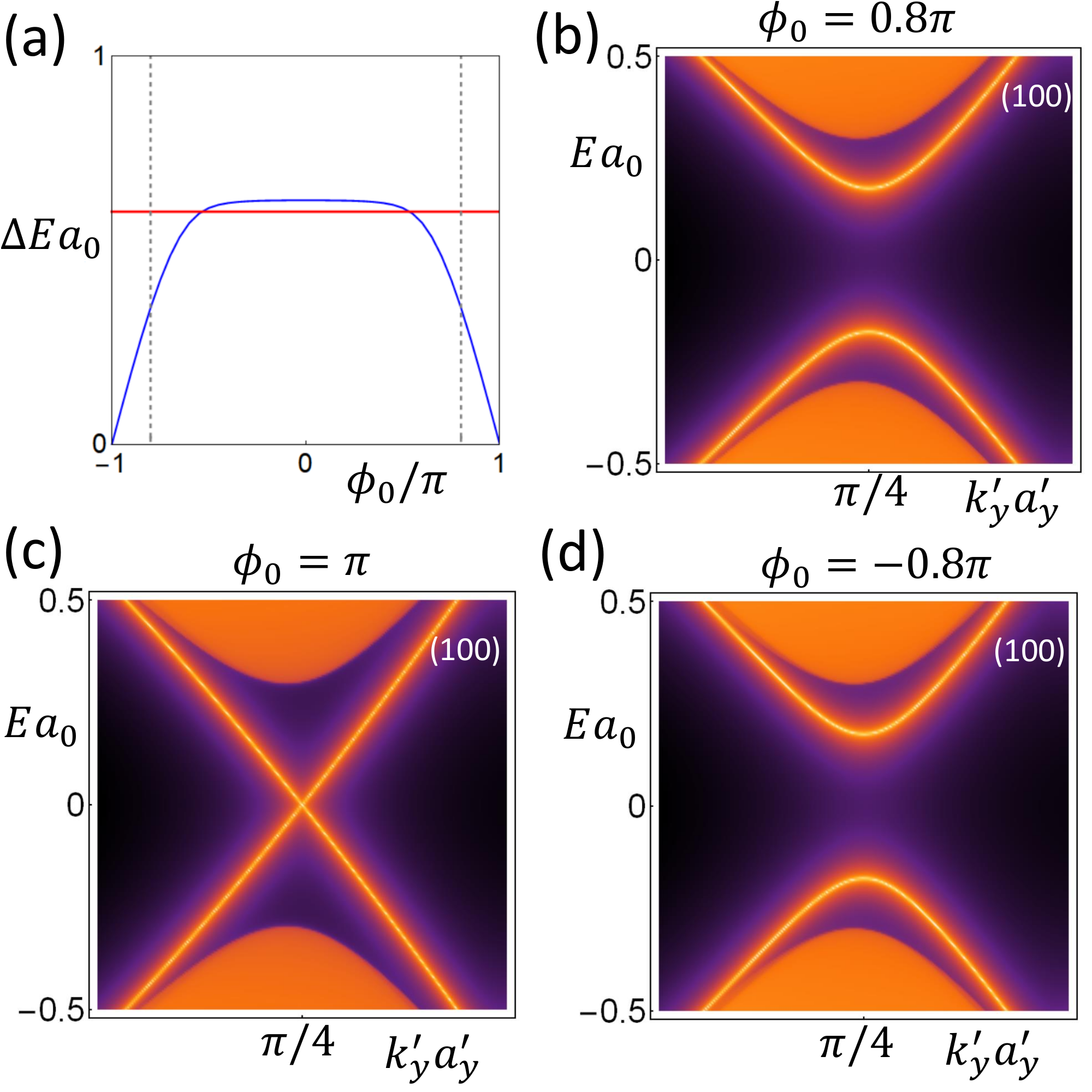}
    \caption{
    (a) The gap of the slab TB model $H_{TB,0}+H_{TB,CDW}$ at $(k_y' a_y', k_z' a_z')=(\pi/4,0)$ as a function of $\phi_0$ for periodic boundary conditions (red) and open boundary conditions (blue) along only $x$ (keeping the $y$ and $z$ directions periodic).
    The boundary gap closes at $\phi_0=\pm\pi$ simultaneously on the top and bottom surfaces, due to the special form of the TB model $H_{TB,0}+H_{TB,CDW}$.
    In more realistic models (see \figref{fig:extra_term}), the gap closing on each surface occurs at a different value of $\phi_0$.
    In (b), (c), and (d), we plot the surface spectral function of the (100) surface of the TB model $H_{TB,0}+H_{TB,CDW}$ at $\phi_0=0.8 \pi$, $\phi_0=\pi$, and $\phi_0=-0.8 \pi$, respectively.
    In (b), (c), and (d), the dispersion is plotted along $k_z'=0$ near $k_y' a_y'=\pi/4$.
    }
    \label{fig:TB_BOTQPT_Surface}
\end{figure}

In this section, we will first show that the slab configuration of $H_{TB,0}+H_{TB,CDW}$ has a boundary gap closing at $\phi_0=\pm \pi$, which we will show to be a boundary TQPT that changes the surface $Z_2$ index and induces a discontinuous change of the DPME.
This boundary gap closing accidentally happens on the two surfaces of the slab at the same value of $\arg(\mu_1 +\ii \mu_2)$, allowing us to fully interpret the discontinuous change of DPME within the low-energy theory.
We will then add an extra term to $H_{TB,0}+H_{TB,CDW}$ to split the accidental simultaneous surface gap closing, resulting in a more realistic model in which the gap closings on the two surfaces occur at different values of $\arg(\mu_1 +\ii \mu_2)$.
Lastly, we will demonstrate that a jump in the DPME still occurs across each surface gap closing, though the low-energy theory is incapable in fully describing the jump due to the unavoidable presence of a gapless boundary helical mode on one side of the jump.
Throughout this section, we will continue to choose $\phi_0=\arg(\mu_1+\ii \mu_2)$ by setting $\varphi=0$ in \eqref{eq:TB_phi_0}, except in \secref{sec:split_surface_gapclosing}.

\subsection{TB Model}

According to \eqnref{eq:TB_phi_0}, $\phi_0$ only appears in the TB model as $\cos(\phi_0)$ and $\sin(\phi_0)$ in $\mu_1$ and $\mu_2$, respectively, and thus any TB result must be periodic in $\phi_0$.
Hence, tuning $\mu_2$ from negative to positive while keeping $\mu_1<0$ should drive $\phi_0$ from $-\pi$ to $\pi$ and give a jump of the magnetization, as shown in \figref{fig:TB_response}(d).
The dramatic difference between the current distributions at $\phi_0=\pm 0.9$ in \figref{fig:TB_response}(a) provides evidence of the expected jump in the bulk magnetization.
\figref{fig:TB_BOTQPT_Surface}(a) suggests that the jump of the DPME at $\phi_0=\pm\pi$ happens along with the boundary gap closing while the bulk stays gapped.
Moreover, the gap closing manifests as one 2D gapless Dirac cone in each valley on each surface perpendicular to $x$, as shown in \figref{fig:TB_BOTQPT_Surface}(b-d).
Because there are two TR-related Dirac cones on one surface at the gap closing, then the surface gap closing has the same form as the 2D $Z_2$ transition that happens at a TR-related pair of generic momenta~\cite{Kane2005Z2,Moore20072DTRI, Murakami2007QSH}.
Because the bulk remains gapped across the transition, then the surface gap closings represent examples of boundary TQPTs, which can be detected by jumps in the DPME.

\begin{figure}[t]
    \centering
    \includegraphics[width=\columnwidth]{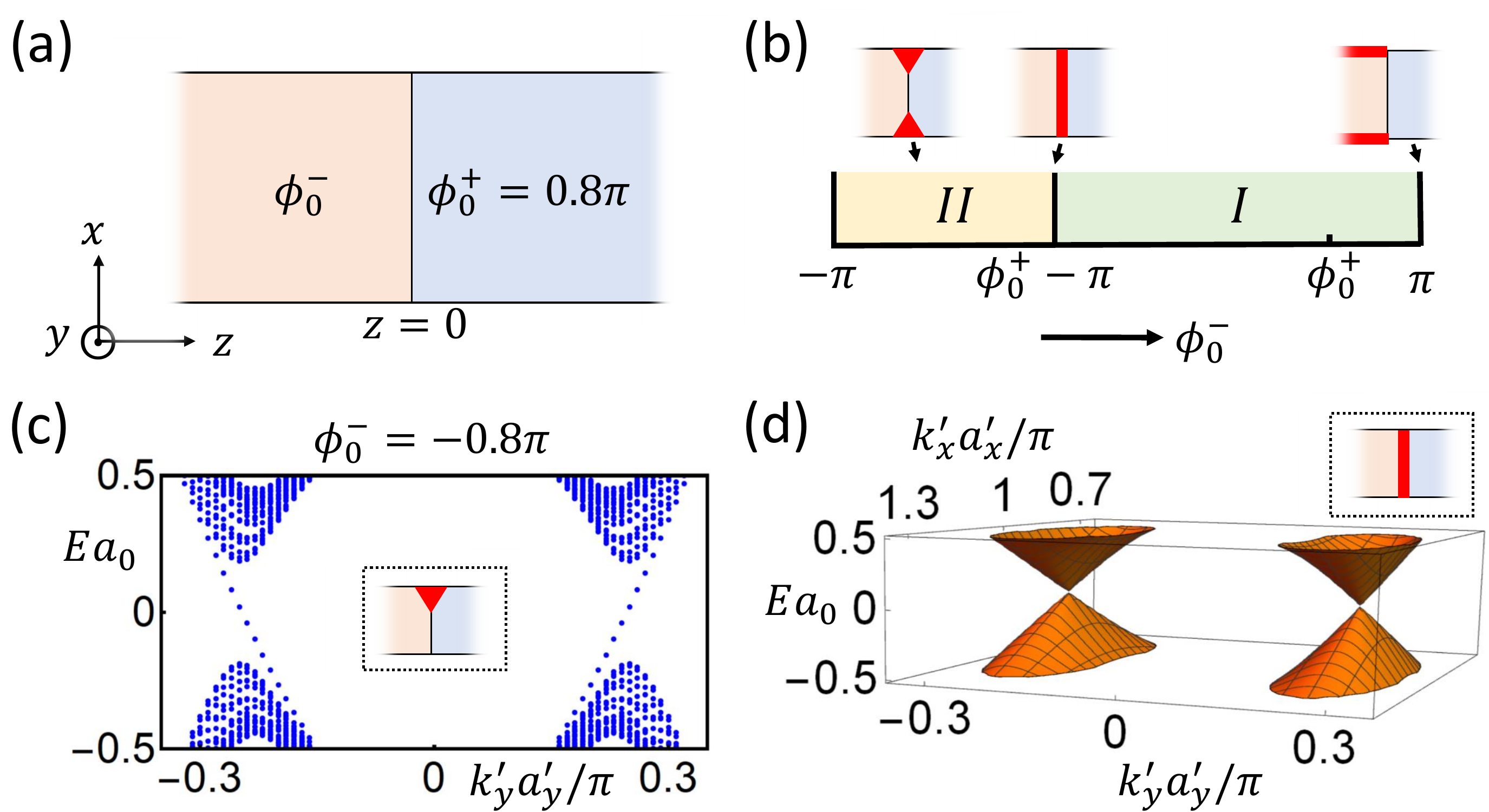}
    \caption{
    (a) A schematic showing the domain wall structure along $z$ with an open boundary condition along $x$ and periodic boundary conditions along $y$ for the TB model $H_{TB,0}+H_{TB,CDW}$.
    The interface lies at $z=0$, and we have omitted the surfaces at large $|z|$.
    The lower panel in (b) shows the phase diagram for the domain wall structure as $\phi_0^-$ is varied while fixing $\phi_0^+=0.8\pi$, which contains two phases $I$ and $II$ and two transition points $\phi_0^-=\pm\pi$ and $\phi_0^-=\phi_0^+-\pi$.
    The upper panel in (b) shows the positions (red) of the gapless modes in the domain wall structure for the corresponding values of $\phi_0^-$.
    (c) The energy dispersion near the top edge (inset) of the interface for $\phi_0^-=-0.8 \pi$ in the phase $II$ of (b).
    (d) The energy dispersion near the interface (inset) for $\phi_0^-=\phi_0^+-\pi=-0.2 \pi$.
    }
    \label{fig:TB_BOTQPT_Domain}
\end{figure}

Another signature of the boundary TQPT appears in the domain wall structure shown in \figref{fig:TB_BOTQPT_Domain}(a).
We consider a slab configuration that is split into two parts along the $z$ direction, where each part exhibits a different value of the CDW phase $\phi_0$ (specifically $\phi_0^{+}$  for $z>0$ and $\phi_0^{-}$ for $z<0$), while all other parameters in the slab are taken to be the same for the two parts.
The sample is set to be periodic in the $y$ direction and open in the $x$ direction.
In our numerical calculations, we have specifically employed a slab with 20 layers along $x$ and have fixed $\phi_0^{+}=0.8\pi$.
We note that we have also chosen $20$ layers along $z$ for $z<0$ and $20$ layers along $z$ for $z>0$, but we do not depict the additional surface modes at large $|z|$ (\ie\ the leftmost and rightmost surfaces in \figref{fig:TB_BOTQPT_Domain}(a)).

We plot the phase diagram of the domain wall structure in \figref{fig:TB_BOTQPT_Domain}(b) by varying $\phi^{-}_0$ from $\phi_0^{+}$ to $\pi$ and then from $-\pi$ back to $\phi_0^{+}$.
As a result, we identify two phases, which we label as $I$ and $II$.
The phase $I$ contains the point $\phi^{-}_0=\phi^{+}_0$, implying that the surfaces of both parts of the slab are not related by boundary TQPTs, such that boundary between the two domains is gapped.
As $\phi_0^{-}$ is varied from $\pi+0^-$ to $-\pi+0^+$, the gap closes on the top and bottom surfaces normal to the $x$ direction in the $z<0$ region, as discussed above and shown in \figref{fig:TB_BOTQPT_Surface}.
This indicates that the $z<0$ region has undergone a pair of boundary TQPTs to enter phase II.
The appearance of 1D gapless helical modes at the edge of the $z=0$ interface in phase II confirms the presence of a nonzero surface relative $Z_2$ index for the two sides of the gap closing (\figref{fig:TB_BOTQPT_Domain}(c)).
The 1D gapless helical modes in phase II persist until $\phi_0^{-}$ reaches $\phi^{+}_0-\pi$, where the gap closes in the 2D bulk of the interface as shown in \figref{fig:TB_BOTQPT_Domain}(d).
The interface gap closing again manifests as two gapless Dirac cones at two valleys and thus changes the $Z_2$ index of the interface, coinciding the disappearance of the helical edge modes in phase I.

\subsection{Low-energy Effective Theory}

\begin{figure}[t]
    \centering
    \includegraphics[width=\columnwidth]{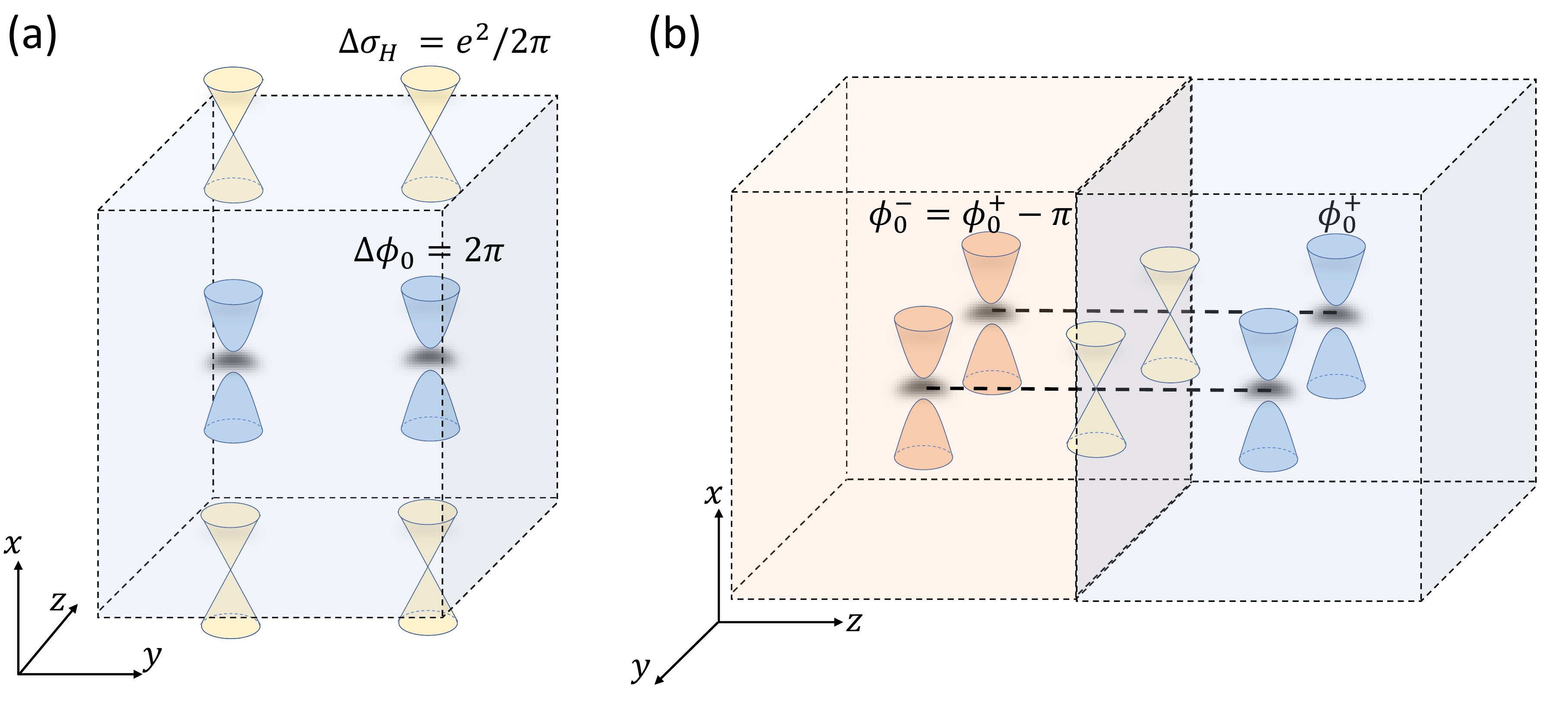}
    \caption{(a) Gap closings on the surfaces perpendicular to $x$ in a pair of boundary TQPTs.
    In this figure, we focus on the (artificial) case where the system is fine tuned such that gap closings simultaneously occur on both surfaces.
    (b) 2D gapless Dirac cones at the interface of the domain when the two sides of the domain wall differ by $\pi$ in $\phi_0$.
    In general, the number of 2D gapless Dirac cones at the interface is $2+4n$ where $n$ is a non-negative integer.
    }
    \label{fig:LE_boundary}
\end{figure}

We now interpret the boundary TQPT in the TB model from the perspective of the low-energy theory.
According to \eqnref{eq:L_tot}, one bulk Dirac cone has two mass terms, and thus the bulk gap closing for \eqnref{eq:L_tot} requires fine-tuning at least two parameters, which typically does not occur in a realistic model or material.
On the (100) surface, the projections of the valleys are along the $m_y\TR$-invariant line, and the gap closing along this line only requires fine-tuning one parameter, according to \refcite{Yu2019PETTQPT}.
The analysis in \refcite{Yu2019PETTQPT} further suggests that the gap closing appears as one gapless surface Dirac cone for each valley (\figref{fig:LE_boundary}(a)) and is thus a surface $Z_2$ transition, coinciding with the TB results in \figref{fig:TB_BOTQPT_Surface} and \figref{fig:TB_BOTQPT_Domain}(c).
The parameter values for which the gap closings appear depend on the boundary condition that we choose in \eqnref{eq:L_tot} (see \appref{app:S_eff} for a special boundary condition that realizes both surface gap closings at $\phi_0=\pi$).
Nevertheless, the codimension-1 nature of the gap closing indicates that,  even if the boundary conditions are varied, it is still difficult to remove the gap closing point.
When the boundary conditions are changed, the gap closing instead shifts to a different value of $\phi_{0}$.
Indeed, as we will shortly show in using a TB model with an extra term that splits the simultaneous boundary gap closing, the boundary phase transitions are movable in $\arg(\mu_1+\ii \mu_2)$, but globally unremovable.
This agrees with the picture presented in \refcite{Wieder2020AxionCDWWeyl}, in which tuning $\phi_0$ pumps 2D TI layers in the WTI phase until a layer reaches the system boundary, causing a surface gap closing.
The same argument can also be applied to the $(\bar{1}00)$ surface.

In general, the gap closings on the $(100)$ and $(\bar{1}00)$ surfaces do not happen at the same critical value of $\phi_0$.
We find that simultaneous surface gap closings only occur when the system configurations (parameter values or boundary conditions) are designed in a fine-tuned manner such that unrealistic (\emph{i.e.} artificial) effective symmetries appear in the effective action (such as an effective TR symmetry within one valley after omitting the CDW wavevector).
This suggests that the presence of simultaneous surface gap closings in the above TB model $H_{TB,0}+H_{TB,CDW}$ is accidental.
In this accidental (fine-tuned) case, a simultaneous gap closing changes the surface valley Hall conductance by $\pm e^2/2\pi$, which, according to the bulk-boundary correspondence of the axion term, results in a change of the $\theta^{bulk}_a$ -- or equivalently $\phi_0$ -- by $2\pi$.
Combined with \eqnref{eq:M_bulk}, the $2\pi$ jump of $\phi_0$ further results in a jump of the magnetization
\eq{
\Delta\bsl{M}_{bulk}=\frac{e \xi_y}{\pi v_y}\sgn{v_x v_y v_z} \dot{u}_{zz} \bsl{e}_y=-\frac{1}{2 \pi \sqrt{2} } M_0 \bsl{e}_y\ ,
}
in which the Fermi velocities have been restored (see \appref{app:S_eff}) and the parameter values derived from the projection of the TB model (\eqnref{eq:TB_v} and \eqnref{eq:TB_xi}) have been used in the second equality.
The predicted $\Delta\bsl{M}_{bulk}$ precisely matches the jump given by the TB model in \figref{fig:TB_response}(d).
Therefore, the boundary TQPT and the induced jump of the DPME can be captured within the low-energy theory when valleys are well-defined and when the gap closings happen simultaneously on both surfaces.

Lastly, we will use the low-energy theory to explain the gap closing in the 2D bulk of the interface of the domain wall (\figref{fig:TB_BOTQPT_Domain}(d)).
Within the low-energy theory, if two gapped Dirac cones have a mass phase difference $\pi$ and form a domain wall structure, then there must be an odd number of 2D gapless Dirac cones localized at the interface~\cite{Sehayek2020CDWWeyl}.
Therefore, when $\phi_0^{\pm}$ in \figref{fig:TB_BOTQPT_Domain}(a) differ by $\pi$, an odd number of 2D gapless Dirac cones appear at the interface for each valley, as shown in \figref{fig:LE_boundary}(b), and the gap closing correspondingly changes the $Z_2$ index of the interface.
Indeed, the description given by the low-energy theory coincides with the TB result shown in \figref{fig:TB_BOTQPT_Domain}(b) and (d).

\subsection{Separate Gap Closings on Two Surfaces}
\label{sec:split_surface_gapclosing}

\begin{figure*}
\includegraphics[width=1.9\columnwidth]{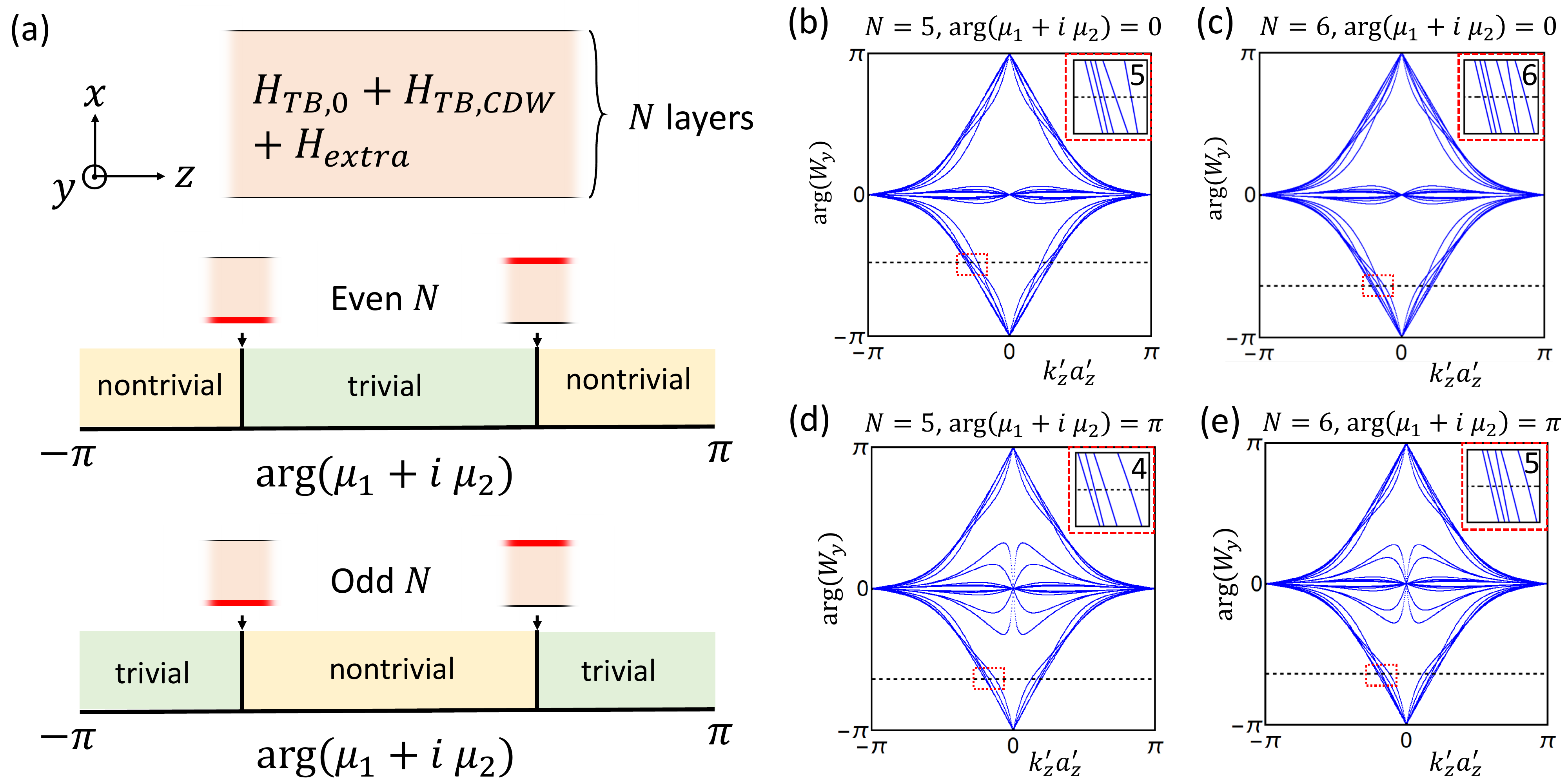}
\caption{Slab geometry for a realistic model of a TR-invariant Weyl-CDW.
    The top panel in (a) shows a slab of $H_{TB,0}+H_{TB,CDW}+H_{extra}$ that is open in $x$ and periodic in $y$ and $z$.
    The middle panel in (a) schematically illustrates the case of a generic TR-invariant minimal Weyl-CDW where the gap closings on the top and bottom surfaces happen separately for an even number of layers.
    In the case of an even number of layers, the slab has two phases: (i) a $Z_2$-trivial phase that includes $\arg(\mu_1+\ii \mu_2)=\pi$, and (ii) a $Z_2$-nontrivial phase that includes $\arg(\mu_1+\ii \mu_2)=\pi$.
    The bottom panel in (a) corresponds to the case in which the slab has an odd number of layers, which causes the $Z_{2}$-trivial and nontrivial phases to flip relative to the middle panel in (a).
    (b-e) The y-directed slab Wilson loop~\cite{Wieder2018AXIFragile} as a function of $k_z'a_z'$ for the slab configuration with $5$ and $6$ layers and $\arg(\mu_1+\ii\mu_2)=0,\pi$.
    In (b-e), $W_y$ is the eigenvalue of the Wilson loop evaluated along $k_y'$.
The dashed line lies at a Wilson energy of $-1.5$ in (b) and $-2$ in (c-e).
	In the inset panels in (b-e), we show the number of Wilson loop bands passing through the dashed line in half of the 1BZ, which are 5, 6, 4 and 5 for (b,c,d,e), respectively.
 	An odd (even) number of Wilson crossings in half of the 1BZ at a fixed Wilson energy indicates that the slab $Z_{2}$ index is nontrivial (trivial)~\cite{Soluyanov2011WannierZ2,Dai2011Z2WilsonLoop}.
}
\label{fig:extra_term}
\end{figure*}

In the final part of this section, we will discuss the more general (and also more realistic) case in which the gap closings on the $(100)$ and $(\bar{1}00)$ surfaces occur at different values of $\arg(\mu_1+\ii \mu_2)$.
In this subsection, we do not set $\varphi=0$ in \eqnref{eq:TB_phi_0} and thus in general $\phi_0\neq \arg(\mu_1+\ii \mu_2)$.
We can shift the gap closings on two surfaces in opposite directions in $\arg(\mu_1+\ii \mu_2)$ by adding an extra TR- and $m_{y}$-symmetry-preserving term $H_{extra}$ in the TB model, as schematically shown in \figref{fig:extra_term}(a). (The explicit form of $H_{extra}$ is provided in \appref{app:TB}).
Because each surface gap closing is a $Z_2$ TQPT, then the $Z_2$ index of the entire slab is changed across the transition, resulting in two phases with different $Z_2$ indices for the whole slab (\ie\ the slab as a whole, for varying $\arg(\mu_1+\ii \mu_2)$, is or is not a 2D $Z_2$ TI).
Exactly which phase of the slab has nontrivial $Z_2$ index is determined by the number of layers, owing to the nontrivial weak $Z_2$ index in the bulk (\figref{fig:extra_term}(a)).

To numerically model this more generic case, we choose appropriate parameter values for the model (\appref{app:TB}) to split the simultaneous boundary phase transition into two boundary transitions: one at $\arg(\mu_1+\ii\mu_2)=0.9\pi$ on the (100) surface and another at $\arg(\mu_1+\ii\mu_2)=-0.9\pi$ on the $(\bar{1}00)$ surface.
This results in the appearance of two phases in the slab: one phase that includes $\arg(\mu_1+\ii \mu_2)=0$, and another that includes $\arg(\mu_1+\ii \mu_2)=\pi$.
As shown through slab Wilson-loop calculations~\cite{Wieder2020AxionCDWWeyl} in \figref{fig:extra_term}(b,d), the 5-layer slab is $Z_2$ trivial in the $\arg(\mu_1+\ii \mu_2)=\pi$ phase, and is nontrivial in the $\arg(\mu_1+\ii \mu_2)=0$ phase.
On the other hand, \figref{fig:extra_term}(c,e) indicate that the 6-layer slab is $Z_2$ trivial in the $\arg(\mu_1+\ii \mu_2)=0$ phase and nontrivial in the $\arg(\mu_1+\ii \mu_2)=\pi$ phase.
The 5-layer and 6-layer results can be generalized for all odd-layer and even-layer slabs, respectively, as long as the number of layers is large enough to avoid any additional layer-dependent gap closings.
In general, this is consistent with the early recognition of odd-even boundary modes in WTIs~\cite{Stern2012WTI}, and with the picture established in \refcite{Wieder2020AxionCDWWeyl} in which a TR-invariant Weyl-CDW phase can be captured by a stack of 2D TIs whose normal vectors lie parallel to the wavevector $\bsl{Q}$, where the position of the 2D TI in each cell is set by $\arg(\mu_1+\ii \mu_2)$.

The DPME predicted by the effective action is valid only when the slab is $Z_2$ trivial, as the nontrivial $Z_2$ index of the slab necessarily indicates the presence of a gapless helical mode on the side surface, violating the gapped boundary requirement (\ie, the validity of DPME predicted by the low-energy effective action requires a gapped and symmetry-preserving boundary).
The failure of the effective action can also be seen from the bulk-boundary correspondence.
The bulk-boundary correspondence of axion electrodynamics implies that an unambiguous bulk value of the axion field (in the units of $2\pi$) should be equal to the Hall conductance of every gapped surface (in unit of $e^2/(2\pi)$)~\cite{Qi2008TFT}.
When one surface undergoes a $Z_2$ transition, the valley Hall conductance on that surface changes by $e^2/(2\pi)$, while the valley Hall conductance on the other surface remains constant.
As a result, at least on one side of each surface transition, different surfaces infer different bulk values of the valley axion field, indicating the incapability of the effective action in predicting the DPME~\cite{Qi2008TFT}.
The underlying physical reason for the failure of the effective action is that there is a contribution from the gapless helical mode to the DPME that cannot be captured by the low-energy effective action.
Therefore, in general, it is not always appropriate to set $\varphi=0$ in \eqnref{eq:TB_phi_0}.
Instead, one should choose $\varphi$ such that the DPME predicted from the low-energy action matches the TB result when the slab is $Z_2$-trivial.

Nevertheless, the discontinuous change of the DMPE should still exist across the gap closing on one surface; the jump will just contain two contributions in the more realistic case.
One contribution arises from the appearance of an extra gapless helical mode.
The other contribution is given by the discontinuous change of the surface valley Hall conductance (or more directly, the discontinuous change of the strain-induced surface current).
The slab configuration in \figref{fig:extra_term}(a) allows us to demonstrate the second contribution to the DPME jump (\figref{fig:exp}(a)), because the periodic boundary conditions along $y$ and $z$ avoid the contribution from the side-surface helical modes.
As long as the effective action is valid in the slab-$Z_2$-trivial region, the total change of the DPME over the slab-$Z_2$-nontrivial region can still be predicted.
Moreover, in a domain wall structure like \figref{fig:TB_BOTQPT_Domain}(a), the gapless surface domain wall mode still appears across the gap closing on one surface of the $z<0$ side, where the domain wall mode is \emph{exactly} the extra gapless helical mode that corresponds to the change in the slab $Z_2$ index on the $z<0$ side of the domain wall.

\section{Conclusion and Discussion}
\label{sec:conclusion}

Using low-energy theory and TB calculations, we have in this work introduced the DPME of TR-invariant WSMs in the presence of a bulk-constant (static and homogeneous in the bulk) CDW that gaps the bulk Weyl points.
The DPME is a fundamentally 3D strain effect that specifically originates from a valley axion field.
We further demonstrate a discontinuous change of the DPME across a boundary $Z_2$ TQPT by tuning the phase of the CDW order parameter.
The discontinuous change of the DPME can serve as a bulk experimental signature of the boundary TQPT.
Although we have only considered a pair of TR-related valleys, the analysis performed in this work can straightforwardly be generalized to multiple pairs of TR-related Weyl points, as long as one does not enforce additional crystalline symmetries that restrict the total DPME to be zero.

\begin{figure}[t]
    \centering
    \includegraphics[width=\columnwidth]{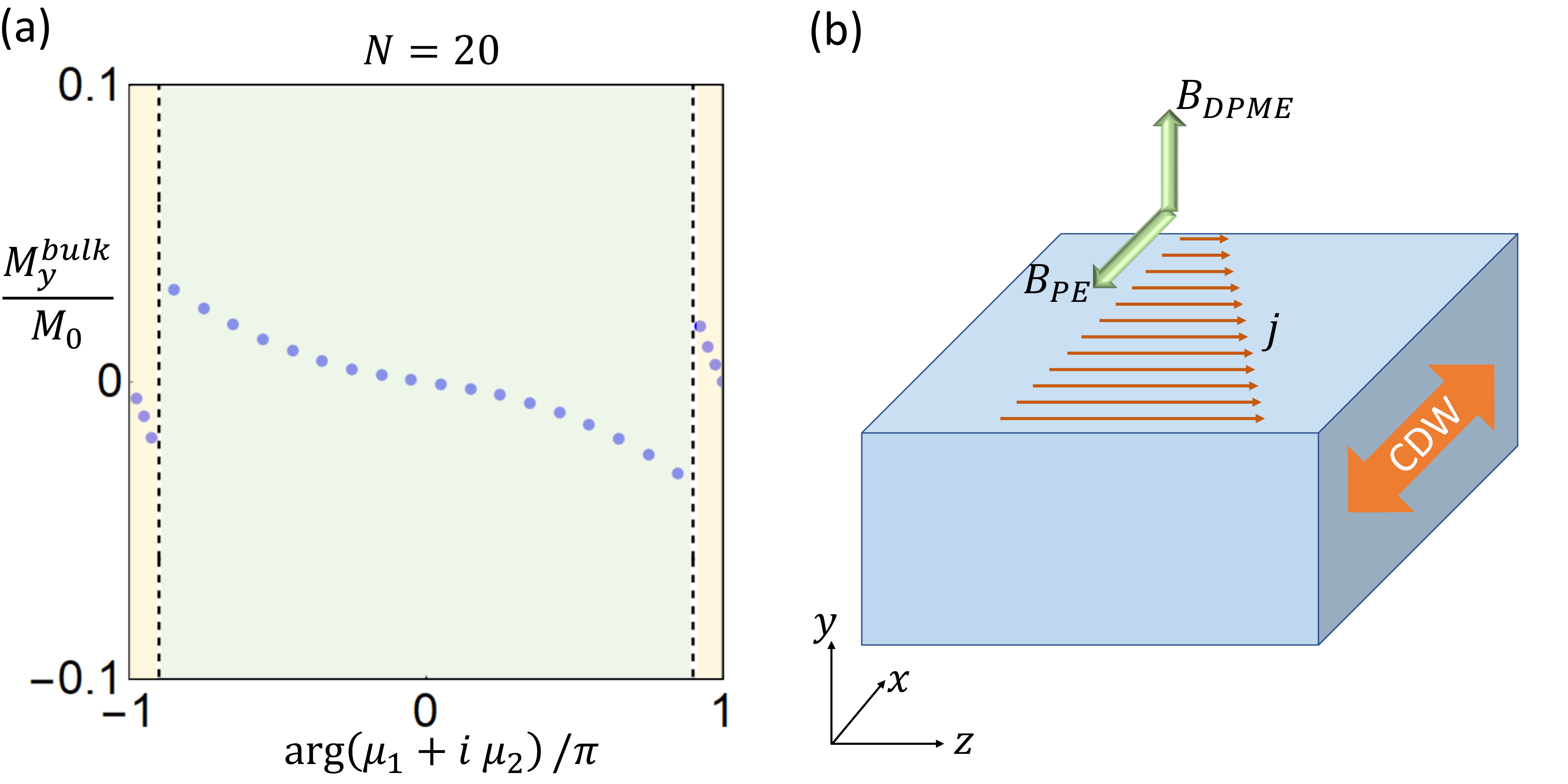}
    \caption{
     (a) The strain-induced bulk magnetization calculated from the realistic slab configuration of $H_{TB,0}+H_{TB,CDW}+H_{extra}$ (see \figref{fig:extra_term}).
The number of layers is chosen to be 20.
The two black dashed lines indicate the two surface transitions at $\arg(\mu_1+\ii\mu_2)=\pm 0.9\pi$.
    (b) A sample with finite size along three directions.
    Outside of the $(010)$ surface, the uniform piezoelectric current generates a magnetic field $B_{PE}$ along $x$, while the magnetic field $B_{DPME}$ given by the DPME is directed along $y$.
    The CDW wavevector is oriented along the $x$ direction.
    }
    \label{fig:exp}
\end{figure}

To probe the DPME, one can measure the induced magnetic field outside of a Weyl-CDW sample.
However, the uniform piezoelectric current also generates a magnetic field outside of the sample, and thus it is important to devise a means of distinguishing the uniform piece from the DPME.
%
One solution is to measure the magnetic field just outside of the (010) surface of a sample with a (100)-directed CDW that gaps the bulk Weyl points.
As shown in \figref{fig:exp}(b), the magnetic field given by the DPME in this geometry is directed along $y$, whereas the magnetic field induced by the uniform piezoelectric current is directed along $x$.
Additionally, the helical modes on the surfaces perpendicular to $y$ and $z$ can be gapped out by finite-size effects (similar as the gapped side surfaces of the axion insulators~\cite{Xiao2018AI} and antiferromagnetic topological insulators~\cite{Zhang2018AIMnBi2Te4,Vishwanath2019MnBiTeAI}), when the phase of the CDW order parameter and the number of layers are chosen to guarantee that the 2D $Z_2$ index of the slab is trivial.
The helical modes can alternatively be removed by a side-surface CDW~\cite{Liu2012HalfQSH}.
Based on \eqnref{eq:M_bulk_TB_LE}, we estimate the order of magnitude of the response coefficient for the DPME to be $|\partial\bsl{M}/\partial\dot{u}_{zz}|\sim 0.8 e/\AA$ for $\xi_y\sim 1 eV$~\cite{Rostami2018PiE}, $\phi_0\sim \pi$, and $v_y\sim 10^{-4} c$~\cite{Shi2019CDWWSMTasSeI}.
We find that the DPME response coefficient has the same units as the 2D piezoelectric coefficient, which is reasonable, because the 3D bulk magnetization in the DPME manifests as the strain-induced 2D surface current density.
Hence, we may directly compare the value of the DPME response coefficient with typical experimental values of 2D piezoelectric coefficients ($\sim 10^{-20} C/\AA$)~\cite{Zhu2014PET}, suggesting that the above estimated value of the DPME response coefficient is experimentally observable.
An important direction of future study is to formulate the contribution from the helical mode to the DPME, which is relevant in the case in which there is one gapless helical mode left on the side-surface for certain values of the phase of the CDW order parameter with respect to the number of layers (see \figref{fig:extra_term}).

In addition to a direct probe of the DPME, our theory predicts the existence of a boundary TQPT and the appearance of 1D gapless helical modes along surface domain walls of the CDW phase.
The gapless helical domain wall fermions can in principle be probed through scanning tunneling microscopy.
Moreover, it is intriguing to ask whether the boundary TQPT separates two 3D phases with different boundary-obstructed topology~\cite{Khalaf2019BoundaryObstructedTopo}, and to elucidate the precise relationship between the boundary TQPT and symmetry-enhanced topological surface anomalies~\cite{Liu2012HalfQSH,Wieder2018WallPaperFermion,Fang2019RotationAnomaly}.

In this work, we have focused on TR-invariant gapped Weyl-CDWs.
TR-invariant WSMs have been realized in a number of non-centrosymmetric systems, including NbAs~\cite{Xu2015WSMNbAs}, TaAs~\cite{Xu2015WSMTaAs,Yang2015WSMTaAs,Ding2015WSMTaAsNatPhys,Ding2015WSMTaAsPRX}, and (TaSe$_4$)$_2$I~\cite{Shi2019CDWWSMTasSeI,Li2019Ta2Se8I}, and an axionic Weyl-CDW phase has recently been demonstrated in (TaSe$_4$)$_2$I~\cite{Gooth2019WSMCDWAxion}.
We emphasize that the intuitive picture of the DPME given in \secref{sec:intuitive_picture} is applicable as long as the low-energy physics of a given system is well-captured by 3D Dirac fermions with complex mass terms.
This implies that the DPME may also exist in other 3D Dirac materials~\cite{Zhang2016NematicDSM}.
In the current work, we have treated the axion field (or the CDW phase) as a fixed background field.
After taking into account the dynamics of CDW (such as a phason), the effective action in \eqnref{eq:S_eff_a_uzz} suggests a nonzero coupling between the phason of the CDW and the strain field, implying the intriguing possibility of strain engineering the CDW phase angle in TR-invariant WSMs.

\section{Acknowledgements}
J.Y. and C.X.L thank Wladimir A. Benalcazar and Radu Roiban, B.J.W. thanks Barry Bradlyn, and all authors thank B. Andrei Bernevig for helpful discussions.
The work done at Penn State, including most analytical derivation and numerical calculation, is primarily supported by the DOE grant (DE-SC0019064).
%
%
B.J.W. acknowledges support from B. Andrei
Bernevig through Department of Energy Grant No. DESC0016239,
Simons Investigator Grant No. 404513, BSF
Israel US Foundation Grant No. 2018226, ONR Grant
No. N00014-20-1-2303, and the Gordon and Betty Moore
Foundation through Grant No. GBMF8685 towards the
Princeton theory program.

During the final stages of preparing this work, an updated version of \refcite{Wieder2020AxionCDWWeyl} demonstrated that TR-invariant Dirac-CDWs are topologically equivalent to $\phi_0$-dependent WTIs.
The results of \refcite{Wieder2020AxionCDWWeyl} are complementary to and in complete agreement with the results of this work.

\bibliography{bibfile_references}

\begin{widetext}
\newpage
\appendix

\section{Derivation of Mean-Field CDW Term}
\label{app:CDW_MF}

In this section, we derive the mean-field CDW term in \eqnref{eq:L_CDW_MF}.
Here, we use the imaginary time and allow the temperature to be nonzero.

We first convert \eqnref{eq:L_WP} to the imaginary time, resulting in
\eq{
\label{eq:S0_free_tau}
S_0=\sum_{a,\alpha}\int \frac{d^4 q}{(2\pi)^4} c^\dagger_{q,a,\alpha}(\ii \omega+ \alpha \sum_{i}v_i q_i \sigma_i)c_{q,a,\alpha}\ ,
}
where $q=(\omega,\bsl{q})$, $\omega=(2n+1)\pi/\beta$ is the fermionic Matsubara frequency, $\int d\omega=(2\pi/\beta)\sum_{\omega}$ if temperate $T$ is not zero, $\beta=1/(k_B T)$, $x=(\tau,\bsl{r})$, $\psi_{x,a,\alpha}=\int \frac{d^4 k}{(2\pi)^4} e^{\ii x q + \ii \bsl{r} \cdot \bsl{k}_{a,\alpha}} c_{q,a,\alpha}$, and $x q = \omega \tau + \bsl{q}\cdot\bsl{r}$.

We consider two channels of the interaction
\eq{
S_{int,1}=-g_1 \sum_{a} \int d^4 x (\psi^\dagger_{x,a,+}\psi_{x,a,-})(\psi^\dagger_{x,a,-}\psi_{x,a,+})
}
and
\eq{
S_{int,2}=-g_2 \int d^4 x \left[ (\psi^\dagger_{x,1,+}\psi_{x,2,-})(\psi^\dagger_{x,2,+}\psi_{x,1,-})+(\psi^\dagger_{x,2,-}\psi_{x,1,+})(\psi^\dagger_{x,1,-}\psi_{x,2,+}) \right]\ ,
}
where $g_1>0$ and $g_2<0$, and $g_2$ is a perturbation with $|g_2|\sim 0$.
The $g_1$ term is just a double copy of that used in \refcite{Zhang2013WSMCDWAxion}.
As shown below, $g_1$ accounts for the nonzero CDW magnitudes while $g_2$ determines their relative phase.

To derive the mean-field CDW term, we first perform the Hubbard-Stratonovich transformation on $S_{int,1}$
\eq{
e^{-S_{int,1}}=\prod_a \int D \widetilde{m}_a^* D \widetilde{m}_a \exp\left[\int d^4 x \left( -\frac{|\widetilde{m}_{a}(x)|^2}{g_1} - \widetilde{m}_a(x) \psi_{a,+}^\dagger \psi_{a,-} - \widetilde{m}_a^*(x) \psi_{a,-}^\dagger \psi_{a,+}\right)\right]\ .
}
Then, we have
\eq{
Z=\int D \widetilde{m}^* D \widetilde{m}  \int D\psi^\dagger D\psi  e^{-S_0-S_{int,2}+\int d^4 x \sum_a\left(-\frac{|\widetilde{m}_{a}(x)|^2}{g_1}- \widetilde{m}_a(x) \psi_{a,+}^\dagger \psi_{a,-}-\widetilde{m}_a^*(x) \psi_{a,-}^\dagger \psi_{a,+}\right)}= \int D \widetilde{m}^* D \widetilde{m} e^{-S_{MF}[\widetilde{m}]}
}
Now, we perform the mean-field approximation.
We neglect the quantum fluctuation of $\widetilde{m}_a(x)$ and only consider a classical $\widetilde{m}_a(x)$ that minimizes $S_{MF}[\widetilde{m}]$, meaning that $\widetilde{m}_a(x)\propto \langle \psi^\dagger_{x,a,-}\psi_{x,a,+} \rangle$.
Comparing $\widetilde{m}_a(x)$ with \eqnref{eq:L_CDW_MF} of the main text, we can define $\widetilde{m}_a(x)= m_a e^{\ii (-1)^{a-1}\bsl{Q}\cdot \bsl{r}}$ where $m_a$ is the CDW parameter mentioned in main text.
Since we only care about the CDW order parameter that is constant in the bulk of the system, we choose $m_a$ to be independent of $x$.
As a result, we arrive at a simplified $Z$:
\eq{
Z=\int D\psi^\dagger D\psi  e^{-S_{int,2}-\sum_{a}\int \frac{d^4 q}{(2\pi)^4} \psi^\dagger_{q,a} G_a^{-1}(q)\psi_{q,a} -\sum_{a}\beta \mathcal{V}\frac{|m_{a}|^2}{g_1}}=  e^{-S_{MF}}\ ,
}
where $G_a^{-1}(q)=\ii \omega+ \sum_{i}v_i q_i\tau_z\sigma_i + M_a$, $M_a= m_a \tau_+\sigma_0+m_a^* \tau_-\sigma_0$, $\tau_\pm=(\tau_x\pm\ii\tau_y)/2$, and $\mathcal{V}$ is the total volume of the system.

Next, we derive $S_{MF}$ to the first order of $g_2$.
The $g_2$-independent part of $S_{MF}$, labeled as $S_{MF,0}$, reads
\eq{
e^{-S_{MF,0}}=\prod_a \int D\psi^\dagger_a D\psi_a  \exp\left[-\int \frac{d^4 q}{(2\pi)^4} \psi^\dagger_{q,a} G_a^{-1}(q)\psi_{q,a} - \beta \mathcal{V}\frac{|m_{a}|^2}{g_1}\right]\ ,
}
which gives
\eqa{
& S_{MF,0}=-\beta \mathcal{V}\sum_{a}\left\{ \int \frac{d^4 q}{(2\pi)^4} \log\det{[G^{-1}_a (q)]} -\frac{|m_{a}|^2}{g_1}\right\}+const.\\
&= -\beta \mathcal{V}\sum_{a}\left\{ \int \frac{d^4 q}{(2\pi)^4} 2\log(\omega^2+\sum_{i}v_i^2q_i^2+|m_a|^2) -\frac{|m_{a}|^2}{g_1}\right\}+const.\\ .
}
Define $m_a=|m_a| e^{\ii \phi_a}$, then we see that $S_{MF,0}$ does not depend on $\phi_a$.

For the first order of $g_2$, we have
\eqa{
S_{MF,1}&=\beta \mathcal{V} g_2\sum_{\delta=\pm} \int \frac{d^4 q}{2\pi^4} \int \frac{d^4 q'}{2\pi^4} \Tr[ G_1 (q) \tau_\delta\sigma_0 G_2(q')\tau_\delta\sigma_0]\\
&= 2\beta \mathcal{V} g_2(m_1 m_2+m_1^* m_2^*)  \int \frac{d^4 q}{2\pi^4} \int \frac{d^4 q'}{2\pi^4} \frac{1}{(\omega^2+\sum_i q_i^2 v_i^2+ |m_1|^2) (\omega'^2+\sum_i q_i'^2 v_i^2+ |m_2|^2)}\\
&= 4\beta \mathcal{V} g_2 |m_1| |m_2| \cos(\phi_1+\phi_2) I_1 I_2\ ,
}
where $I_a=\int \frac{d^4 q}{(2\pi)^4}\frac{1}{\omega^2+\sum_i v_i^2 q_i^2+|m_a|^2}$.

As a result, we have $S_{MF}=S_{MF,0}+S_{MF,1}$ to the first order of $g_2$.
Next, we minimize $S_{MF}$.
First, for $\phi_a$, we have
\eq{
\frac{\partial}{\partial (\phi_1+\phi_2)}S_{MF}=0\Rightarrow \sin(\phi_1+\phi_2)=0\Rightarrow \phi_1+\phi_2=n \pi\ .
}
Since $g_2<0$ and $I_a>0$, $\phi_1+\phi_2=2 n \pi$ minimizes $S_{MF}$.
The $2\pi$ ambiguity will disappear after introducing a gapped boundary, and a symmetry-preserving boundary would give \eq{\phi_1=-\phi_2=\phi_0}
in the bulk.
Second, for $|m_a|$, we have
\eq{
\frac{\partial}{\partial |m_a|}S_{MF}=0\Rightarrow -\frac{1}{2 g_1} +\int \frac{d^4 q}{(2\pi)^4} \frac{1}{\omega^2+\sum_i q_i^2 v_i^2+|m_a|^2} + O(g_2) =0\ ,
}
resulting in
\eq{
-\frac{|v_x v_y v_z|}{g_1}+\frac{\Lambda^2}{8\pi^2 }= \frac{1}{8\pi^2} |m_a|^2 \log(\frac{|m_a|^2+\Lambda^2}{|m_a|^2})+O(g_2)\ ,
}
where $\omega^2+\sum_i q_i^2 v_i^2\leq \Lambda^2$ is used.
The equation for $|m_a|$ with $v_x=v_y=v_z=1$ matches that in \refcite{Zhang2013WSMCDWAxion}, which indicates that we need to have a large enough $g_1$ to have the nonzero CDW magnitude.
The solution to the above equation has the form
\eq{
|m_a|=|m_0| + O(g_2)\ ,
}
where $|m_0|$ is independent of $g_2$.
In the main text, we directly neglect the $g_2$ in $|m_a|$ and choose $|m_a|=|m_0|$.
Then, $m_0=|m_0|e^{\ii \phi_0}$.

\section{Incorporating the Effects of Strain}
\label{app:el-str}
In this section, we discuss the effect of strain on the crystals in details.
To discuss strain, we need to introduce the displacement gradient $u_{ij}=\frac{\partial{u_i}}{\partial{r_j}}$, where $u_i$ is the $i$th component of the displacement of the point at $\bsl{r}$.
The strain tensor is just the symmetric part of the tensor $(u_{ij}+u_{ji})/2$, while the anti-symmetric part $(u_{ij}-u_{ji})/2$ is the rotation.
By setting the strain to be adiabatic, homogeneous, and infinitesimal, we mean to choose $u_{ij}$ to have these properties.

In the following, we describe the theory for $u_{ij}$, which contains the strain as a special case.
We first discuss the general formalism for crystals, then the TB model, and at last the low-energy model.
Throughout the work, $u(t)$ is treated as a real fixed background, which acts as a constant under symmetry operators, \eg, $\TR u(t) \TR^{-1}=u^*(t)=u(t)$ for TR symmetry.

\subsection{General Formalism}

We first discuss a generic single-particle Hamiltonian for electrons in a crystal:
\eq{
\label{eq:H_0}
H_0=\frac{\bsl{p}^2}{2m_e}+ \lambda \bsl{S}\cdot (\bsl{\nabla}_{\bsl{x}}V(\bsl{x})\times \bsl{p})+V(\bsl{x})\ ,
}
where $\lambda\in \mathds{R}$ labels the spin-orbit coupling, $\bsl{\nabla}_{\bsl{x}} V(\bsl{x})=\ii [\bsl{p},V(\bsl{x})]$,
\eq{
V(\bsl{x})=\sum_{\bsl{R},i}V_{i}(\bsl{x}-\bsl{R}-\bsl{\tau}_i)\ ,
}
$\bsl{R}$ is the lattice vector, and $\bsl{\tau}_i$ labels the sublattice.
We adopt the clamped-ion approximation~\cite{Vanderbilt2000BPPZ}, and then the ions exactly follow the homogeneous deformation $\bsl{R}+\bsl{\tau}_i\rightarrow (1+u)(\bsl{R}+\bsl{\tau}_i)$.
With the homogeneous infinitesimal $u$, the Hamiltonian becomes
\eq{
\label{eq:H_u}
H_u=\frac{\bsl{p}^2}{2m_e}+ \lambda \bsl{S}\cdot (\bsl{\nabla}_{\bsl{x}}V_u(\bsl{x})\times \bsl{p})+V_u(\bsl{x})\ ,
}
where
\eq{
V_u(\bsl{x})=\sum_{\bsl{R},i}V_{i}(\bsl{x}-(1+u)(\bsl{R}+\bsl{\tau}_i))\ .
}

$H_0$ has the lattice translation symmetry $[H_0,T_{\bsl{R}}]=0$ with $T_{\bsl{R}}=e^{-\ii \bsl{p}\cdot\bsl{R}}$.
Then, $H_0$ can be rewritten as
\eq{
\label{eq:H_0_k}
H_0=\int_{1BZ} \frac{d^d k}{(2\pi)^d} \sum_{\bsl{G},\bsl{G}',s,s'} c^\dagger_{\bsl{k}+\bsl{G},s} [h_0(\bsl{k})]_{\bsl{G}s,\bsl{G}'s'} c_{\bsl{k}+\bsl{G}',s'}=\int \frac{d^d k}{(2\pi)^d} c^\dagger_{\bsl{k}} h_0(\bsl{k}) c_{\bsl{k}}\ ,
}
where $\bsl{k}\in$1BZ, and $\bsl{G}$ is the reciprocal lattice vector.
$c^\dagger_{\bsl{k}+\bsl{G},s}$ is the creation operator for $\ket{\bsl{k}+\bsl{G},s}$, and satisfies
\eq{
\label{eq:c_k_comm}
\{c^\dagger_{\bsl{k}+\bsl{G},s},c_{\bsl{k}'+\bsl{G}',s'}\}=(2\pi)^d \delta(\bsl{k}-\bsl{k}')\delta_{\bsl{G}\bsl{G}'}\delta{ss'}\ .
}
Moreover, $c^\dagger_{\bsl{k}}$ is a vector operator with $c^\dagger_{\bsl{k}+\bsl{G},s}$ its the $(\bsl{G},s)$ component.

In the presence of $u$, the lattice translation of $H_u$ becomes $[H_u,T_{\bsl{R}_u}]=0$ with $\bsl{R}_u=(1+u)\bsl{R}$.
As a result, the reciprocal lattice vectors and Bloch momenta become $\bsl{G}_u=(1-u^T)\bsl{G}$ and $\bsl{k}_u=(1-u^T)\bsl{k}$.
Then, $H_u$ can be rewritten as
\eq{
H_u=\int_{1BZ_u} \frac{d^d k_u}{(2\pi)^d} \sum_{\bsl{G},\bsl{G}',s,s'} c^\dagger_{\bsl{k}_u+\bsl{G}_u,s} [h_u(\bsl{k})]_{\bsl{G}s,\bsl{G}'s'} c_{\bsl{k}_u+\bsl{G}'_u,s'}\ ,
}
where $\sum_{\bsl{G}_u}$ is equivalent to $\sum_{\bsl{G}}$ since $\bsl{G}_u$ has a one-to-one relation to $\bsl{G}$.
Moreover, $H_{u=0}=H_0$ means that $h_{u=0}(\bsl{k})=h_0(\bsl{k})$.
The anticommutation relation for $c^\dagger_{\bsl{k}_u+\bsl{G}_u,s}$ reads
\eq{
\{c^\dagger_{\bsl{k}_u+\bsl{G}_u,s},c_{\bsl{k}'_u+\bsl{G}'_u,s'}\}=(2\pi)^d \delta(\bsl{k}_u-\bsl{k}'_u)\delta_{\bsl{G}\bsl{G}'}\delta{ss'}=(2\pi)^d \delta(\bsl{k}-\bsl{k}')\frac{1}{|det(1-u^T)|}\delta_{\bsl{G}\bsl{G}'}\delta{ss'}\ .
}
We can define $\widetilde{c}^\dagger_{\bsl{k},\bsl{G},s}(u)=|det(1-u^T)|^{1/2} c^\dagger_{\bsl{k}_u+\bsl{G}_u,s}$, and then it has the same anti-commutation relation as \eqnref{eq:c_k_comm}:
\eq{
\label{eq:c_tilde_k_comm}
\{\widetilde{c}^\dagger_{\bsl{k},\bsl{G},s},\widetilde{c}_{\bsl{k}',\bsl{G}',s'}\}=(2\pi)^d \delta(\bsl{k}-\bsl{k}')\delta_{\bsl{G}\bsl{G}'}\delta{ss'}\ .
}
As a result, $H_u$ can be further re-expressed in terms of $\widetilde{c}$ as
\eq{
\label{eq:H_u_k}
H_u=\int_{1BZ} \frac{d^d k}{(2\pi)^d} \widetilde{c}^\dagger_{\bsl{k}} h_u(\bsl{k}) \widetilde{c}_{\bsl{k}}\ .
}

Comparing \eqnref{eq:H_0_k} and \eqnref{eq:H_u_k}, it is clear that the deformation induces two changes: (i) $c^\dagger_{\bsl{k}}\rightarrow \widetilde{c}^\dagger_{\bsl{k}}(u)$, and (ii) $h_0(\bsl{k})\rightarrow h_u(\bsl{k})$.
According to \eqnref{eq:c_k_comm} and \eqnref{eq:c_tilde_k_comm}, $c^\dagger_{\bsl{k}}$ and $\widetilde{c}^\dagger_{\bsl{k}}(u)$ have the same anti-commutation relations.
The similarity between them can also be reflected by the equivalent representations furnished for the corresponding symmetry operators.

For the lattice translation, we have
\eq{
\label{eq:c_k_lattice_translation}
T_{\bsl{R}}c^\dagger_{\bsl{k}}T_{\bsl{R}}^{-1}=c^\dagger_{\bsl{k}} e^{-\ii \bsl{k}\cdot \bsl{R}}\ ,\ T_{(1+u)\bsl{R}}\widetilde{c}^\dagger_{\bsl{k}}T_{(1+u)\bsl{R}}^{-1}=\widetilde{c}^\dagger_{\bsl{k}} e^{-\ii \bsl{k}\cdot \bsl{R}}\ .
}
More generally, for a generic space group operator $g=\{R|\bsl{t}\}$, the representation furnished by $c^\dagger_{\bsl{k}}$ reads
\eq{
\label{eq:c_k_g}
g c^\dagger_{\bsl{k}} g^{-1}=c^\dagger_{R \bsl{k}} e^{-\ii R \bsl{k}\cdot\bsl{t}} U_g\ ,
}
where $[U_g]_{\bsl{G}'s',\bsl{G}s}=\delta_{\bsl{G}',R\bsl{G}}e^{-\ii \bsl{G}'\cdot\bsl{t}} (R_S)_{s's}$ and $R_S$ is the matrix representation of $R$ in the spin subspace.
Then, we can define $g_u=\{ R | \bsl{t}_u=(1+u)\bsl{t}\}$ and have
\eq{
\label{eq:c_k_u_g}
g_u \widetilde{c}^\dagger_{\bsl{k}}(R^{-1} u R) g_u^{-1}=\widetilde{c}^\dagger_{R \bsl{k}}(u) e^{-\ii R \bsl{k}\cdot\bsl{t}} U_g\ .
}
For the TR symmetry that acts on the Bloch states, we have
\eq{
\label{eq:c_k_TR}
\TR c^\dagger_{\bsl{k}} \TR^{-1}=c^\dagger_{-\bsl{k}}U_{\TR}\ , \TR \widetilde{c}^\dagger_{\bsl{k}}(u(t)) \TR^{-1}=\widetilde{c}^\dagger_{-\bsl{k}}(u(t))U_{\TR}
}
where $[U_{\TR}]_{\bsl{G}'s',\bsl{G}s}=\delta_{\bsl{G}',-\bsl{G}}(\ii \sigma_y)_{s's}$.

With these symmetry representations, we can derive the symmetry properties of $h_u(\bsl{k})$ from those of $h_0(\bsl{k})$, which are useful for the change (ii).
Suppose $[g,H_0]=0$. Then, we have $g_u (H_{R^{-1} u R } )g_u^{-1}=H_u$ based on \eqnref{eq:H_0} and \eqnref{eq:H_u}.
As a result, we have
\eq{
\label{eq:h_Ug}
U_g h_0(\bsl{k}) U_g^\dagger= h_{0}(R\bsl{k})\ ,\ U_g h_u(\bsl{k}) U_g^\dagger= h_{R u R^{-1}}(R\bsl{k})\ .
}
For TR symmetry, suppose $[\TR,H_0]=0$, and then we have $\TR H_{u(t)} \TR^{-1}=H_{u(t)}$, which gives
\eq{
\label{eq:h_UTR}
U_{\TR} h_0^*(\bsl{k}) U_{\TR}^\dagger= h_{0}(-\bsl{k})\ ,\ U_{\TR} h_{u(t)}^*(\bsl{k}) U_{\TR}^\dagger= h_{u(t)}(-\bsl{k})\ .
}

\subsection{Tight-binding Model}

The above formalism in general is hard to deal with analytically.
More commonly, we deal with the TB model, which reads
\eq{
\label{eq:H_TB_0_app}
H_{TB,0}=\sum_{\bsl{R},\bsl{R}',i,i'}c^\dagger_{\bsl{R}+\bsl{\tau}_{i}} M_{ii'}(\bsl{R}+\bsl{\tau}_{i}-\bsl{R}'-\bsl{\tau}_{i'}) c_{\bsl{R}'+\bsl{\tau}_{i'}}
}
in the absence of deformation, or reads
\eq{
\label{eq:H_TB_u_app}
H_{TB,u}=\sum_{\bsl{R},\bsl{R}',i,i'}c^\dagger_{(1+u)(\bsl{R}+\bsl{\tau}_{i})} M_{ii'}[(1+u)(\bsl{R}+\bsl{\tau}_{i}-\bsl{R}'-\bsl{\tau}_{i'})] c_{(1+u)(\bsl{R}'+\bsl{\tau}_{i'})}
}
with deformation.
Here $c^\dagger_{\bsl{R}+\bsl{\tau}_{i}}$ is a vector whose components stand for orbital, spin, etc, and are labeled by $\beta_i$ as $c^\dagger_{\bsl{R}+\bsl{\tau}_{i},\beta_i}$.
Note that $\beta_i$ can take different ranges of values for different sublattices.
In this work, we approximate $M_{ii'}[(1+u)(\bsl{R}+\bsl{\tau}_{i}-\bsl{R}'-\bsl{\tau}_{i'})]$ as~\cite{Li2016StrainTMD}
\eq{
M_{ii'}[(1+u)(\bsl{R}+\bsl{\tau}_{i}-\bsl{R}'-\bsl{\tau}_{i'})]\approx \left(1-\frac{\bsl{\delta}^T u \bsl{\delta}}{|\bsl{\delta}|^2}\right) M_{ii'}(\bsl{R}+\bsl{\tau}_{i}-\bsl{R}'-\bsl{\tau}_{i'})\ ,
}
where $\bsl{\delta}=\bsl{R}+\bsl{\tau}_{i}-\bsl{R}'-\bsl{\tau}_{i'}$ is treated as a column vector.
Since the above expression only involves the symmetric part of $u$, this only takes in to account the effect of strain.
It is reasonable sicne a global rotation of the system cannot induce any response.

Let us define
\eq{
\label{eq:c_k_TB}
c^\dagger_{\bsl{k},i,\beta_i}=\frac{1}{\sqrt{N}}\sum_{\bsl{R}}e^{\ii \bsl{k}\cdot(\bsl{R}+\bsl{\tau}_i)} c^\dagger_{\bsl{R}+\bsl{\tau}_i,\beta_i}\ ,\  \widetilde{c}^\dagger_{\bsl{k},i,\beta_i}(u)=\frac{1}{\sqrt{N}}\sum_{\bsl{R}}e^{\ii \bsl{k}\cdot(\bsl{R}+\bsl{\tau}_i)} c^\dagger_{(1+u)(\bsl{R}+\bsl{\tau}_i),\beta_i}\ ,
}
with $N$ redefined as the total number of lattice sites.
From $\{c^\dagger_{\bsl{r},\beta},c_{\bsl{r}',\beta'}\}=\delta_{\bsl{r}\bsl{r}'}\delta_{\beta\beta'}$, we can derive that $c^\dagger_{\bsl{k},i,\beta_i}$ and $\widetilde{c}^\dagger_{\bsl{k},i,\beta_i}$ have the same anti-commutation relation
\eq{
\left\{ c^\dagger_{\bsl{k},i,\beta_i} , c_{\bsl{k}',i',\beta'_{i'}}\right\} =\left\{ \widetilde{c}^\dagger_{\bsl{k},i,\beta_i} , \widetilde{c}_{\bsl{k}',i',\beta'_{i'}}\right\}=\delta_{\bsl{k}\bsl{k}'}\delta_{ii'}\delta_{\beta_i\beta'_{i'}}\ .
}
With \eqnref{eq:c_k_TB}, the Hamiltonian can be re-expressed as
\eq{
H_{TB,0}=\sum_{\bsl{k}}c^\dagger_{\bsl{k}} h_0(\bsl{k}) c_{\bsl{k}}\ ,\ H_{TB,u}=\sum_{\bsl{k}}\widetilde{c}^\dagger_{\bsl{k}} h_u(\bsl{k}) \widetilde{c}_{\bsl{k}}
}
with
\eq{
[h_0(\bsl{k})]_{i\beta_i,i'\beta'_{i'}}=\sum_{\Delta \bsl{R}} e^{-\ii (\Delta \bsl{R}+\bsl{\tau}_i-\bsl{\tau}_{i'})\cdot\bsl{k}} \left[M_{ii'}(\Delta\bsl{R}+\bsl{\tau}_i-\bsl{\tau}_{i'})\right]_{\beta_i\beta'_{i'}}
}
and
\eq{
[h_u(\bsl{k})]_{i\beta_i,i'\beta'_{i'}}=\sum_{\Delta \bsl{R}} e^{-\ii (\Delta \bsl{R}+\bsl{\tau}_i-\bsl{\tau}_{i'})\cdot\bsl{k}} \left[M_{ii'}((1+u)(\Delta\bsl{R}+\bsl{\tau}_i-\bsl{\tau}_{i'}))\right]_{\beta_i\beta'_{i'}}\ .
}
Therefore, similar as the general formalism, the strain effect to the TB model includes (i) $c^\dagger_{\bsl{k}}\rightarrow \widetilde{c}^\dagger_{\bsl{k}}$ and (ii) $h_0(\bsl{k})\rightarrow h_u(\bsl{k})$\ , where $c^\dagger_{\bsl{k}}$ and $\widetilde{c}^\dagger_{\bsl{k}}$ have the same commutation relation, and $h_{u=0}(\bsl{k})=h_0(\bsl{k})$.

The similarity also exists for the symmetry properties.
First, \eqnref{eq:c_k_lattice_translation} for the lattice translations still holds here.
Second, if $[g,H_0]=0$ for a space group operation $g$, then $c^\dagger_{\bsl{R}+\bsl{\tau}_i}$ furnishes a representation of $g$, \ie,
\eq{
g c^\dagger_{\bsl{R}+\bsl{\tau}_i} g^{-1}=c^\dagger_{R(\bsl{R}+\bsl{\tau}_i)+\bsl{t}} M_g^{i_g i}=c^\dagger_{\bsl{R}_g+\bsl{\tau}_{i_g}} M_g^{i_g i}
}
with $M_g^{i_g i}$ the representation of $g$ in the $\beta$ space.
The existence of $M_g^{i_g i}$ means $i_g$ and $i$ are the same kind of atoms with the same orbitals.
Then, \eqnref{eq:c_k_g}, \eqnref{eq:c_k_u_g}, and \eqnref{eq:h_Ug} hold here for a different definition of $U_g$:
\eq{
[U_g]_{i'\beta'_{i'},i\beta_i}=\delta_{i' i_g}[M_g^{i' i}]_{\beta'_{i'}\beta_i}\ .
}
$M_g^{i' i}$ is defined to be zero for $i'$ and $i$ being different kinds of atoms.
Third, if $[\TR,H_0]=0$ for the TR operation $\TR$, then
\eq{
\TR c^\dagger_{\bsl{R}+\bsl{\tau}_i}\TR^{-1}=c^\dagger_{\bsl{R}+\bsl{\tau}_i} M_{\TR}^i
}
with $M_{\TR}^i$ the representation of $\TR$ in the $\beta$ space, and \eqnref{eq:c_k_TR} and \eqnref{eq:h_UTR} hold here for a different definition of $U_{\TR}$:
\eq{
[U_{\TR}]_{i'\beta'_{i'},i\beta_i}=\delta_{ii'}[M_{\TR}^i]_{\beta'_{i'}\beta_i}\ .
}

Therefore, the strain effect is formally the same for the general Hamiltonian and the TB model.

\subsection{Low-Energy Model}

In this part, we project the general Hamitonian or the TB model to the low energy subspace. We consider a group of orthonormal vectors $v_{\alpha}(\bsl{k}_a)$ that satisfy
\eq{
h_0(\bsl{k}_a)v_{\alpha}(\bsl{k}_a)=E_\alpha(\bsl{k}_a) v_{\alpha}(\bsl{k}_a)\ ,
}
where $a,\alpha$ are re-defined to label the valley and energies.
We choose $v_{\alpha}(\bsl{k}_a)$'s so that they furnish a representation of the symmetry group of $H_0$:
\eq{
U_g v_\alpha(\bsl{k}_a)=v_{\alpha'}(\bsl{k}_{a'})\delta_{\bsl{k}_{a'},R\bsl{k}_a}  [W_g^a]_{\alpha'\alpha}
}
if $g$ is a symmetry of $H_0$, and
\eq{
U_{\TR}v^*_\alpha(\bsl{k}_a)=v_{\alpha'}(\bsl{k}_{a'}) \delta_{\bsl{k}_{a'},-\bsl{k}_a} [W_{\TR}^a]_{\alpha'\alpha}
}
if $\TR$ is a symmetry of $H_0$.

Let us define
\eq{
b^\dagger_{\bsl{q},a,\alpha}=c^\dagger_{\bsl{q}+\bsl{k}_a}v_\alpha(\bsl{k}_a)\ ,\
\widetilde{b}^\dagger_{\bsl{q},a,\alpha}(u)=\widetilde{c}^\dagger_{\bsl{q}+\bsl{k}_a}(u)v_\alpha(\bsl{k}_a)\ ,
}
where $\bsl{q}$ only takes a small symmetric neighborhood around $\bsl{k}_a$.
As a result,  $b^\dagger_{\bsl{q}}$ and $\widetilde{b}^\dagger_{\bsl{q}}$ have the same commutation relation
\eq{
\{ b^\dagger_{\bsl{q},a,\alpha}, b_{\bsl{q}',a',\alpha'} \}=\{ \widetilde{b}^\dagger_{\bsl{q},a,\alpha}, \widetilde{b}_{\bsl{q}',a',\alpha'} \}=(2\pi)^d\delta(\bsl{q}-\bsl{q}') \delta_{aa'}\delta_{\alpha\alpha'}\ ,
}
and the effective Hamiltonian reads
\eq{
H_0^{eff}=\int \frac{d^d q}{(2\pi)^d} b^\dagger_{\bsl{q}} h^{eff}_0(\bsl{q}) b_{\bsl{q}}\ ,\
H_u^{eff}=\int \frac{d^d q}{(2\pi)^d} \widetilde{b}^\dagger_{\bsl{q}} h^{eff}_u(\bsl{q}) \widetilde{b}_{\bsl{q}}\ .
}
Here we use the form for the general Hamiltonian, since the derivation for the TB model is equivalent.
Clearly, the effect of deformation again includes (i) $b^\dagger_{\bsl{k}}\rightarrow \widetilde{b}^\dagger_{\bsl{k}}$ and (ii) $h_0^{eff}(\bsl{k})\rightarrow h_u^{eff}(\bsl{k})$ .
Furthermore, the symmetry properties of $\widetilde{b}^\dagger_{\bsl{q}}$ are
\eq{
g_u \widetilde{b}_{\bsl{q}}^\dagger(R^{-1} u R) g_u^{-1} = \widetilde{b}_{R\bsl{q}}^\dagger(u) U_g^{eff} e^{-\ii R \bsl{q} \cdot \bsl{t}}
}
with $(U_g^{eff})_{a'\alpha',a\alpha}=[W^a_g]_{\alpha'\alpha}\delta_{\bsl{k}_{a'},R\bsl{k}_a} e^{-\ii R\bsl{k}_a\cdot \bsl{t}}$ if $g$ is a symmetry of $H_0$, and
\eq{
\TR \widetilde{b}_{\bsl{q}}^\dagger(u(t)) \TR^{-1} = \widetilde{b}_{-\bsl{q}}^\dagger(u(t)) U_{\TR}^{eff}
}
with $(U_{\TR}^{eff})_{a'\alpha',a\alpha}=[W^a_{\TR}]_{\alpha'\alpha}\delta_{\bsl{k}_{a'},-\bsl{k}_a}$ if $\TR$ is a symmetry of $H_0$.
The symmetry properties of $\bsl{b}_{\bsl{q}}$ can be derived by limiting $u\rightarrow 0$ in the above expression.

In general, $h^{eff}_0(\bsl{q})$ and $h^{eff}_u(\bsl{q})$ can be derived from  $h_0(\bsl{q})$ and $h_u(\bsl{q})$ using the perturbation theory, respectively.
However, this is not always straightforward to be done analytically, so sometimes we derive their form from symmetries.
Note that $H_0^{eff}$ and $H_u^{eff}$ should have the same symmetry properties as $H_0$ and $H_u$, respectively.
Then,we have
\eq{
U_g^{eff} h_0^{eff}(\bsl{q}) (U_g^{eff})^\dagger = h_0^{eff}(R \bsl{q})\ ,\ U_g^{eff} h_u^{eff}(\bsl{q}) (U_g^{eff})^\dagger = h_{R u R^{-1}}^{eff}(R \bsl{q})
}
if $[g,H_0]=0$, and
\eq{
U_{\TR}^{eff} [h_0^{eff}(\bsl{q})]^* (U_{\TR}^{eff})^\dagger = h_0^{eff}(- \bsl{q})\ ,\ U_{\TR}^{eff} [h_{u(t)}^{eff}(\bsl{q})]^* (U_{\TR}^{eff})^\dagger = h_{u(t)}^{eff}(-\bsl{q})
}
if $[\TR,H_0]=0$.

Now we restore the original definition of $a,\alpha$, and consider the case discussed in the main text.
With this scheme and the following symmetry representations
\eqa{
\label{eq:c_sym_rep}
& m_y c^\dagger_{\bsl{q},1,\alpha} m_y^{-1}= c^\dagger_{m_y\bsl{q},2,-\alpha} (-\ii \sigma_y)\ ,\\
& m_y c^\dagger_{\bsl{q},2,\alpha} m_y^{-1}= c^\dagger_{m_y\bsl{q},1,-\alpha} (-\ii \sigma_y)\ ,\\
& \TR c^\dagger_{\bsl{q},1,\alpha} \TR^{-1}= c^\dagger_{-\bsl{q},2,\alpha} (\ii \sigma_y)\ ,\\
& \TR c^\dagger_{\bsl{q},2,\alpha} \TR^{-1}= c^\dagger_{-\bsl{q},1,\alpha} (\ii \sigma_y)\ ,
}
we can obtain the leading-order electron-strain coupling as
\eqa{
H_{str}=\int \frac{d^3 q}{(2\pi)^3} \sum_{a}\widetilde{c}^\dagger_{\bsl{q},a} [\xi_0\tau_0\sigma_0+(-1)^{a-1}(\tau_0\sigma_x \xi_x+\tau_z\sigma_y \xi_y +\tau_0\sigma_z \xi_z)]\widetilde{c}_{\bsl{q},a}u_{zz} \ .
}
Converting to the field operator $\widetilde{c}_{t,\bsl{q},a,\alpha}$ and using
\eq{
\label{eq:Fourier_psi_c}
\psi_{t,\bsl{r},a,\alpha}=e^{\ii \bsl{k}_{a,\alpha}\cdot \bsl{r}}\int \frac{d^3 q}{(2\pi)^3} e^{i\bsl{q}\cdot\bsl{r}} \widetilde{c}_{t,\bsl{q},a,\alpha}\ ,
}
we can obtain \eqnref{eq:L_str} of the main text.
Here the low-energy approximation allows us to extend the range of $\bsl{q}$ to $\mathds{R}^3$ and treat $a,\alpha$ as internal indices.
Since the strain cannot change the commutation relation of the field operator, the measure of the functional integral in the partition function does not change with the strain.

\section{More Details on the Low-Energy Theory}
\label{app:S_eff}

In the section, we provide more details on the effective action and the boundary condition of the low-energy theory.

\begin{figure}[t]
    \centering
    \includegraphics[width=0.9\columnwidth]{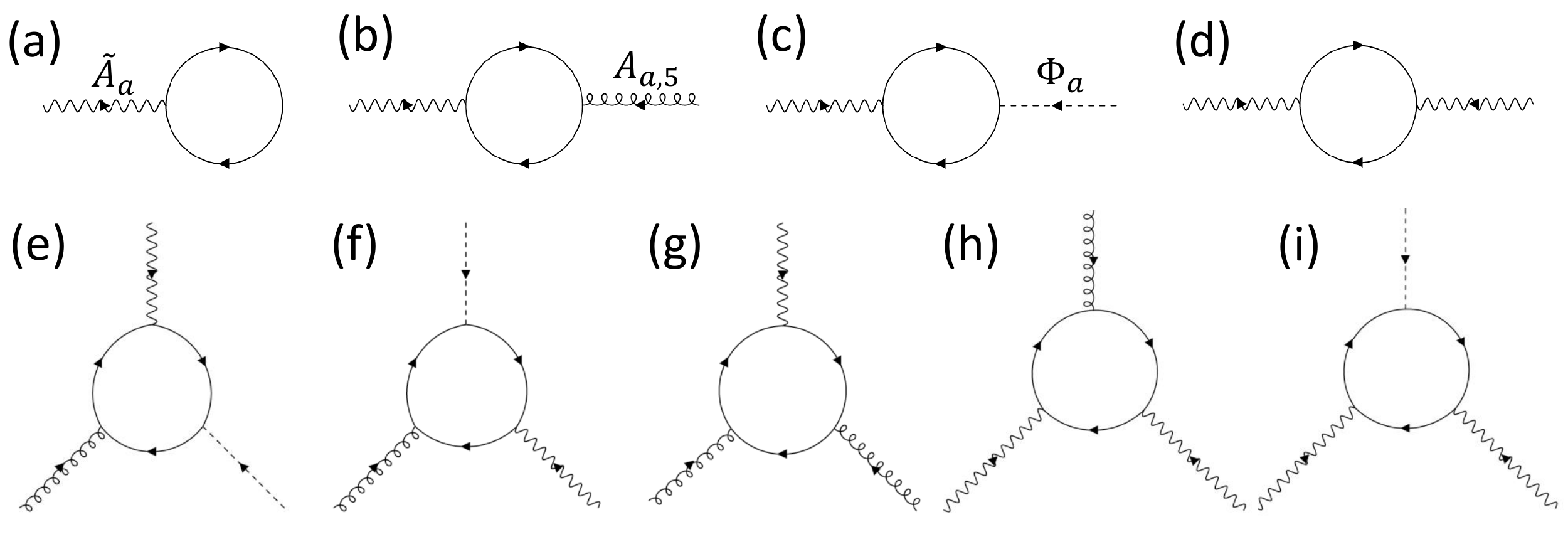}
    \caption{
    This figure shows Feynman diagrams that might contribute to the effective action of \eqnref{eq:L_tot_m0_Phi} up to the leading order.
    The solid line stands for the Fermion propagator \eqnref{eq:Fermion_Prop}, and the meanings of other lines are labeled in the graph.
    }
    \label{fig:Effective_Action_FD}
\end{figure}

\subsection{Effective Action}

In this part, we derive \eqnref{eq:S_eff_a_leading} in the main text from \eqnref{eq:L_tot_sim} and \eqnref{eq:Z_a} of the main text.

Let us first simplify \eqnref{eq:L_tot_sim} of the main text.
As mentioned in the main text, $|m|=|m(\bsl{r})|$ equals to a constant $|m_0|$ in the bulk, and then we can define $M(\bsl{r})$ such that (s.t.) $|m(\bsl{r})|=|m_0|+M(\bsl{r})$.
Since $M(\bsl{r})$ is constantly zero in the bulk regardless of the gapped boundary, it cannot contribute to any bulk response, and thus we can neglect it in the following derivation, simplifying \eqnref{eq:L_tot_sim} of the main text to
\eq{
\label{eq:L_tot_m0}
\mathcal{L}_a= \overline{\psi}_a \left[\ii (\slashed{\partial}+\ii e \slashed{\widetilde{A}_a}- \ii \slashed{A}_{a,5}\gamma^5) -|m_0| e^{-\ii \Phi_a \gamma^5}\right]\psi_{a}\ .
}

From the above equation, a generic term in the effective action has a specific nonnegative powers of $\widetilde{A}_a$, $A_{a,5}$, and $\Phi_a$, labeled as $n_1$, $n_2$, and $n_3$, respectively.
Among all these terms, the charge response comes from those with $n_1\geq 1$, since the functional derivative of the action with respect to $A$ is required to derive the current.
We next adopt the leading order approximation, which consists of three parts.
First, since we only consider the currents that are linear in $u_{zz}$ or $A$, the terms in the effective action that we are interested in must have $n_1\leq 2$.
Second, we only keep terms to the linear order of $\phi$ and $\bsl{Q}$ for simplicity, and then we can simplify \eqnref{eq:L_tot_m0} to
\eq{
\label{eq:L_tot_m0_Phi}
\mathcal{L}_a= \overline{\psi}_a \left[\ii (\slashed{\partial}+\ii e \slashed{\widetilde{A}_a}- \ii \slashed{A}_{a,5}\gamma^5) -|m_0| +\ii |m_0| \Phi_a \gamma^5\right]\psi_{a}\ .
}
The first two approximations further require $n_2\leq 2$ and $n_3\leq 1$ and forbid $(n_1,n_2,n_3)=(1,2,1), (2,1,1), (2,2,0), (2,2,1)$.
As a result, we have only eight possible values of $(n_1,n_2,n_3)$ that might have nonzero contribution to the effective action within the first two approximations, whose Feynman diagrams are summarized in \figref{fig:Effective_Action_FD}.
The third approximation is that since $u_{zz}$ and $A$ are chosen to slowly vary along with $t$ and/or $\bsl{r}$, we keep at most two space-time derivatives of them.

In the Feynman diagrams, the fermion propagator reads $\frac{1}{\ii} S(k)$ with
\eq{
\label{eq:Fermion_Prop}
S(k)=(\slashed{k}+|m_0|)^{-1}\ ,
}
and the vertices are defined according to \eqnref{eq:L_tot_m0_Phi}.
Note that \eqnref{eq:Fermion_Prop} is defined based on a new definition of the Fourier transformation
\eq{
\psi_{x,a,\alpha}=\int \frac{d^4 q}{(2\pi)^4} e^{\ii q x} \psi_{q,a,\alpha}\ ,
}
since this form leaves the measure invariant.
The Fourier transformation of $\widetilde{A}_a$, $A_{a,5}$, and $\Phi_a$ follows the same rule as above.
In this section, we adopt the real time: $x^\mu=(t,\bsl{r})_\mu$ and define $q^\mu=(\nu,\bsl{q})$.
In the following, we evaluate each graph.
For the derivation, we do not use the dimensional regularization due to the existence of Levi-Civita symbol, but adopt the classic Adler's method for the chiral anomaly, which does not choose a regularization scheme as discussed in \refcite{Bertlmann2000Anomalies,Srednicki2007QFT}.
\eqnref{eq:L_tot_m0_Phi} has effective Lorentz invariance and U(1) gauge invariance, and we preserve it to all orders of quantum correction.

Before evaluating each diagram, we would like to discuss a subtlety.
The $U(1)$ gauge field is added as $\partial_{x^\mu}\rightarrow \partial_{x^\mu} + \ii e A_\mu(x)$ for $\psi_{t,\bsl{r}}$.
According to \eqnref{eq:Fourier_psi_c}, $\psi_{t,\bsl{r}}$ actually corresponds to a fermion at $x^{phy}=(t,\bsl{r}^{phy})=(t,(1+u)\bsl{r})$.
Therefore, $A_\mu(x)$ is related to  the actual physical $U(1)$ gauge field $A^{phy}_\mu(x^{phy})$ as
\eq{
\label{eq:A_phy}
A_0(x)=A^{phy}_0(t, (1+u) \bsl{r})\ ,\ \bsl{A}(x)=(1+u)\bsl{A}^{phy}(t, (1+u) \bsl{r})\ .
}
Nevertheless, we can directly replace $A(x)$ by the physical U(1) gauge field $A^{phy}(x)$ in this work as discussed below.
All current responses can be split into two classes depending on whether the response involves the electron-strain coupling parameter $\xi_i$ or not.
For the response that involves $\xi_i$, it must at least involve the electron-strain coupling $\xi_i u_{zz}$ of power one, which means the strain effect in $A(x)$ (if appears) would make the response non-linear in $u_{zz}$.
Since we at most consider the response to the first order of $u_{zz}$, $A(x)$ should be directly replaced by $A^{phy}(x)$ for the response that involves $\xi_i$.
The $\xi_i$-independent response would stay unchanged even if the electron-strain coupling limits to zero.
If we keep the strain effect in $A(x)$ for this type of response, a $\xi_i$-independent strain-induced current might appear.
Such a strain-induced response corresponds to the motions of the electrons that exactly follow the homogeneous deformation, which is ambiguous at the linear order for the infinitely large system according to \refcite{Vanderbilt2000BPPZ} and thus must be neglected.
Therefore, we should also directly replace $A(x)$ by $A^{phy}(x)$ for the response that does not involve $\xi_i$.
In sum, we can treat $A(x)$ as $A^{phy}(x)$ in our work.
In other words, all the strain-induced linear current responses derived here characterize to what degree the electrons fail to follow the homogeneous deformation, which is what we should consider according to \refcite{Vanderbilt2000BPPZ}.

\subsubsection{$\widetilde{A}_a$ Term}

The contribution of \figref{fig:Effective_Action_FD}(a) to $\ii S_{eff,a}$ reads
\eq{
\label{eq:A}
-\int d^4 x \widetilde{A}_{x,a,\mu} \int \frac{d^4 k}{(2\pi)^4}\Tr\left[-\ii e \gamma^\mu \frac{1}{\ii} S(k)\right]=-e \int d^4 x \widetilde{A}_{x,a,\mu} \Tr\left[  \gamma^\mu \gamma^\nu\right] \int \frac{d^4 k}{(2\pi)^4} \frac{k_{\nu}}{k^2+|m_0|^2}=0\ ,
}
where we use
\eq{
\label{eq:tr_gamma_n}
\Tr\left[\gamma^{\mu_{1}}\gamma^{\mu_{2}}...\gamma^{\mu_{n}}\right]=0 \text{ for odd }n\ ,
}
and
\eq{
\int d^4 k k_{\mu_1} k_{\mu_2} ... k_{\mu_n} f(k^2)=0 \text{ for odd $n$}\ .
}

However, there is a tricky part here.
\eqnref{eq:A} is just one way to assign momentum to the graph \figref{fig:Effective_Action_FD}(a), and there are infinite many other ways as
\eq{
-\int d^4 x \widetilde{A}_{x,a,\mu} \int \frac{d^4 k}{(2\pi)^4}\Tr\left[-\ii e \gamma^\mu \frac{1}{\ii} S(k+p)\right]
}
with $p$ independent of $k$.
As $\Tr\left[ \gamma^\mu S(k)\right]\xrightarrow{k^2\rightarrow \infty} 1/k$ and we do not choose any regularization scheme, we have $\int d^4k \Tr\left[ \gamma^\mu  S(k)\right]$ has $UV$ divergence $\int d^4 k 1/k$.
As a result, $\int \frac{d^4 k}{(2\pi)^4}\Tr\left[ \gamma^\mu  S(k+p)\right]\neq \int \frac{d^4 k}{(2\pi)^4}\Tr\left[ \gamma^\mu  S(k)\right]$ for nonzero $p$, meaning that the expression for \figref{fig:Effective_Action_FD}(a) is ambiguous~\cite{Bertlmann2000Anomalies,Srednicki2007QFT}.
This ambiguity only appears for the UV divergence faster than logrimathiric divergence, which also appears for some other graphs in \figref{fig:Effective_Action_FD} as discussed below.

Nevertheless, we can use symmetry and physics to remove this ambiguity for \figref{fig:Effective_Action_FD}(a).
\eqnref{eq:A}, if nonzero, breaks the effective Lorentz invariance and suggests that a nonzero current would exist without any external perturbation.
Therefore, it should be restricted to zero, meaning that \eqnref{eq:A} is the only allowed way of assigning momentum.

\subsubsection{$\widetilde{A}_a A_{a,5}$ Term}

\figref{fig:Effective_Action_FD}(b) reads
\eq{
-\int \frac{d^4 q}{(2\pi)^4} \widetilde{A}_{q,a,\mu}A_{-q,a,5,\nu} \int \frac{d^4 k}{(2\pi)^4} \Tr\left[\frac{1}{\ii} S(k+q) (-\ii e \gamma^\mu) \frac{1}{\ii} S(k) (\ii \gamma^\nu \gamma^5)\right]\ .
}
With
\eqa{
\label{eq:gamma_5_tr}
& \Tr\left[\gamma^5 \gamma^\mu\gamma^\nu\gamma^\rho\gamma^\delta\right]=-4\ii\varepsilon^{\mu\nu\rho\delta} \\
& \Tr\left[\gamma^5 \gamma^{\mu_1}\gamma^{\mu_2}...\gamma^{\mu_n}\right]=0 \text{ for an odd number $n$ of $\gamma^\mu$ matrices}\\
& \Tr\left[\gamma^5 \gamma^\mu\gamma^\nu\right]=\Tr\left[\gamma^5 \right]=0\ ,
}

\figref{fig:Effective_Action_FD}(b) can be simplified to
\eq{
\label{eq:A_A5}
\int \frac{d^4 q}{(2\pi)^4} e \widetilde{A}_{q,a,\mu} A_{-q,a,5,\nu}(-4\ii) \varepsilon^{\mu\nu\rho\delta}q_{\delta} R_\rho\ ,
}
where
\eq{
R_\rho(q)=\int \frac{d^4 k}{(2\pi)^4}   \frac{k_\rho}{((k+q)^2+|m_0|^2)(k^2+|m_0|^2)}\ ,
}
and $\varepsilon^{\mu\nu\rho\delta}$ is the Levi-Civita symbol.
The divergence of $R_\rho$ is of order $k$, and thus is ambiguous under the shift of the $k$.
In general, for any $f(k)$ that of order $O(1/k^3)$ for large $k^2$, the ambiguity can be evaluated as~\cite{Bertlmann2000Anomalies,Srednicki2007QFT}
\eq{
\label{eq:shift_k_gen}
\int \frac{d^4 k}{(2\pi)^4} (f(k+p)-f(k))=\lim_{k^2\rightarrow \infty}\int \frac{d \Omega_{3+1}}{(2\pi)^4}k^2 (k p) f(k)\ ,
}
where $d\Omega_{3+1}$ is the solid angle in the 3+1D space-time manifold and invariant under of the Lorentz transformation.
Then, for $f_{\rho}(k)\xrightarrow{k^2\rightarrow \infty} k_\rho/k^4$, such as the integrand of $R_\rho$, we have
\eq{
\label{eq:shif_k_A_A5}
\int \frac{d^4 k}{(2\pi)^4} (f_\rho(k+p)-f_\rho(k))=\ii \frac{2\pi^2}{(2\pi)^4}\frac{p_\rho}{4}= \frac{\ii}{32\pi^2}p_\rho\ ,
}
where we use
\eq{
\int d \Omega_{3+1}=\ii 2\pi^2\ ,\ \int d \Omega_{3+1} k_\mu k_\nu f(k^2)=\int d \Omega_{3+1} k^2 f(k^2)\frac{g_{\mu\nu}}{4}\ ,
}
which can be derived with Wick rotation.
Here $g_{\mu\nu}=diag(-1,1,1,1)_{\mu\nu}$ is the metric.

Owing to \eqnref{eq:shif_k_A_A5}, the expression of $R_\rho(q)$ in \eqnref{eq:A_A5} should be
\eq{
R_\rho(q)=\int \frac{d^4 k}{(2\pi)^4}   \frac{k_\rho}{((k+q)^2+|m_0|^2)(k^2+|m_0|^2)} \text{ mod $k$ shift}\ ,
}
after taking into account the ambiguity induced by the momentum shift.
Nevertheless, we can evaluate $R_\rho$ up to a momentum shift.
To do so, we need to introduce the widely used trick called Feynman parametrization~\cite{Srednicki2007QFT}:
\eq{
\label{eq:Feyn_Para}
\prod_{i=1}^n\frac{1}{A_i}=\int d F_n \frac{1}{\left(\sum_{i=1}^n \sum_{A_i} x_i\right)^n}
}
with
\eq{
\int d F_n=(n-1)! \int_0^1 dx_1 \int_0^1 dx_2 ... \int_0^1 dx_n\  \delta\left(-1+\sum_{i=1}^n x_i\right)\ .
}
With this trick, $R_\rho$ becomes
\eq{
R_\rho(q)=\int_0^1 dx \int \frac{d^4 l}{(2\pi)^4} \frac{l_\rho - x q_\rho}{(l^2+D)^2}\text{ mod $l$ shift}\Rightarrow R_{\rho}(q)=B_0(q^2) q_\rho+ \frac{\ii }{32\pi^2} p_\rho\ ,
}
where $l=k+ x q $, $D=x q^2-x^2 q^2+|m_0|^2$, $B_0=\int_0^1 dx \int \frac{d^4 l}{(2\pi)^4} \frac{ - x }{(l^2+D)^2}$, $p_\rho$ marks the ambiguity given  by the shift of the momentum, including the change from $k$ to $l$ in the integral also contributes the momentum shift.
The effective Lorentz in-variance requires $R_\rho(\Lambda q)=\Lambda_\rho^{\ \rho'} R_{\rho'}(q)$ for Lorentz transformation $\Lambda$.
This symmetry requirement imposes a constraint on the momentum shift of the $R_\rho(q)$, $p_\rho$ must have the form $p_\rho= B_1(q^2) q_\rho$, which partially removes the ambiguity.
As a result, we have
\eq{
R_\rho(q)= \left[B_0(q^2) + \frac{\ii }{32\pi^2} B_1 (q^2) \right] q_\rho\ .
}
Substituting the above expression into \eqnref{eq:A_A5}, we find \eqnref{eq:A_A5} is zero.

\subsubsection{$\widetilde{A}_a \Phi_a$ Term}

\figref{fig:Effective_Action_FD}(c) reads
\eq{
-\int \frac{d^4 q}{(2\pi)^4} \widetilde{A}_{q,a,\mu}\Phi_{-q,a} \int \frac{d^4 k}{(2\pi)^4} \Tr\left[\frac{1}{\ii} S(k+q) (-\ii e \gamma^\mu) \frac{1}{\ii} S(k) (-|m_0| \gamma^5)\right]\ .
}
Using \eqnref{eq:gamma_5_tr}, the above expression can be evaluated to zero.

\subsubsection{$\widetilde{A}_a \widetilde{A}_a $ Term}

\figref{fig:Effective_Action_FD}(d) formally is the same as the loop correction to the photon propagator for quantum electrodynamics~\cite{Srednicki2007QFT}, which has the form
\eq{
\ii C_0 \frac{e^2}{2}\int d^4 x \widetilde{F}_{a,\mu\nu}\widetilde{F}_{a}^{\mu\nu}\ ,
}
where $C_0$ is a constant.
According to the expression of $\widetilde{F}_a$, summing the above expression over $a$ gives
\eq{
\ii C_0 e^2\int d^4 x F_{\mu\nu}F^{\mu\nu}\ ,
}
where the terms that cannot contribute to the charge response have been neglected.
Therefore, this term stands for the trivial correction of the permittivity and permeability inside the material.

\subsubsection{$\Phi_a \widetilde{A}_a A_{a,5}$ Terms}

\figref{fig:Effective_Action_FD} (e) and (f)  together give rise to the $\Phi_a \widetilde{A}_a A_{a,5}$ term, which reads
\eqa{
\label{eq:A_A5_Phi}
&-\int \frac{d^4 q_1}{(2\pi)^4} \frac{d^4 q_2}{(2\pi)^4} \widetilde{A}_{q_1,a,\mu} A_{q_2,a,5,\nu} \Phi_{q,a} \int \frac{d^4 k}{(2\pi)^4} \left\{\Tr\left[\frac{1}{\ii}S(k+q_1)(-\ii e \gamma^\mu)\frac{1}{\ii}S(k) (\ii \gamma^\nu\gamma^5)\frac{1}{\ii} S(k-q_2)(-|m_0|\gamma^5)\right] \right.\\
& + \left. \Tr\left[\frac{1}{\ii}S(k)(-\ii e \gamma^\mu)\frac{1}{\ii}S(k-q_1) (-|m_0|\gamma^5)\frac{1}{\ii} S(k+q_2)(\ii \gamma^\nu\gamma^5)\right]\right\}\\
&=\ii\int \frac{d^4 q_1}{(2\pi)^4} \frac{d^4 q_2}{(2\pi)^4} e |m_0|\widetilde{A}_{q_1,a,\mu} A_{q_2,a,5,\nu} \Phi_{q,a} \left(R_1^{\mu\nu}-R_2^{\nu\mu}\right) \\
}
where $q=-q_1-q_2$,
\eq{
R_1^{\mu\nu}=\int \frac{d^4 k}{(2\pi)^4} \Tr\left[S(k+q_1)\gamma^\mu S(k) \gamma^\nu\gamma^5 S(k-q_2)\gamma^5\right]
}
and
\eq{
R_2^{\nu\mu}= \int \frac{d^4 k}{(2\pi)^4}\Tr\left[ \gamma^5 S(k+q_2) \gamma^5\gamma^\nu S(k) \gamma^\mu S(k-q_1)\right]
}
At large $k^2$, we have
\eq{
\Tr\left[S(k+q_1)\gamma^\mu S(k) \gamma^\nu\gamma^5 S(k-q_2)\gamma^5\right] \sim O(\frac{1}{k^4})
}
and
\eq{
\Tr\left[ \gamma^5 S(k+q_2) \gamma^5\gamma^\nu S(k) \gamma^\mu S(k-q_1)\right] \sim O(\frac{1}{k^4})\ ,
}
where we use
\eq{
\Tr\left[\gamma^{\mu_1}\gamma^{\mu_2}...\gamma^{\mu_n}\right]=0\ \text{for odd } n\ .
}
It means $R_1^{\mu\nu}$ and $R_2^{\nu\mu}$ are logarithmically divergent and thus unambiguous under the shift of $k$ in the integral.

With the Feynman parametrization (\eqnref{eq:Feyn_Para}), we can derive the expression for $R_1$ and $R_2$ as
\eq{
R_1^{\mu\nu}=R_2^{\nu\mu}=\int \frac{d^4 l}{(2\pi)^4} \int d F_3 \frac{ 4 |m_0|}{(l^2  + D)^3} \left[(-l^2-|m_0|^2)g^{\mu \nu} - 2 (Q^\mu q^\nu_1+Q^\mu q_2^\nu) +q_1^\mu q_2^\nu-q_2^\mu q_1^\nu+(2q_1Q-Q^2+q_1 q_2)g^{\mu\nu}\right]\ ,
}
where $Q=x_1 q_1 - x_3 q_2$, $D=x_1 q_1^2+x_3 q_2^2-Q^2+|m_0|^2$, and $l=k+Q$.
We also use
\eq{
\Tr\left[\gamma^\mu\gamma^\nu\gamma^\rho\gamma^\delta\right]=4(g^{\mu\nu} g^{\rho\delta}-g^{\mu\rho} g^{\nu\delta}+g^{\mu\delta} g^{\rho\nu})\ \text{and}\ \Tr\left[\gamma^\mu\gamma^\nu\right]=-4 g^{\mu\nu}\ .
}

As a result, \eqnref{eq:A_A5_Phi} is zero.

\subsubsection{$ \widetilde{A}_a A_{a,5} A_{a,5}$ Terms}

\figref{fig:Effective_Action_FD}(g) reads
\eqa{
&-\int \frac{d^4 q_1}{(2\pi)^4} \frac{d^4 q_2}{(2\pi)^4} \widetilde{A}_{-q_1-q_2,a,\mu} A_{q_1,a,5,\nu} A_{q_2,a,5,\rho} \int \frac{d^4 k}{(2\pi)^4} \Tr\left[\frac{1}{\ii}S(k+q_1)(\ii \gamma^\nu \gamma^5)\frac{1}{\ii} S(k) (\ii \gamma^\rho\gamma^5)\frac{1}{\ii}S(k-q_2)(-\ii e \gamma^\mu)\right]\\
&=-e\int \frac{d^4 q_1}{(2\pi)^4} \frac{d^4 q_2}{(2\pi)^4} \widetilde{A}_{-q_1-q_2,a,\mu} A_{q_1,a,5,\nu} A_{q_2,a,5,\rho}   U^{\mu\nu\rho}_{q_1,q_2}\ ,
}
where
\eq{
\label{eq:A_A5_A5_U}
U^{\mu\nu\rho}_{q_1,q_2}=\frac{1}{2}\int \frac{d^4 k}{(2\pi)^4} f_U^{\mu\nu\rho}(k,q_1,0,-q_2)+(q_1\leftrightarrow q_2,\rho \leftrightarrow \nu)\ ,
}
and
\eq{
f_U^{\mu\nu\rho}(k,p_1,p_2,p_3)=\Tr\left[S(k+p_1) \gamma^\nu \gamma^5 S(k+p_2)  \gamma^5\gamma^\rho S(k+p_3) \gamma^\mu\right]\ .
}
Since $f_U^{\mu\nu\rho}(k,p_1,p_2,p_3)$ is of $O(1/k^3)$ order as
\eq{
f_U^{\mu\nu\rho}(k,p_1,p_2,p_3)\xrightarrow{k^2\rightarrow \infty} \frac{\Tr[\gamma^\mu \slashed{k} \gamma^\nu \slashed{k} \gamma ^\rho \slashed{k}]}{k^6}\ ,
}
the shift of $k$ in the integral of $f_U^{\mu\nu\rho}(k,q_1,q_2)$ causes ambiguity as
\eq{
\label{eq:A_A5_A5_momentum_shit}
\int \frac{d^4 k}{(2\pi)^4}\left[f_U^{\mu\nu\rho}(k+p,p_1,p_2,p_3)- f_U^{\mu\nu\rho}(k,p_1,p_2,p_3)\right]=\frac{\ii}{24\pi^2}(p^\nu g^{\mu\rho}+p^\rho g^{\mu\nu}+p^\mu g^{\nu\rho})\ .
}
Here we use
\eq{
\label{eq:int_k^4}
\int d\Omega_{3+1} k_{\mu} k_{\nu}k_{\rho}k_{\delta} f(k^2)=\int d\Omega_{3+1} k^4 k_{\delta} f(k^2)\frac{1}{24}\left(g_{\mu\nu}g_{\rho\delta}+g_{\mu\rho}g_{\nu\delta}+g_{\mu\delta}g_{\rho\nu}\right)
}
and
\eq{
\gamma^\mu \gamma^\nu\gamma_\mu=2 \gamma^\nu\ .
}
Then, we can evaluate the integral of $f_U^{\mu\nu\rho}(k,p_1,p_2,p_3)$ up to a shift of integration variable:
\eqa{
&\int \frac{d^4 k}{(2\pi)^4} f_U^{\mu\nu\rho}(k,p_1,p_2,p_3)\\
&= \int dF_3 \int \frac{d^4 l}{(2\pi)^4} \frac{\Tr[\gamma^\mu (-\slashed{l}+\slashed{Q}_1+|m_0|)\gamma^\nu (\slashed{l}-\slashed{Q}_2+|m_0|)\gamma^\rho (-\slashed{l}+\slashed{Q}_3+|m_0|)]}{(l^2+D)^3}\text{ mod $l$ shift}\\
&= \int d F_3 \int \frac{d^4 l}{(2\pi)^4}\frac{1}{(l^2+|m_0|^2)^3}\left\{ -2 l^2 [g^{\mu\nu} (Q_3-Q_1+Q_2)^\rho)+g^{\mu\rho} (-Q_3+Q_1+Q_2)^\nu+g^{\nu\rho} (Q_3+Q_1-Q_2)^\mu] \right.\\
&\left.+ 4 |m_0|^2 [g^{\mu\nu} (Q_3-Q_1-Q_2)^\rho+g^{\mu\rho} (-Q_3+Q_1-Q_2)^\nu+g^{\nu\rho} (Q_3+Q_1+Q_2)^\mu]\right\}+O(p^3) \text{ mod $l$ shift}
}
where
\eqa{
\label{eq:l_D_Q_A_A5_A5}
l=k+\sum_{i=1}^3 x_i p_i\ ,\ D=\sum_{i=1}^3 x_i q_i^2-(x_1 q_1+x_2 q_2 +x_3 q_3)^2 + |m_0|^2\ ,\ Q_i= \sum_{i=1}^3 x_i p_i-p_i\ ,
}
and the Taylor expansion with respect to the $p_i$ is used for the last equality.

Substitute the $p_1=q_1$, $p_2=0$, and $p_3=-q_2$ in the above equation and use
\eq{
\label{eq:F3_int_x_i}
\int d F_3 x_i=1/3\ ,
}
we arrive at
\eqa{
&\int \frac{d^4 k}{(2\pi)^4} f_U^{\mu\nu\rho}(k,q_1,0,-q_2)= \int \frac{d^4 l}{(2\pi)^4} \frac{1}{(l^2+|m_0|^2)^3}\left\{ -\frac{4}{3} l^2 (g^{\mu\nu} (2q_1+q_2)^\rho+g^{\mu\rho}(-q_1-2q_2)^{\nu}+g^{\nu\rho}(-q_1+q_2)^\mu)\right.\\
&\left.+\frac{8}{3}|m_0|^2 (g^{\mu\nu}(q_1+2q_2)^\rho+g^{\mu\rho}(- q_2-2q_1)^\nu)\right\}+O(q^3)\text{ mod $l$ shift}\ ,
}
where the leading order term would vanish after symmetrizing by $q_1\leftrightarrow q_2$ and $\nu\leftrightarrow \rho$.
We can neglect the $O(q^3)$ terms in the above equation since the leading order approximation allows at most two space-time derivatives on the fields, meaning that we only need to consider the terms up to $q^2$ order.
Then, since \eqnref{eq:A_A5_A5_momentum_shit} indicates that \eqnref{eq:A_A5_A5_U} only holds up to a shift in integration variable, we have
\eq{
U^{\mu\nu\rho}_{q_1,q_2}=\frac{\ii}{24\pi^2}(p^\nu g^{\mu\rho}+p^\rho g^{\mu\nu}+p^\mu g^{\nu\rho})\ ,
}
where $p^\mu$ labels the total ambiguity, and $p^\mu$ can depend on $q_1$ or $q_2$ but must be invariant under $q_1\leftrightarrow q_2$.
This ambiguity is removed by the $U(1)$ gauge invariance (Ward identity) that requires
\eq{
(-q_1-q_2)_\mu U^{\mu\nu\rho}_{q_1,q_2}=0\ ,
}
which further requires that $p$ should be set to zero.
As a result, we conclude that \figref{fig:Effective_Action_FD}(g) can be neglected within the leading-order approximation.

\subsubsection{$\widetilde{A}_{a} \widetilde{A}_{a} A_{a,5}$ term}

\figref{fig:Effective_Action_FD}(h) reads
\eqa{
&-\int \frac{d^4 q_1}{(2\pi)^4} \frac{d^4 q_2}{(2\pi)^4} \widetilde{A}_{q_1,a,\mu} \widetilde{A}_{q_2,a,\nu} A_{-q_1-q_2,a,5,\rho} \int \frac{d^4 k}{(2\pi)^4} \Tr\left[\frac{1}{\ii}S(k+q_1)(-\ii e \gamma^\mu )\frac{1}{\ii} S(k) (-\ii e \gamma^\nu)\frac{1}{\ii}S(k-q_2)(\ii \gamma^\rho \gamma^5)\right]\\
&=\ii e^2 \int \frac{d^4 q_1}{(2\pi)^4} \frac{d^4 q_2}{(2\pi)^4} \widetilde{A}_{q_1,a,\mu} \widetilde{A}_{q_2,a,\nu} A_{-q_1-q_2,a,5,\rho}V^{\mu\nu\rho}_{q_1,q_2}\ ,
}
where
\eq{
\label{eq:A_A_A5_V}
V^{\mu\nu\rho}_{q_1,q_2}=\frac{\ii}{2}\int \frac{d^4 k}{(2\pi)^4} f_V^{\mu\nu\rho}(k,q_1,0,-q_2)+(q_1\leftrightarrow q_2,\mu \leftrightarrow \nu)\ ,
}
and
\eq{
f_V^{\mu\nu\rho}(k,p_1,p_2,p_3)=\Tr\left[\gamma^5 S(k+p_1) \gamma^\mu S(k+p_2)  \gamma^\nu S(k+q_3) \gamma^\rho \right]\ .
}

Since $f_V^{\mu\nu\rho}(k,p_1,p_2,p_3)$ decays as $1/k^3$ for large $k^2$ as
\eq{
f_V^{\mu\nu\rho}(k,p_1,p_2,p_3)\xrightarrow{k^2\rightarrow \infty} -\frac{\Tr[\gamma^5\slashed{k}\gamma^\mu\slashed{k}\gamma^\nu\slashed{k}\gamma^\rho]}{k^6}\ ,
}
its integral is ambiguous as
\eq{
\label{eq:A_A_A5_fV_k_shift}
\int \frac{d^4 k}{(2\pi)^4} \left[f_V^{\mu\nu\rho}(k+p,p_1,p_2,p_3)- f_V^{\mu\nu\rho}(k,p_1,p_2,p_3)\right]=\frac{1}{8\pi^2}\varepsilon^{\mu\nu\rho\delta}p_{\delta}\ ,
}
where \eqnref{eq:gamma_5_tr} and \eqnref{eq:int_k^4} are used.
It means \eqnref{eq:A_A_A5_V} holds up to a shift of momentum.

Now we evaluate the integral of $f_V^{\mu\nu\rho}(k,p_1,p_2,p_3)$:
\eqa{
& \int \frac{d^4 k}{(2\pi)^4}f_V^{\mu\nu\rho}(k,p_1,p_2,p_3) \\
& =\int d F_3 \int \frac{d^4 l}{(2\pi)^4} \frac{\Tr\left[\gamma^5(-\slashed{l}+\slashed{Q}_1+|m_0|)\gamma^\mu(-\slashed{l}+\slashed{Q}_2+|m_0|)\gamma^\nu(-\slashed{l}+\slashed{Q}_3+|m_0|)\gamma^\rho\right]}{(l^2+D)^3} \text{ mod $l$ shift}\\
& = \int d F_3 \int \frac{d^4 l}{(2\pi)^4} \frac{-4\ii}{(l^2+|m_0|^2)^3} \varepsilon^{\delta\mu\nu\rho}\left[(Q_1+Q_2+Q_3)_\delta \frac{l^2}{2}+(Q_1-Q_2+Q_3)_\delta |m_0|^2 \right] + O(p^3) \text{ mod $l$ shift}\\
& = \frac{1}{8 \pi^2}\varepsilon^{\mu\nu\rho\delta}\frac{2}{3}(p_1-2 p_2 + p_3)_{\delta}+O(p^3) \text{ mod $l$ shift}\\
& = 0 + O(p^3) \text{ mod $l$ shift}\ ,
}
where the first equality uses \eqnref{eq:Feyn_Para} and \eqnref{eq:l_D_Q_A_A5_A5}, the second equality uses \eqnref{eq:gamma_5_tr} and the Tyler expansions with respect to $p_i$, the third equality uses \eqnref{eq:F3_int_x_i} and
\eq{
\int \frac{d^4 l}{(2\pi)^4} \frac{1}{(l^2+D)^3}\xrightarrow{\text{Wick Rotation}} \int \ii \frac{d^4 \bar{l}}{(2\pi)^4} \frac{1}{(\bar{l}^2+D)^3}=\frac{\ii}{32\pi^2 D}\ ,
}
and the last equality uses \eqnref{eq:A_A_A5_fV_k_shift}.
Substituting $p_1=q_1$, $p_2=0$ and $p_3=-q_2$ into the above equation and neglect $O(q^3)$ terms according to the leading order approximation, we can derive the expression of $V$ from \eqnref{eq:A_A_A5_V}as
\eq{
V^{\mu\nu\rho}_{q_1,q_2}=\frac{\ii}{8\pi^2}\epsilon^{\mu\nu\rho\delta}p_{\delta}\ ,
}
where $p_{\delta}$ should depend on $q_1,q_2$ and changes its sign under $q_1\leftrightarrow q_2$.
Then, in general $p_{\delta}=q_{1,\delta} C_1(q_1,q_2)+ q_{2,\delta} C_2(q_1,q_2)$, where $C_{2}(q_2,q_1)=-C_1(q_1,q_2)$.
Again, the ambiguity is removed by the $U(1)$ gauge invariance
\eq{
q_{1,\mu}V^{\mu\nu\rho}_{q_1,q_2}=q_{2,\nu}V^{\mu\nu\rho}_{q_1,q_2}=0\ ,
}
which requires $C_1=C_2=0$.
Eventually, we know \figref{fig:Effective_Action_FD}(h) is negligible within the leading order approximation.

\subsubsection{$\widetilde{A}_{a} \widetilde{A}_{a} \Phi_{a}$ Term}

\figref{fig:Effective_Action_FD}(i) reads
\eqa{
\label{eq:nontrivial_response_app}
&-\int \frac{d^4 q_1}{(2\pi)^4} \frac{d^4 q_2}{(2\pi)^4} \widetilde{A}_{q_1,a,\mu} \widetilde{A}_{q_2,a,\nu} \Phi_{-q_1-q_2,a} \int \frac{d^4 k}{(2\pi)^4} \Tr\left[\frac{1}{\ii}S(k+q_1)(-\ii e \gamma^\mu )\frac{1}{\ii} S(k) (-\ii e \gamma^\nu)\frac{1}{\ii}S(k-q_2)(-|m_0| \gamma^5)\right]\\
&= -\ii \int \frac{d^4 q_1}{(2\pi)^4} \frac{d^4 q_2}{(2\pi)^4} \widetilde{A}_{q_1,a,\mu} \widetilde{A}_{q_2,a,\nu} \Phi_{-q_1-q_2,a} e^2 |m_0| T^{\mu\nu}(q_1,q_2)\ ,
}
where
\eq{
T^{\mu\nu}(q_1,q_2)=\frac{1}{2} \int \frac{d^4 k}{(2\pi)^4} f_T^{\mu\nu}(k,q_1,0,-q_2)+(\mu\leftrightarrow \nu, q_1\leftrightarrow q_2)\ ,
}
and
\eq{
f_T^{\mu\nu}(k,p_1,p_2,p_3)=\Tr\left[S(k+p_1)\gamma^\mu S(k+p_2) \gamma^\nu S(k+p_3) \gamma^5\right]\ .
}
Since
\eq{
f_T^{\mu\nu}(k,p_1,p_2,p_3)\xrightarrow{k^2\rightarrow\infty} O(1/k^4)\ ,
}
the integral of $f_T$ has at most logarithmic divergence, and thus is unambiguous under the shift of integration variable.
Then, we can evaluate the integral
\eqa{
&\int \frac{d^4k}{(2\pi)^4} f_T^{\mu\nu}(k,p_1,p_2,p_3)\\
&= \int d F_3\int \frac{d^4 l}{(2\pi)^4} \frac{\Tr\left[\gamma^5(-\slashed{l}+\slashed{Q}_1+|m_0|)\gamma^\mu(-\slashed{l}+\slashed{Q}_2+|m_0|)\gamma^\nu(-\slashed{l}+\slashed{Q}_3+|m_0|)\gamma^5\right]}{(l^2+D)^3} \\
&= \int d F_3\int \frac{d^4 l}{(2\pi)^4} \frac{|m_0|}{(l^2 +|m_0|^2)^3} (-4\ii) \varepsilon^{\mu\nu\rho\delta} (-Q_{1,\rho} Q_{2,\delta} -Q_{2,\rho} Q_{3,\delta}-Q_{3,\rho} Q_{1,\delta})+O(p^3)\\
&= -\frac{1}{8\pi^2 |m_0|} \varepsilon^{\mu\nu\rho\delta} (p_{1,\rho} p_{2,\delta} +p_{2,\rho} p_{3,\delta}+p_{3,\rho} p_{1,\delta})+O(p^3)\ ,
}
where the first equality uses \eqnref{eq:Feyn_Para} and \eqnref{eq:l_D_Q_A_A5_A5}, the second equality uses \eqnref{eq:gamma_5_tr} and the Tyler expansion with respect to $p_i$, and the third equality uses $\int d F_n\ 1=1$.

Substituting $p_1=q_1$, $p_2=0$, and $p_3=-q_2$ and neglecting the $O(q^3)$ terms, we have
\eq{
T^{\mu\nu}(q_1,q_2)=-\frac{1}{8\pi^2 |m_0|}\varepsilon^{\mu\nu\rho\delta}q_{1\rho}q_{2\delta}\ ,
}
which means the leading-order contribution from \figref{fig:Effective_Action_FD}(i) to the effective action has the form of the axion term:
\eq{
\int d^4 x \frac{e^2}{32\pi^2} \Phi_a \epsilon^{\mu\nu\rho\delta} \widetilde{F}_{a,\mu\nu}\widetilde{F}_{a,\rho\delta}\ .
}
Only this term contains nontrivial contribution to the leading-order linear response.

\subsubsection{Restoring Fermi Velocities}

As shown above, only \figref{fig:Effective_Action_FD}(d) and (i) have nonzero contribution to the leading-order linear response.
\figref{fig:Effective_Action_FD}(d) only gives the correction to the permittivity and permeability in the material and thus is trivial.
The only nontrivial leading-order linear response comes from \figref{fig:Effective_Action_FD}(i), which gives \eqnref{eq:S_eff_a_leading}.

Note that the following transformation is used to derive \eqnref{eq:L_tot} and \eqnref{eq:L_tot_sim} of the main text,
\eqa{
\label{eq:para_trans}
& r_i\rightarrow r_i v_i\ ,q_i\rightarrow q_i/v_i, k_{a,\alpha,i}\rightarrow k_{a,\alpha,i}/v_i,\\
& A_i\rightarrow A_i/v_i, \psi_{a,\alpha}\rightarrow |v_x v_y v_z|^{-1/2} \psi_{a,\alpha}\ .
}
Then, we can perform the inverse transformation on \eqnref{eq:S_eff_a_leading} to restore the Fermi velocities.

As a result, \eqnref{eq:E_pse} of the main text becomes
\eq{
\label{eq:E_pse_RFV}
\bsl{E}^{pse}_a=(-1)^a \frac{\xi_y}{e v_y} \dot{u}_{zz} \bsl{e}_y\ ,
}
\eqnref{eq:theta_a} of the main text becomes
\eq{
\label{eq:theta_a_RFV}
\theta_a=(-1)^{a-1} \sgn{v_x v_y v_z} \phi\ ,
}
\eqnref{eq:sigma_H_a} of the main text becomes
\eq{
\label{eq:sigma_H_a_RFV}
\bsl{\Sigma}_{H,a}=(-1)^a \sgn{v_x v_y v_z}  \frac{\bsl{Q}}{2\pi} \frac{e^2}{2\pi}\ .
}
\eqnref{eq:S_eff_a_uzz} of the main text becomes
\eq{
\label{eq:S_eff_a_uzz_RFV}
S_{eff,a,u_{zz}}=\sgn{v_x v_y v_z} \frac{e \xi_y}{4\pi^2 v_y}\int dt d^3 r   (\phi+\bsl{Q}\cdot\bsl{r})\dot{u}_{zz} B_y\ ,
}
\eqnref{eq:LE_chi} of the main text becomes
\eq{
\label{eq:LE_chi_RFV}
\chi_{izz}=\frac{\partial j_{PE,i}}{\partial \dot{u}_{zz}}= \sgn{v_x v_y v_z}\frac{e}{2\pi^2} \frac{\xi_y}{v_y} [\bsl{Q}\times \bsl{e}_y]_i\ ,
}
and \eqnref{eq:M_bulk} of the main text becomes
\eq{
\label{eq:M_bulk_RFV}
\bsl{M}^{bulk}=-\sum_a \frac{e^2}{2 \pi}\frac{\theta_a^{bulk}}{2\pi} \bsl{E}^{pse}_{a} \ .
}

\subsubsection{Massless Limit}

The derivation here suggests that the axion term comes from $\ii |m_0| \Phi_a \bar{\psi}_a\gamma^5 \psi_a$ term, while the $A_{a,5,\mu} \bar{\psi}_a \gamma^\mu \gamma^5 \bar{\psi}$ has zero contribution.
This seems to contradict the fact that $A_{a,5,\mu} \bar{\psi}_a \gamma^\mu \gamma^5 \psi_a$ in the $|m_0|=0$ case should account for the axion term of the chiral anomaly.
The reason for this seeming contradiction is that the Taylor expansion with respect to the momentum of $\widetilde{A}_a$/$A_{a,5}$/$\Phi_a$ that we perform above is invalid in the $|m_0|\rightarrow 0$ limit.
In another word, the method used here does not have the proper $|m_0|\rightarrow 0$ limit and thus cannot restore the massless chiral anomaly.
However, we still adopt this method since it can reproduce the previous experimentally-verified results~\cite{Zhang2013WSMCDWAxion,Gooth2019WSMCDWAxion} and the results derived above are verified by the TB model as discussed in \appref{app:TB}.

\subsubsection{Momentum-Cutoff Correction to \eqnref{eq:nontrivial_response_app}}

The nontrivial contribution to the leading order response is given by \eqnref{eq:nontrivial_response_app}.
In the above, we have derived the response without imposing a finite momentum cutoff.
In this part, we will impose a momentum cutoff $|\bsl{k}| < \Lambda$ in \eqnref{eq:nontrivial_response_app} to discuss the possible correction.
$\Lambda$ here is different from that in \appref{app:CDW_MF}.

The key quantity in \eqnref{eq:nontrivial_response_app} is $T^{\mu\nu}(q_1,q_2)$, which reads
\eq{
T^{\mu\nu}(q_1,q_2)=\frac{1}{2} I^{\mu \nu}(q_1,q_2)+(\mu\leftrightarrow \nu, q_1\leftrightarrow q_2)\ ,
}
where
\eq{
I^{\mu \nu}(q_1,q_2)=\int \frac{d^4 k}{(2\pi)^4} \Tr\left[S(k+q_1)\gamma^\mu S(k) \gamma^\nu S(k-q_2) \gamma^5 \right]\ .
}
With \eqnref{eq:gamma_5_tr}, we have
\eq{
I^{\mu \nu}(q_1,q_2)=|m_0| \int \frac{d^4 k}{(2\pi)^4}\frac{-4\ii \epsilon^{\rho\mu\delta\nu} q_{1,\rho} q_{2,\delta}}{((k+q_1)^2+|m_0|^2)(k^2+|m_0|^2)((k-q_2)^2+|m_0|^2)}\ .
}
Performing the Wick rotation and neglecting the $O(q^3)$ order since $\widetilde{A}$ is a slow field, we arrive at
\eqa{
I^{\mu \nu}(q_1,q_2)& =4 \epsilon^{\rho\mu\delta\nu} q_{1,\rho} q_{2,\delta}\frac{4 \pi}{(2\pi)^4} \int^\Lambda_0 d|\bsl{k}| |\bsl{k}|^2 \int d \bar{k}_0 \frac{|m_0|}{( \bar{k}_0^2 +|\bsl{k}|^2 + |m_0|^2 )^3} \\
& = \epsilon^{\rho\mu\delta\nu} q_{1,\rho} q_{2,\delta} \frac{1}{8\pi^2}
\frac{1}{(\frac{|m_0|^2}{\Lambda^2}+1)^{3/2}} \frac{1}{|m_0|}\ .
}
As a result, the corresponding terms in the action reads
\eq{
\label{eq:Lambda_correction}
\left[\frac{1}{(\frac{|m_0|^2}{\Lambda^2}+1)^{3/2}} \right]\int d^4 x \frac{e^2}{32\pi^2} \Phi_a \epsilon^{\mu\nu\rho\delta} \widetilde{F}_{a,\mu\nu}\widetilde{F}_{a,\rho\delta}=\left[1+ O(\frac{|m_0|^2}{\Lambda^2}) \right]\int d^4 x \frac{e^2}{32\pi^2} \Phi_a \epsilon^{\mu\nu\rho\delta} \widetilde{F}_{a,\mu\nu}\widetilde{F}_{a,\rho\delta}\ ,
}
implying that imposing a momentum cutoff $\Lambda$ would lead to a $O(\frac{|m_0|^2}{\Lambda^2})$ correction to the response coefficient of \eqnref{eq:S_eff_a_gen}.

In the above derivation, we use \eqnref{eq:L_tot_sim}, in which the momentum origin of the fermion field is typically not at the Weyl point.
However, the order of magnitude of the derived correction to the response coefficient will be left invariant under a shift of the origin to the Weyl points, since such a shift can only change the forms of $A_{a,5}$ and $\Phi_a$, and therefore cannot affect the form of the response coefficient according to \eqnref{eq:Lambda_correction}.

\subsection{A Symmetry Preserving Boundary Condition}

\begin{figure}[t]
    \centering
    \includegraphics[width=\columnwidth]{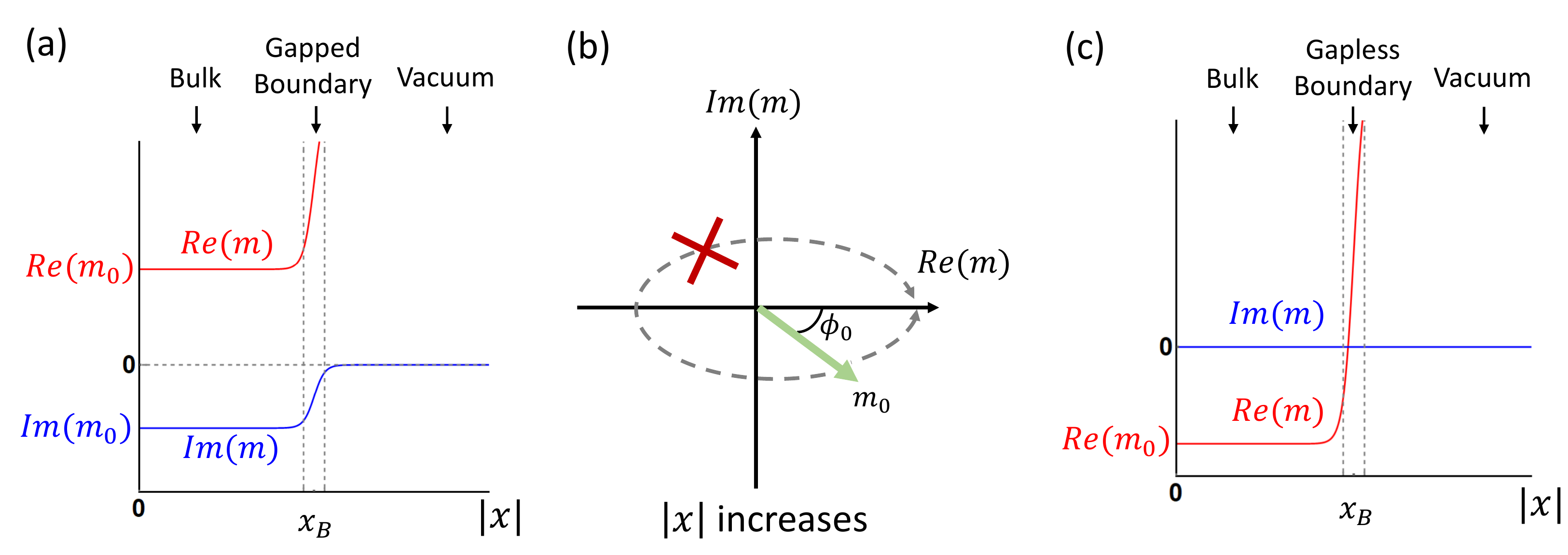}
    \caption{
    (a) shows the spacial dependence of $m$ in \eqnref{eq:L_tot_fermion_sim} for a slab configuration with boundaries perpendicular to $x$.
    Both $\Im(m)$ and $\Re(m)$ are smooth and monotonic functions of $|x|$.
    (b) shows the spatial dependence of the phase $\phi$ for (a).
    Owing to the monotonicity of $\Im(m)$, the $\phi_0$ should take values in $(-\pi,\pi)$ and the continuous $\phi$ cannot take the path that passes $\pm\pi$ as $|x|$ increases.
    (c) shows the spatial dependence of $m$ for $\Im(m_0)=0$ and $\Re(m_0)<0$, where a gapless mode appears at the boundary.
    }
    \label{fig:BC}
\end{figure}

In this part, we present a symmetry-preserving boundary condition that can realize the boundary TQPT. We consider a slab sample with open boundary perpendicular to $x$ and thickness $2 x_B$, \ie, $|x|<x_B$, while the momenta along $y$ and $z$ are kept as good quantum numbers.
For the discussion of the boundary condition, we can omit the gauge field since it has no effect on the boundary condition.
Moreover, we can also neglect strain since we choose it to be homogeneous in the entire space, meaning that the difference between the material and vacuum is solely accounted for by the spatial dependence of the CDW order parameters.
We choose the boundary to preserve the TR and mirror symmetries, and thereby we may focus on one valley since the other one is related by the TR symmetry.
In this case, we can freely rotate the fermion bases to cancel the $\varphi_a$ and $\bsl{Q}$ terms in \eqnref{eq:L_tot} of the main text, leaving us
\eq{
\label{eq:L_tot_fermion_sim}
\mathcal{L}_a= \overline{\psi}_a \left[\ii \slashed{\partial} - |m| e^{-\ii (-1)^{a-1}\phi \gamma^5}\right]\psi_{a}\ .
}
As mentioned in the main text, $m$ equals to a constant $m_0$ deep in the bulk of the sample, \ie, $m(|x|=0)=m_0$; the vacuum outside the sample is approximated as a Dirac fermion with infinitely large real mass for each valley, \ie, $\Re[m(|x|\rightarrow\infty)]\rightarrow \infty$ and $\Im[m(|x|\rightarrow\infty)]=0$.
Between these two limits, we choose both $\Re[m]$ and $\Im[m]$ to be smooth and monotonic for simplicity (see \figref{fig:BC}(a) for an example), and thus $|m|$ is always continuous.

Owing to $\Im[m(|x|\rightarrow\infty)]=0$, we can always set $\phi(|x|\rightarrow\infty)=0$ for the vacuum.
Furthermore, we always try to choose a continuous $\phi$.
When $\Im[m(|x|=0)]=\Im[m_0]\neq 0$, the monotonicity requires $\Im[m]$ cannot take zero values for finite $|x|$.
As a result, a continuous $\phi$ can only take values in $(-\pi,\pi)$ and thus requires $\phi(|x|=0)=\arg(m_0)\equiv \phi_0$ to only take values in $(-\pi,\pi)$, since $\phi$ otherwise must pass $\pm \pi$ as $|x|$ increases and breaks the monotonicity of $\Im[m]$.(See \figref{fig:BC}(b).)
When $\Im[m_0]=0$, the monotonicity requires $\Im[m(|x|)]=0$ for any $x$ and $\phi$ must be a step function with $\phi(|x|<x_B)=n\pi$ with $n$ odd and $\phi(|x|>x_B)=0$.
This discontinuity of $\phi$ for $\Im[m_0]=0$ comes from the gapless boundary mode given by the real mass domain wall $\Re[m_0]<0$ and $\Re[m(|x|\rightarrow \infty)]\rightarrow +\infty$, as well as the vanishing $\Im[m]=0$, as shown in \figref{fig:BC}(c).
When $\phi_0\neq \pi$ (\eg, \figref{fig:BC}(a)), the nonvanishing $\Im[m]$ guarantees the boundary to be gapped and makes sure that $\phi$ is continuous.
Since the derivation of the response from the effective action is only valid when the system is gapped everywhere and $\phi$ is continuous, \eqnref{eq:M_bulk} of the main text is only valid for $\phi_0\neq \pi$, which gives us a uniform strain induced magnetization deep in the bulk of the system.
If we keep $\Re[m_0]<0$ and tune $\Im[m_0]$ from $0^-$ to $0^+$, $\phi_0$ should jump from $-\pi+0^+$ to $\pi + 0^-$, leading to a jump of bulk magnetization.
This jump is induced by the gap closing at the boundary of the system, while the bulk of the system stay gapped.

In the above discussion, we choose smooth monotonic functions for $m$.
The existence of the gapless boundary mode for each valley and the magnetization jump is stable against any symmetry-preserving perturbation, as long as the two valleys are well defined.
It is because the gapless mode for one valley on one surface is a Weyl point in the $(q_y,q_z,\phi_0)$ space, meaning that the perturbations can only shift the appearance of the gapless mode and the magnetization jump to other values of $\phi_0$ instead of removing them.
It coincides with the fact that the gap closing only needs 1 fine-tuning parameter.

\section{Details on TB Model}
\label{app:TB}

\begin{figure}[t]
    \centering
    \includegraphics[width=0.8\columnwidth]{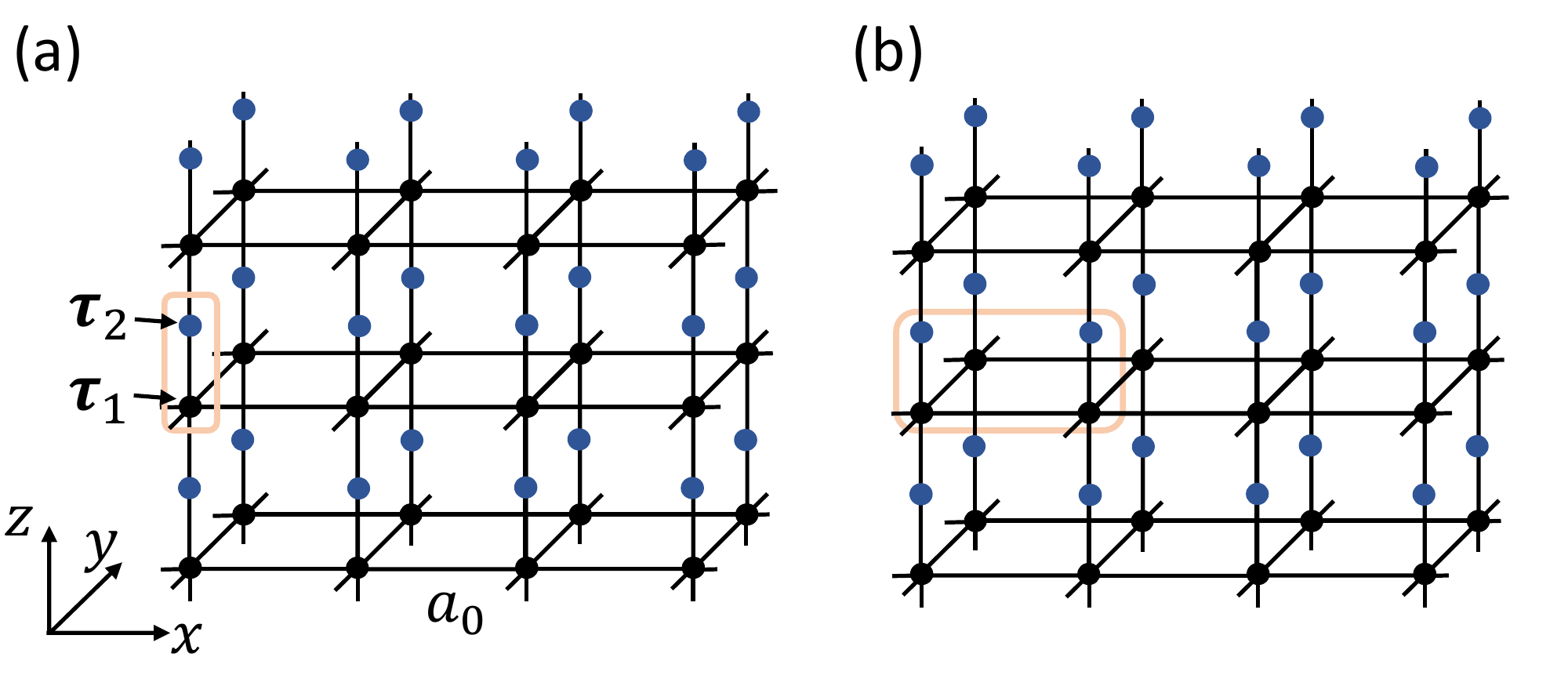}
    \caption{
    (a) shows the lattice of the TB model without CDW \eqnref{eq:H_TB_u}.
    This is a cubic lattice with lattice constant $a_0$ and there are two sublattices in one unit cell.
    (b) demonstrates the choice of the unit cell for the TB model with CDW.
    Now the unit cell contains four sublattices.
    }
    \label{fig:TB_Lattice}
\end{figure}

In this part, we build a TB model to reproduce the results derived from the effective action.

\subsection{Without CDW}

We first consider the case without CDW.
The model is built on a square lattice with lattice constant set to $a_0$, and each lattice site contains two sub-lattice atoms, one at $\bsl{\tau}_1=(0,0,0)$ and the other at $\bsl{\tau}_{2}=(0,0,1/2)a_0$, as shown in \figref{fig:TB_Lattice}(a).
We put a spinful $s$ orbital at $\bsl{\tau}_1$ and a spinful $p_y$ orbital at $\bsl{\tau}_2$, and the bases read $\ket{\bsl{R}+\bsl{\tau}_i,s}$ with $s=\uparrow\downarrow$ the spin index.
According to the Fourier transformation
\eq{
c^\dagger_{\bsl{k},i,s}=\frac{1}{\sqrt{N}}\sum_{\bsl{R}}e^{\ii (\bsl{R}+\bsl{\tau}_i)\cdot\bsl{k}}c^\dagger_{\bsl{R}+\bsl{\tau}_i,s}\ ,
}
the representations of the symmetry operations read
\eq{
m_y c^\dagger_{\bsl{k}}m_y^{-1}=c^\dagger_{m_y\bsl{k}}(-\ii \tau_z\sigma_y)\ ,\ \TR c^\dagger_{\bsl{k}}\TR^{-1}=c^\dagger_{-\bsl{k}}(\ii \tau_0\sigma_y)\ ,
}
where $c^\dagger_{\bsl{R}+\bsl{\tau}_i,s}$ is the creation operator for $\ket{\bsl{R}+\bsl{\tau}_i,s}$, and $\tau$ and $\sigma$ label the sublattice and spin indices, respectively.

With certain nearest-neighbor hopping terms, we choose the following symmetry-allowed form for the strained TB model:
\eq{
\label{eq:H_TB_u}
H_{TB,u}=\sum_{\bsl{k}}c^\dagger_{\bsl{k}} h_{TB,u}(\bsl{k}) c_{\bsl{k}}\ ,
}
\eq{
h_{TB,u}(\bsl{k})=\frac{1}{a_0}\left[ d_1 \tau_z\sigma_0+d_2\tau_y\sigma_0+d_3\tau_x\sigma_x+d_4\tau_x\sigma_z+d_5\tau_y\sigma_x\right]\ ,
}
where the $\bsl{k}$ dependence of $d$'s is implied, the strain-induced redefinition of $c_{\bsl{k}}$ discussed in \appref{app:el-str} is implied,  and
\eqa{
& d_1=n_0-1+\cos(k_x a_0)+n_2 \cos(k_y a_0) + (1-u_{zz}) \cos(k_z a_0) \\
& d_2=(1 - \frac{u_{zz}}{5})\sin(k_y a_0)\cos(k_z a_0/2)\\
& d_3=(1 - u_{zz})\sin(k_z a_0/2) \\
& d_4=(1 - \frac{u_{zz}}{5})\cos(k_x a_0)\sin(k_z a_0/2)\\
& d_5=(1 - \frac{u_{zz}}{5})n_1 \cos(k_z a_0 /2)\cos(k_y a_0)+ (1 - \frac{u_{zz}}{5}) n_3 \cos(k_z a_0/2) \cos(k_x a_0)\ ,
}
where $u_{zz}$ stands for the normal strain along $z$.
$H_{TB,0}$ is just $H_{TB,u}$ with $u_{zz}=0$.
Eigenenergies take the form $\pm \sqrt{d_1^2+d_3^2+(\sqrt{d_2^2+d_4^2}\pm |d_5|)^2}$, and we consider half filling, resulting in the gapless condition $d_1=d_3=\sqrt{d_2^2+d_4^2}-|d_5|=0$.

For concreteness, we choose
\eq{
\label{eq:n_values}
n_0=-\sqrt{2},\ n_1=1,\ n_2=2,\ n_3=-1\ .
}
Then, without $u_{zz}=0$, the gapless points exist at $\bsl{k}=(\pm \pi/2,\pm \pi/4,0)/a_0$, and the zero-energy eigenvectors at $\bsl{k}=(\pi/2, \pi/4,0)$ read
\eqa{
&v_1 = e^{\ii 5\pi/8}(1,-1,-1,1)^T/2\\
&v_2 =  e^{-\ii \pi/8}(1,-1,1,-1)^T/2\ .
}
The zero-energy eigenvectors at the three other gapless points are related by symmetries to realize \eqnref{eq:c_sym_rep}.
In general, the expression of $v_1$ and $v_2$ allows an arbitrary global $U(1)$ factor, \ie
\eq{
\label{eq:v_U1}
v_1\rightarrow v_1 e^{\ii \varphi}\ ,\ v_2\rightarrow v_2 e^{\ii \varphi}\ ,
}
which can alter the projection of CDW in the following.

\subsection{With CDW}

The CDW-like term that we add in the TB model is shown in \eqnref{eq:TB_CDW} of the main text, where
 \eq{
 M_1=[-\cos(k_z a_0/2)\tau_y\sigma_0+\tau_z\sigma_x]\sin(k_y a_0)\ ,
 }
 and
 \eq{
 M_2=[-\cos(k_z a_0/2)\tau_y\sigma_x+\tau_z\sigma_0]/\sqrt{2}\ .
 }
The CDW term couples Weyl points that are separated by $(\pi,0,0)/a_0$, which is commensurate.
Therefore, we can double the unit cell along $x$ by defining
\eq{
\bar{c}^\dagger_{R_x',k_y',k_z',i_x}=c^\dagger_{R_x'+i_x a_0,k_y',k_z'}
}
 with $R_x'=2 l_x' a_0 $, $i_x=0,1$, and $l_x'$ an integer, to exploit the reduced lattice translation symmetry.
It means that the new lattice constants are $a_x'=2a_0,a_y'=a_0,a_z'=a_0$, as shown in \figref{fig:TB_Lattice}(b).
Using $k_x'$ to label the Bloch momentum conjugate to $R_x'$ and defining $\bsl{k}'=(k_x',k_y',k_z')$, we can re-write the CDW term as
\eq{
H_{TB,CDW}=\sum_{\bsl{k}'} \bar{c}^\dagger_{\bsl{k}'}h_{CDW}(\bsl{k}') \bar{c}_{\bsl{k}'}=\sum_{\bsl{k}'} \bar{c}^\dagger_{\bsl{k}'}\left[\mu_1 \rho_y\sin(k_x' a_x'/2) M_1(k_y',k_z')+\mu_2\rho_z M_2(k_y',k_z')\right] \bar{c}_{\bsl{k}'}\ .
}
$\rho$'s are Pauli matrices for new index $1,2$ introduced by the doubling the unit cell,
\eq{
\bar{c}^\dagger_{\bsl{k}'}=\frac{1}{\sqrt{N_x'}}\sum_{R_x'}e^{\ii R_x' k_x'}\bar{c}^\dagger_{R_x',k_y',k_z'} \mat{1 & \\ & e^{\ii k_x' a_x'/2}}\ ,
}
and $N_x'=N_x/2$.

The previous $H_{TB,u}$ can also be rewritten with $\bar{c}$ as
\eq{
H_{TB,u}=\sum_{k_x',k_y,k_z} \bar{c}^\dagger_{\bsl{k}'}\bar{h}_{TB,u}(\bsl{k}') \bar{c}_{\bsl{k}'}\ ,
}
where
\eqa{
&\bar{h}_{TB,u}(\bsl{k}')=\frac{1}{a_0}\left\{\rho_0\left[(n_0-1+n_2 \cos(k_y'a_y') + (1-u_{zz}) \cos(k_z'a_z'))\tau_z\sigma_0 + d_2(\bsl{k}')\tau_y\sigma_0+d_3(\bsl{k}') \tau_x\sigma_x\right. \right.\\
&\left.+(1 - \frac{u_{zz}}{5})n_1 \cos(k_z' a_z'/2)\cos(k_y' a_y')\tau_y\sigma_x\right]\\
&\left.+\cos(k_x' a_x'/2)\rho_x\left[\tau_z\sigma_0+(1 - \frac{u_{zz}}{5})\sin(k_z'a_z'/2)\tau_x\sigma_z+(1 - \frac{u_{zz}}{5}) \frac{n_3}{2} \cos(k_z' a_z'/2)\tau_y\sigma_x\right]\right\}\ .
}
Then, the total Hamiltonian reads $H_{TB}=H_{TB,u}+H_{TB,CDW}$.

When $\mu_2=0$, the model has an effective $m_x$ symmetry
\eq{
\rho_x [\bar{h}_{TB,u}(k_x',\bsl{k}'_\perp)+h_{CDW}(k_x',\bsl{k}'_\perp)] \rho_x =\bar{h}_{TB,u}(-k_x',\bsl{k}'_\perp)+h_{CDW}(-k_x',\bsl{k}'_\perp)\ .
}

\subsection{Low-Energy Projection}

Suppose $[H_{TB,0},c^{\dagger}_{\bsl{k}} v]=E c^{\dagger}_{\bsl{k}} v$, then we have $[H_{TB,0},\bar{c}^{\dagger}_{\bsl{k}'=\bsl{k}} \bar{v}]=E \bar{c}^{\dagger}_{\bsl{k}} \bar{v}$ with $\bar{v}=(v^T,v^T)^T/\sqrt{2}$.
Therefore, at $\bsl{k}'=(\pi/ a_x', \pi/(4 a_y'),0)$, we can choose zero-energy eigenvectors for $H_{TB,0}$ with \eqnref{eq:n_values} as
\eqa{
v_1'=\frac{1}{\sqrt{2}}\mat{v_1 \\ v_1}\ ,\ v_2'=\frac{1}{\sqrt{2}}\mat{v_2 \\ v_2}\ ,\ v_1''=\frac{1}{\sqrt{2}}\mat{\tau_z\sigma_0 v_1^*\\-\tau_z\sigma_0 v_1^*}\ ,\ v_2''=\frac{1}{\sqrt{2}}\mat{\tau_z\sigma_0 v_2^*\\ - \tau_z\sigma_0 v_2^*}\ ,
}
where the form of $v_1''$ and $v_2''$ are determined by symmetries.
By projecting the whole Hamiltonian $H_{TB}$ to them, we have the following low-energy model to the leading order of $q$ and $u_{zz}$
\eq{
\left(
\begin{array}{cccc}
 \frac{q_{z}}{2} & \sqrt{2} q_{x}+2 \ii q_{y}+\frac{e^{\ii \frac{\pi}{4}}}{a_0} u_{zz} & \mu_1 +\ii \mu_2  & 0 \\
 \sqrt{2} q_{x}-2 \ii q_{y}+\frac{e^{-\ii \frac{\pi}{4}}}{a_0} u_{zz} & -\frac{q_{z}}{2} & 0 & \mu_1 +\ii \mu_2  \\
 \mu_1 -\ii \mu_2  & 0 & -\frac{q_{z}}{2} & \left(-\sqrt{2}\right) q_{x}-2 \ii q_{y}+\frac{e^{-\ii \frac{\pi}{4}}}{a_0} u_{zz} \\
 0 & \mu_1 -\ii \mu_2  & \left(-\sqrt{2}\right) q_{x}+2 \ii q_{y}+\frac{e^{\ii \frac{\pi}{4}}}{a_0} u_{zz} & \frac{q_{z}}{2} \\
\end{array}
\right)\ .
}
The $U(1)$ freedom \eqnref{eq:v_U1} can only rotate $\mu_1+\ii \mu_2$ to $(\mu_1+\ii \mu_2)e^{-\ii 2\varphi}$.
Compared with \eqnref{eq:L_WP}, \eqnref{eq:L_CDW_MF}, and \eqnref{eq:L_str} of the main text, we can get the parameter values listed in \eqnref{eq:TB_v}, \eqnref{eq:TB_xi}, and \eqnref{eq:TB_phi_0} of the main text.

\subsection{Calculation of the 2D Layered Currents}

As discussed in the main text, the strained-induced current distribution like \figref{fig:TB_response}(a) is calculated for a slab configuration of $H_{TB}$ with the open-boundary condition along $x$, labeled as
\eq{
H_{TB}^{slab}=\sum_{k_y',k_z'} \bar{c}^{\dagger}_{k_y',k_z'} h_{slab}(k_y',k_z',u_{zz}) \bar{c}_{k_y',k_z'}\ .
}
Here $\bar{c}^{\dagger}_{k_y',k_z'}$ includes the layer index $\bar{c}^{\dagger}_{k_y',k_z',l_x',i_x,i,s}$ with $l_x'=1,...,20$, and
\eq{
\left[h_{slab}(k_y',k_z',u_{zz})\right]_{l_{x,1}',l_{x,2}'}=\frac{1}{N_x'}\sum_{k_x'}e^{\ii (l_{x,1}'-l_{x,2}') a_x' k_x'} \mat{1 & \\ & e^{\ii k_x' a_x'/2}} \left[\bar{h}_{TB,u}(\bsl{k}')+h_{CDW}(\bsl{k}')\right]\mat{1 & \\ & e^{-\ii k_x' a_x'/2}}\ .
}
$H_{TB}^{slab}$ is effectively a 2D system (with two well-defined momenta), and thus we can calculate its 2D piezoelectric coefficient according to
\eqa{
\label{eq:gamma_bc}
\chi_{izz}^{2D}=-e\int \frac{d^2k'}{(2\pi)^2} \sum_{n\in\ occupied}\left. F_{k_i', u_{zz}}^n\right|_{u_{zz}\rightarrow 0}\ ,
}
with
\eq{
\label{eq:F}
F_{k_i', u_{zz}}^n=(-\ii)\left( \partial_{k_i'} V_{n,k_y',k_z',u_{zz}}\right)^{\dagger} \partial_{u_{zz}} V_{n,k_y',k_z',u_{zz}}-(k_i'\leftrightarrow u_{zz})
}
with $V_{n,k_y',k_z',u_{zz}}$ a eigenvector of $h_{slab}(k_y',k_z',u_{zz})$.
We can rewrite the expression into the Kubo formula form as
\eqa{
 \sum_{n\in\ occupied}F_{k_i', u_{zz}}^n= \sum_{n\in\ occupied,m\in\ empty}(-\ii)\frac{1}{(E_{n}-E_{m})^2} V_{n}^{\dagger} \partial_{k_i'}h_{slab} V_{m}  V_{m}^\dagger \partial_{u_{zz}} h_{slab} V_{n} -(k_i'\leftrightarrow u_{zz})= \Tr\left[\mathcal{F}_{k_i',u_{zz}}\right]\ ,
}
where
\eq{
\mathcal{F}_{k_i',u_{zz}}=\sum_{n\in\ occupied,m\in\ empty}(-\ii)\frac{1}{(E_{n}-E_{m})^2} (V_{n} V_{n}^{\dagger}) \partial_{k_i'}h_{slab} (V_{m}  V_{m}^\dagger) \partial_{u_{zz}} h_{slab}  +h.c.\ .
}
As a result, the expression of $\chi_{izz}^{2D}$ (\eqnref{eq:PET_A} of the main text) can be rewritten as
\eqa{
\chi_{izz}^{2D}=-e\int \frac{d^2k'}{(2\pi)^2} \Tr\left[\mathcal{F}_{k_i',u_{zz}}\right]_{u_{zz}\rightarrow 0}\ .
}

The above form allows us to project the total piezoelectric constant into different layers as
\eq{
\chi_{izz}^{2D}(l_x')=-e\int \frac{d^2k'}{(2\pi)^2} \Tr\left[P_{l_x'}\mathcal{F}_{k_i',u_{zz}}\right]_{u_{zz}\rightarrow 0}\ ,
}
where
\eq{
[P_{l_x'}]_{l_{x,1}',l_{x,2}'}= \delta_{l_{x,1}',l_x'} \delta_{l_{x,2}',l_x'} \mathds{1}_{8\times 8}\ .
}
Clearly,
\eq{
\chi_{izz}^{2D}=\sum_{l_x'=1}^{N}\chi_{izz}^{2D}(l_x')\ ,
}
where $N$ is chosen to be 20 in our numerical calculations. Since the total piezoelectric current of the slab reads
\eq{
j^{2D,tot}_{i}= \chi_{izz}^{2D} \dot{u}_{zz}\ ,
}
the 2D current for each layer should read
\eq{
j^{2D}_{i}(l_x')= \chi_{izz}^{2D}(l_x')\dot{u}_{zz}\ .
}

\subsection{An Extra Term That Splits the Simultaneous Boundary Transitions}
The extra term $H_{extra}$ in the TB model that mentioned in the main text has the form:
\eq{
H_{extra}=\sum_{\bsl{k}'} \tilde{c}^{\dagger}_{\bsl{k}'} \frac{n_4}{a_0} \sin(k_y'a_y') \rho_0\tau_z\sigma_x\ ,
}
where we choose $n_4=0.2$ for the numerical calculation.
$H_{extra}$ preserves the TR and $m_y$ symmetries, as well as the effective $m_x$ symmetry at $\mu_2=0$.

\end{widetext}
\end{document}